\documentclass[aps,pre,showpacs]{revtex4}
\usepackage{amsmath}
\usepackage{amsfonts}
\usepackage{graphicx}

\begin{document}

\title{On the origin of thermality}

\author{Bernard S. Kay}
\email{bernard.kay@york.ac.uk}

\affiliation{Department of Mathematics, University of York, York YO10 5DD, UK}

\begin{abstract}
It is well-known that a small system weakly coupled to a large energy bath will, when the total system is in a microcanonical ensemble, find itself to be in an (approximately) thermal state (i.e.\ canonical ensemble) and, recently, it has been shown that, if the total state is, instead, a random pure state with energy in a narrow range, then the small system will still be approximately thermal with a high probability (defined by `Haar measure' on the total Hilbert space).  Here we ask what conditions are required for something resembling either/both of these `traditional' and `modern' thermality results to still hold when the system and energy bath are {\it of comparable size}.  In Part~1, we show that, for given system and energy-bath densities of states, $\sigma_{\mathrm{S}}(\epsilon)$ and $\sigma_{\mathrm{B}}(\epsilon)$, thermality does not hold in general, as we illustrate when $\sigma_{\mathrm{S}}$ and $\sigma_{\mathrm{B}}$ both increase as powers of energy, but that it does hold in certain approximate senses, in both traditional and modern frameworks, when $\sigma_{\mathrm{S}}$ and $\sigma_{\mathrm{B}}$ both grow as $e^{b\epsilon}$ or as $e^{q\epsilon^2}$ (for constants $b$ and $q$) and we calculate the system entropy in these cases. In their `modern' version, our results rely on new quantities, which we introduce and call the $S$ and $B$ `modapprox' density operators, which are defined for any positively supported, monotonically increasing, $\sigma_{\mathrm{S}}$ and $\sigma_{\mathrm{B}}$, and which,  we claim, will, with high probability, closely approximate the reduced density operators for the system and energy bath when the total state of system plus energy bath is a random pure state with energy in a narrow range.  In Part~2 we clarify the meaning of these modapprox density operators and give arguments for our claim.  

The prime examples of non-small thermal systems are quantum black holes.    Here and in two companion papers, we argue that current string-theoretic derivations of black hole entropy and thermal properties are  incomplete and, on the question of information loss, inconclusive.  However, we argue that these deficiencies are remedied with a modified scenario which relies on the modern strand of our methods and results here and is based on our previous {\it matter-gravity entanglement hypothesis}.
\end{abstract}


\maketitle

\section{\label{Sect:Intro} Introduction}

\subsection{\label{Sect:Background} Background}

This paper is concerned with the general question: ``How do physical systems get to be hot?''. By `hot' here, we do not simply mean `having lots of energy'. We shall reserve the word `energetic' for that. Rather, we mean the more specialized notion of being in what is known, in (quantum) statistical mechanics, as a Gibbs state, i.e.\ a state described by a density operator of form
\begin{equation}
\label{Gibbs}
\rho^{\mathrm{Gibbs}}_\beta=Z_\beta^{-1}e^{-\beta H}
\end{equation}
where $H$ is a suitable (usually, of necessity, approximate) Hamiltonian (assumed to have discrete spectrum) for the system and $\beta$ is related to the system's temperature, $T$, by $\beta=1/kT$ where $k$ is Boltzmann's constant (henceforth set to 1). Here $Z_\beta$ stands for ${\rm tr}(e^{-\beta H})$ and is the normalization constant which ensures that $\rho^{\mathrm{Gibbs}}_\beta$ will have unit trace. (When regarded as a function of $\beta$ it is, of course, the system's `partition function'.) Such states are also known as `canonical states' or `thermal equilibrium states' or `KMS states'. We shall sometimes refer to them simply as `thermal' states. A possible source of confusion here is the fact that it is sometimes found to be convenient to adopt the fiction that a system which is merely energetic is in a Gibbs state at a temperature chosen so as to give it the same mean energy. Additionally, given a system with a density of states $\sigma(\epsilon)$, it can sometimes be convenient to assign to it a `temperature', $T(\epsilon)$, at each energy, $\epsilon$, according to the formula $1/T(\epsilon)=d\log\sigma(\epsilon)/d\epsilon$ \cite{fictTemp}. We wish to underline that we shall not be concerned with such a fiction, nor with such an assignment of an energy-dependent `temperature', here. Rather we are interested in how systems get into states which are {\it actually} Gibbs states. In particular, we are interested in black bodies, and, more particularly, black holes (in suitable boxes; here we refer to the remarkable developments in `Euclidean Quantum Gravity' and in (Quantum) `Black Hole Thermodynamics' which arose from Hawking's pioneering work \cite{HawkingEvap} on `Black Hole Evaporation' -- see e.g.\ the papers on quantum black holes in the collections \cite{HawkingBBBH, GibHawkEuc}).

Of course, one way for a system to get into a Gibbs state is for it to be weakly coupled to a (much larger) heat bath which is already in a Gibbs state at the desired temperature. There is a considerable literature, which, with varying degrees of mathematical rigour and generality, shows that, as one might expect, a typical such system will, more or less irrespective of its initial state, approximately get into a Gibbs state at the same temperature at late times -- see e.g.\ \cite{FordKacMazur}, \cite{DaviesOpen}. However, what we are really interested in when we ask our general question ``How do physical systems get to be hot?'' is:

\smallskip
\noindent
\textit{``How does any physical system ever get to be hot in the first place?''}

\smallskip

Obviously, an explanation of how one system gets to be hot which invokes the existence of another system (the above-mentioned heat bath) which is assumed already to be hot can't help to answer this version of our question!

Another traditional explanation for the propensity of some systems to be in Gibbs states goes along the following lines (see e.g.\ \cite{FeynmanStat} and, for a treatment of some of the related mathematical aspects, e.g.\ \cite{Thirring} as well as the paper \cite{GoldsteinLebowitzetal} which recalls this traditional explanation as a preliminary to its main purpose -- for which see below): One assumes one's system of interest, say described by a Hamiltonian, $H_{\mathrm{S}}$, on a Hilbert space, ${\cal H}_{\mathrm{S}}$, to be weakly coupled to a much larger `energy bath', with Hamiltonian, $H_{\mathrm{B}}$, on a Hilbert space, ${\cal H}_{\mathrm{B}}$ -- both Hamiltonians being assumed to have a finite number of energy levels in any finite energy interval, with the number of states of the energy bath in an energy interval, $\delta$, being approximately given in terms of a `density of states', $\sigma_{\mathrm{B}}$, as $\sigma_{\mathrm{B}}(\epsilon)\delta$ -- $\sigma_{\mathrm{B}}$ being assumed to have some typical, say, power-law form (see below) -- and one assumes the whole system to be in a total microcanonical state.  Before we explain what we mean by this, we pause to remark, first, that, in order to avoid ambiguous usages of the word `system', we shall, from now on, adopt the word {\it totem} (short for `total system') to denote what we referred to above as our `whole system'. So we shall talk about a `totem' which consists of a `system'. `S', and an `energy bath', `B'.   Our assumption of weak coupling is then the assumption that the totem Hamiltonian will take the form
\begin{equation}
\label{coupling}
H=H_{\mathrm{S}} \otimes 1 + 1 \otimes H_{\mathrm{B}} + \hbox{weak coupling term}
\end{equation}
on the totem Hilbert space, ${\cal H}={\cal H}_{\mathrm{S}} \otimes {\cal H}_{\mathrm{B}}$, and we shall assume further that the coupling term is so weak that it can be neglected for state and energy-level counting purposes.  To say that our totem is in a microcanonical state then means to assume it is described by the density operator
\begin{equation}
\label{micro}
\rho_{\mathrm{microc}}=M^{-1}\sum |\epsilon\rangle\langle\epsilon|
\end{equation}
on the totem Hilbert space, $\cal H$, where the sum is over a basis of energy eigenstates for the subspace of $\cal H$ consisting of energy levels with energies in an interval, $[E, E+\Delta]$, which is small, yet large enough for the total number of totem energy eigenstates in this range to be very large, while the normalization constant, $M$ (which is expected to roughly scale with $\Delta$) is the total number of such basis eigenstates.  We further pause to note that we shall assume throughout the present paper, as is usually assumed for `ordinary' physical systems, that both Hamiltonians, $H_{\mathrm{S}}$ and $H_{\mathrm{B}}$, are positive and their densities of states monotonically increasing. We remark though that, as we will discuss further in Section \ref{Sect:moretherm}, were any of these assumptions to be relaxed, then the prospects for systems to become hot become much less constrained and, in particular, there are ways in which a system can be hot while the totem is in a pure state which differ from the `modern' scenarios we discuss below. 

Proceeding with the above assumptions, the states, $|\epsilon\rangle$, in the sum in (\ref{micro}) will each take the form $|\epsilon_{\mathrm{S}}\rangle \otimes |\epsilon_{\mathrm{B}}\rangle$ and the sum over totem energy levels will become (see (\ref{microsysbath}) below) a double sum over system energy levels, $\epsilon_{\mathrm{S}}$, and energy-bath energy levels, $\epsilon_{\mathrm{B}}$, which satisfy the condition $\epsilon_{\mathrm{S}}+\epsilon_{\mathrm{B}} \in [E, E+\Delta]$. The resulting state of the system is then represented mathematically, in the usual way, by the reduced density operator, $\rho^{\mathrm{microc}}_{\mathrm{S}}$ on ${\cal H}_{\mathrm{S}}$ i.e.\ by the partial trace of $\rho_{\mathrm{microc}}$ over ${\cal H}_{\mathrm{B}}$.

To remind ourselves how thermality of our system can then come about in this traditional explanation, it is instructive first to consider an oversimplified model in which our system Hilbert space, ${\cal H}_{\mathrm{S}}$, is two-dimensional with only two energy levels with energies $\epsilon^1_{\mathrm{S}}$ and $\epsilon^2_{\mathrm{S}}$ such that $\epsilon^2_{\mathrm{S}}-\epsilon^1_{\mathrm{S}} \gg \Delta$ and in which the density of states, $\sigma_{\mathrm{B}}$, of the energy bath grows exponentially -- we shall write $\sigma_{\mathrm{B}}(\epsilon)=ce^{b\epsilon}$. (We shall discuss the case where both system and energy bath both have such a density of states in Sections \ref{Sect:Exp} and \ref{Sect:ExpForm}.)

Then we easily see that $\rho^{\mathrm{microc}}_{\mathrm{S}}$ will be approximately
\[
\rho^{\mathrm{microc}}_{\mathrm{S}}=n^{-1}(c\Delta e^{b(E-\epsilon_1)}
|\epsilon^1_{\mathrm{S}}\rangle\langle\epsilon^1_{\mathrm{S}}|+
c\Delta e^{b(E-\epsilon_2)} |\epsilon^2_{\mathrm{S}}\rangle\langle\epsilon^2_{\mathrm{S}}|)
\]
where $n$ denotes the appropriate normalization constant, and this is clearly the same as the Gibbs state
\[
\rho_\beta=Z_\beta^{-1}(e^{-\beta\epsilon_1}
|\epsilon^1_{\mathrm{S}}\rangle\langle\epsilon^1_{\mathrm{S}}|+
e^{-\beta\epsilon_2} |\epsilon^2_{\mathrm{S}}\rangle\langle\epsilon^2_{\mathrm{S}}|)
\]
for $\beta=b$ for a suitable, normalizing, $Z_\beta$.

In the full story, where we now assume that also the states of the system are approximately given by a density of states, $\sigma_{\mathrm{S}}$, it is convenient to assume that $E$ is an integral multiple of $\Delta$ and locally to slightly distort the spectra of system and energy bath so that their energy levels are evenly spaced at intervals $\Delta, 2\Delta, \dots, E$ with each system level having degeneracy
\begin{equation}
\label{degdensS}
n_{\mathrm{S}}(\epsilon)=\sigma_{\mathrm{S}}(\epsilon)\Delta
\end{equation}
and each energy-bath level having degeneracy
\begin{equation}
\label{degdensB}
n_{\mathrm{B}}(\epsilon)=\sigma_{\mathrm{B}}(\epsilon)\Delta.
\end{equation}
If, as we shall further assume, this can be done in such a way as to maintain the same `smoothed out' densities of states, then it will not seriously alter the values of any quantities of interest.  Choosing a basis within the degeneracy subspace of ${\cal H}_{\mathrm{S}}$ with each energy, $\epsilon$, and labelling its elements $|\epsilon, i\rangle$, where, for each $\epsilon$, $i=1, \dots, n_{\mathrm{S}}(\epsilon)$ while $\epsilon$ ranges from $\Delta$ to $E$ in integer steps of $\Delta$ (and similarly for the energy bath) we then easily have that $\rho_{\mathrm{microc}}$ (\ref{micro}) can be rewritten as
\begin{equation}
\label{microsysbath}
\rho_{\mathrm{microc}}=
M^{-1}\sum_{\epsilon_{\mathrm{S}}}\sum_{\epsilon_{\mathrm{B}}}\sum_i\sum_j
|\epsilon_{\mathrm{S}}, i\rangle \otimes |\epsilon_{\mathrm{B}} , j\rangle
\langle \epsilon_{\mathrm{S}}, i|\otimes\langle \epsilon_{\mathrm{B}}, j|
\end{equation}
where the sum over $i$ goes from $1$ to $n_{\mathrm{S}}(\epsilon_{\mathrm{S}})$, the sum over $j$ goes from $1$ to $n_{\mathrm{B}}(\epsilon_{\mathrm{B}})$ and the sums over $\epsilon_{\mathrm{S}}$ and $\epsilon_{\mathrm{B}}$ are over values which are positive-integer multiples of $\Delta$ and are constrained to have $\epsilon_{\mathrm{S}} + \epsilon_{\mathrm{B}} = E$, while the normalization constant, $M$, defined after (\ref{micro}), is also given by
\begin{equation}
\label{sumnorm}
M=\sum_{\epsilon=\Delta}^E n_{\mathrm{S}}(\epsilon)n_{\mathrm{B}}(E-\epsilon)
\end{equation}
or, roughly equivalently \cite{spurious}, by making the replacement
\begin{equation}
\label{continuum}
\sum_{\epsilon=\Delta}^E \ \  \hbox{by} \ \ \Delta^{-1}\int_0^E d\epsilon
\end{equation}
by the approximate formula
\begin{equation}
\label{intnorm}
M=\Delta\int_0^E \sigma_{\mathrm{S}}(\epsilon)\sigma_{\mathrm{B}}(E-\epsilon)d\epsilon.
\end{equation}
Moreover, $\delta/\Delta$ times the summand in (\ref{sumnorm}) or $\delta\Delta$ times the integrand in (\ref{intnorm}) is, for suitable (small but not too small) $\delta$ (approximately) the number of energy eigenstates for which the energy of the totem lies in the interval $[E, E+\Delta]$ while the energy of the system lies in the interval $[\epsilon, \epsilon + \delta]$. When our totem is in the microcanonical state (\ref{micro}), (\ref{microsysbath}), this summand divided by $M$ may thus be interpreted as the probability that the system energy lies in this latter interval. We shall denote it by $P_{\mathrm{S}}(\epsilon)\delta$ and call $P_{\mathrm{S}}(\epsilon)$ the system's \textit{energy probability density} so we have
\begin{equation}
\label{enprobdens}
P_{\mathrm{S}}(\epsilon)=
\frac{\Delta}{M}\sigma_{\mathrm{S}}(\epsilon)\sigma_{\mathrm{B}}(E-\epsilon)
\simeq \frac{1}{M\Delta}n_{\mathrm{S}}(\epsilon)n_{\mathrm{B}}(E-\epsilon),
\end{equation}
and we notice, in passing, that
\[
P_{\mathrm{B}}(\epsilon)=P_{\mathrm{S}}(E-\epsilon).
\]
The reduced density operator, $\rho^{\mathrm{microc}}_{\mathrm{S}}$, of  $\rho_{\mathrm{microc}}$ on ${\cal H}_{\mathrm{S}}$ will clearly be
\begin{equation}
\label{microreduced}
\rho^{\mathrm{microc}}_{\mathrm{S}}=M^{-1}\sum_{\epsilon=\Delta}^E
n_{\mathrm{B}}(E-\epsilon) \sum_{i=1}^{n_{\mathrm{S}}(\epsilon)}
|\epsilon, i\rangle\langle \epsilon, i|.
\end{equation}
(Here and below, to avoid cluttering up our formulae, we drop the `s' suffix on $\epsilon$ -- also in $|\epsilon, i\rangle$ -- when there can be no ambiguity.)

One can then show, for a wide range of `realistic' energy-bath models that, in the limit as the energy bath gets large while the system remains unchanged, $\rho_{\mathrm S}^{\mathrm{microc}}$ will converge to a thermal state at an inverse temperature $\beta$ given, \cite{Ftnt4}, in terms of the large-size behaviour of the energy bath's density of states.

In particular, and specializing now to a case (cf.\ again e.g.\ \cite{FeynmanStat}) that will interest us further below, if the density of states, $\sigma_{\mathrm{B}}$, has the typical power-law form of ordinary (radiationless) matter:
\begin{equation}
\label{sigmabpower}
\sigma_{\mathrm{B}}(\epsilon) = A_{\mathrm{B}}\epsilon^{N_{\mathrm{B}}},
\end{equation}
where $A_{\mathrm{B}}$ is a constant and $N_{\mathrm{B}}$ is an `Avogadro-sized' number which could stand e.g.\ for `3/2 times the number of molecules' in the energy bath or the `number of oscillators' in the energy bath (see again e.g.\ \cite{FeynmanStat} for the origin of the $3/2$ etc.)\ etc.\ then, in the limit as the total energy, $E$, of the totem gets larger while the size of the energy bath gets larger -- in the sense that $N_{\mathrm{B}}$ gets larger -- while the system remains unaltered and $N_{\mathrm{B}}/E$ converges to a constant, $\beta$, $\rho^{\mathrm{microc}}_{\mathrm{S}}$ will converge to a thermal state at inverse temperature $\beta$ -- i.e.\ to the $\rho^{\mathrm{Gibbs}}_{{\mathrm S},\beta}$ of Equation (\ref{sGibbs}) below. In the special case that the system has a density of states also of power-law form (see (\ref{sigmasbpower})) we shall provide a proof of this result, in passing, in Section \ref{Sect:Power} below which is particularly instructive in relation to our present purposes. See the last paragraph in Section \ref{Sect:Power}. So, in this way, one shows that a small system in contact with a large energy bath with a suitable density of states will approximately be in a Gibbs state when the totem is in a microcanonical state.

Above, a Gibbs state (\ref{Gibbs}) of our system will obviously take the form (assuming again the spectrum to be slightly distorted as explained before equation (\ref{microsysbath}))
\begin{equation}
\label{sGibbs}
\rho^{\mathrm{Gibbs}}_{{\mathrm S},\beta}=Z_{S,\beta}^{-1}
\sum_{\epsilon=\Delta}^\infty e^{-\beta\epsilon} \sum_{i=1}^{n_{\mathrm{S}}(\epsilon)}
|\epsilon,i\rangle\langle\epsilon,i|
\end{equation}
where (approximating the obvious sum by an integral as we did when we passed from (\ref{sumnorm}) to (\ref{intnorm}))
\begin{equation}
\label{Zs}
Z_{S,\beta}=\int_0^\infty \sigma_{\mathrm{S}}(\epsilon)\exp(-\beta\epsilon)d\epsilon.
\end{equation}

However, this traditional explanation of the origin of thermality (of a small system) is also unsatisfactory since it still begs the question of how the totem got into a microcanonical state. What would really be desirable would be an explanation of the origin of thermality consistent with the basic assumption of standard quantum mechanics that the total state of a closed system (in our case, our totem) is a pure state -- i.e.\ in the language of density operators, the projector, $|\Psi\rangle\langle\Psi|$, onto a single vector, $\Psi$, in the closed system's (/our totem's) Hilbert space.

Such an explanation has, in fact, recently been given by a number of authors again for the case of a small system in contact with a large energy bath. See especially the paper \cite{GoldsteinLebowitzetal} entitled `Canonical Typicality' by Goldstein, Lebowitz et al.\ and also the references therein. The result of that paper -- when specialized to our power-law density of states model (\ref{sigmabpower}) -- amounts to the statement that if, for a `system' and `energy bath' as considered above, one takes a random pure state with energy in the energy range $[E, E+\Delta]$, then, again imagining the energy bath to get larger while $N_{\mathrm{B}}/E$ converges to $\beta$, for sufficiently large $E$, the reduced density operator of the system, $\rho^{\mathrm{modern}}_{\mathrm S}$, will, with very high probability, be very close to a Gibbs state (i.e.\ the $\rho^{\mathrm{Gibbs}}_{{\mathrm S},\beta}$ of (\ref{sGibbs})) at inverse temperature $\beta$.

We shall also re-obtain this result ourselves as a limiting case of one of our main new results in Section \ref{Sect:Ans}.

The precise mathematical statement can be inferred by inspecting the paper \cite{GoldsteinLebowitzetal} and/or see the more general rigorous result proved by Popescu et al.\ \cite{Popescuetal}.

Goldstein, Lebowitz et al.\ define what they mean here by `random' and by `probability' by taking the natural measure on the set of unit vectors of the relevant Hilbert space -- assumed to have large, but finite, dimension $M$ -- by thinking of it as a ($2M-1$)-dimensional real unit sphere and taking the natural invariant measure induced on that by Haar measure on the orthogonal group. In doing so, they follow pioneering work of Lubkin \cite{Lubkin} who, in 1978, after introducing \cite{Ftnt1} this use of this measure (following Lubkin and subsequent authors, we shall simply call it `Haar' measure from now on) showed that a randomly chosen pure density operator, $\rho^{mn}=|\Psi\rangle\langle\Psi|$ (without any restriction on energy or anything else) on the tensor-product Hilbert space, ${\cal H}_m\otimes {\cal H}_n$, of a pair of quantum systems -- ${\cal H}_m$ being $m$-dimensional and ${\cal H}_n$ being $n$-dimensional -- will, for fixed $m$ and $n\gg m$, have, with high probability, a reduced density operator, $\rho^{mn}_m$, on ${\cal H}_m$, which is close to the maximally mixed density operator -- with components, in any Hilbert space basis, ${\rm diag}(1/m, \dots, 1/m)$. We shall discuss further this result of Lubkin and some related developments in Section \ref{Sect:Prelim} at the beginning of Part~2 since they will be needed as a preliminary towards our argument for Equation (\ref{purereduced}) and the related claimed proposition in Section \ref{Sect:Ans}.

In essence, one might characterize the relation between Lubkin's work and the work, \cite{GoldsteinLebowitzetal}, of Goldstein, Lebowitz et al.\ by saying that Lubkin obtained microcanonicality of a small subsystem from randomness of a totem pure state while Goldstein, Lebowitz et al.\ obtained canonicality of a small subsystem when an, otherwise random, totem pure state is constrained to have a definite energy. (Popescu et al.\ \cite{Popescuetal} then generalized these developments by allowing for more general constraints, and also made them mathematically rigorous.)

The modern (see Endnote \cite{modern}) results, \cite{GoldsteinLebowitzetal, Popescuetal}, of Goldstein, Lebowitz et al.\ and of Popescu et al.\ are an advance on the traditional results in that they replace the assumption of a total microcanonical state by the assumption of a total pure state. However, they still share the limitation of the traditional approach of still only being capable of explaining how, at most, only a small subsystem of a given `large' totem can get to be (approximately) thermal. The main purpose of the present paper will be to explore to what extent, and/or under what altered circumstances, this limitation can be overcome. Our main motivation relates to the theory of quantum black holes. Black holes are a puzzle in relation to the above results if one believes, as seems compelling, that the totem consisting of a black hole in equilibrium with its atmosphere in a box at approximately fixed energy is completely (approximately) thermal \cite{Ftnt5}.

\subsection{{\label{Sect:BlackHoles}}Quantum black holes}

In such black hole equilibrium states we may roughly (albeit not exactly, see Endnote (iii) in \cite{KayAbyaneh}) identify the black hole itself with `gravity' and the atmosphere with `matter'.  In an earlier proposal  (see \cite{Kay1, Kay2} and especially Endnotes (i), (ii), (iii) and (v) in \cite{KayAbyaneh})  of the author (which predated the work \cite{GoldsteinLebowitzetal, Popescuetal} in a more general, but non-gravitational \cite{Ftnt6}, context of Goldstein, Lebowitz et al.\ and of Popescu et al.\ by around seven years) a radically-different-from-usual hypothesis was put forward as to the nature of quantum black hole equilibrium states according to which the total state is a pure state (in line with what we are calling here the `modern' approach -- see Endnote \cite{modern} -- but in contrast to the usual assumption in work on quantum black holes that it is a Gibbs state at the Hawking temperature) while the reduced state of the gravitational field alone and also the reduced state of the matter fields alone are each thermal (i.e.\ each Gibbs states) at the appropriate Hawking temperature (see below). (Here we use the word `matter' to include e.g.\ the electromagnetic field.) Below, we shall sometimes call such a total pure state {\it bithermal}. This hypothesis formed, in turn, just a part of our wider hypothesis \cite{Kay1, Kay2, KayAbyaneh} (which we shall sometimes refer to here as our \textit{matter-gravity entanglement hypothesis}) according to which, quite generally, one should always take into account the quantum gravitational field as well as all matter fields in describing the full dynamics of any physically closed totem, and that, while the state of the totem is always pure and evolves unitarily, the `physically relevant' quantum state is to be identified with the reduced density operator of the matter alone and, concomitantly (see Section \ref{Sect:Entropy} and, in particular, Endnote \cite{Ftnt3}), the physical entropy of a closed totem is to be identified with its {\it matter-gravity entanglement entropy}. Interpreted according to this wider hypothesis, our hypothesis that quantum black hole equilibrium states are bithermal then implies that, {\it physically}, such states are {\it completely thermal}. We remark that, given our wider hypothesis, what is required for this complete thermality is, of course, just thermality of the reduced state of the matter. However, there are strong reasons (particularly the fact \cite{GibHawkEuc} that the Euclideanized Schwarzschild metric is periodic in imaginary time with period $8\pi G{\cal M}$) for believing that the mathematical nature of the reduced state of gravity will also be thermal and this is what we have assumed above and will continue to assume in the remainder of this section and in Section  \ref{Sect:BHimplications}. 

To summarize and also to recall the relevant formulae: While we accept the (conventional) belief that, in black hole equilibria, both matter and gravity are each separately thermal at the Hawking temperature, $T_{\mathrm H}$, we propose (unconventionally in comparison to other work on quantum black holes) that the total state of matter-gravity is pure (rather than itself being a thermal state).  The thermality of each of the reduced states (i.e.\ of matter and of gravity separately) will then arise as the result of entanglement between matter and gravity in the pure totem state.  We shall refer to this picture of black hole equilibrium states as our \textit{entanglement picture of black hole equilibrium}.   (We shall assume in Section \ref{Sect:BHimplications} and in \cite{Kaycompanion, Kayprefactor} that, in this picture, the overall (i.e.\ totem) state of black hole equilibrium is not only pure but also close to an energy eigenstate.)  We further emphasize that while this proposal is unconventional when compared to other work on quantum black holes, it seems to fit well with modern approaches (such as those of \cite{GoldsteinLebowitzetal, Popescuetal}) towards understanding the origin of thermality which have recently been proposed in non-gravitational contexts.  Here, we recall that the Hawking temperature, $T_{\mathrm H}$, is given \cite{HawkingBBBH, GibHawkEuc}, in the case of a Schwarzschild (i.e.\ spherical, uncharged) black hole of mass $\cal M$, by $T_{\mathrm H}=1/8\pi G{\cal M}$ (in general the surface gravity multiplied by $2\pi$). Here, $G$ denotes Newton's constant and we set $c$ and $\hbar$ to 1. Moreover, we accept the conventional belief that the physical entropy -- again in the spherical, uncharged case -- has the Hawking value of $4\pi G{\cal M}^2$ (in general, one quarter of the area of the event horizon, divided by $G$) and what is new about our proposal is our claim that this entropy-value should ultimately be explainable as the matter-gravity entanglement entropy of a pure state of the overall matter-gravity totem.

Finally, we note that our matter gravity entanglement hypothesis and our entanglement picture of black hole equilibrium also offer a natural resolution to the Information Loss Puzzle \cite{HawkingInfo}.  This puzzle arose because, as long as it was believed that black holes were correctly described by mixed states, then, in a dynamical process in which black holes were formed from collapsing stars etc., it appeared that an initial pure state would dynamically evolve into a mixed state, contradicting unitarity.  On the other hand, there is no difficulty in reconciling our matter-gravity entanglement hypthesis with a unitary quantum mechanical time evolution and, once we identify entropy as matter-gravity entanglement entropy, this is entirely consistent with increasing entropy (i.e.\ information loss).  We note that this proposed resolution to the Information Loss Puzzle is, in fact, just a special case of our proposed resolution to the Second Law Puzzle \cite{Kay1, KayAbyaneh, Kaycompanion}.

\subsection{\label{Sect:Spec} Our specific question}

The specific question we shall endeavour to answer in this paper assumes, as its basic setting, that a totem be given which consists of a pair of weakly coupled systems, S and B, each with its own Hilbert space, ${\cal H}_{\mathrm{S}}$ and ${\cal H}_{\mathrm{B}}$, and each with its own density of states, $\sigma_{\mathrm{S}}$ and $\sigma_{\mathrm{B}}$.

 Our specific question is then:

\smallskip

\textit{If the systems, S and B, are} {\bf of comparable size} \cite{comparable}{\it ,} \textit{what modifications need to be made either to the traditional `total microcanonical state' approach or, more relevantly since we believe it to be a step closer to the right answer, to the more modern `total pure state' approach of Goldstein, Lebowitz et al.\ and of Popescu et al.\ and others, as described above, so as to ensure that when the totem has a total state with energy in an interval $[E, E+\Delta]$, the reduced states of S and B will each likely be approximately thermal states? (and, in particular, in the `total-pure state approach', the total state will likely be approximately bithermal).}

\smallskip
\noindent
(What is meant here by `comparable size' has, of course, to be encoded into the functional form of the densities of states $\sigma_{\mathrm{S}}(\epsilon)$ and $\sigma_{\mathrm{B}}(\epsilon)$. How this is done will be clear from the specific examples we discuss.)

We hope the answers we obtain below may be of interest in their own right and that the formalism we deploy to answer them may find a variety of other applications.   But the immediate application we have in mind is to the theory of quantum black holes.   In Section \ref{Sect:BHimplications} and in our two companion papers, \cite{Kaycompanion, Kayprefactor}, we shall argue that our answers help to strengthen the case for, and give concrete form to, our matter-gravity entanglement hypothesis and particularly our entanglement picture of black hole equilibrium discussed in Section \ref{Sect:BlackHoles}.

\subsection{\label{Sect:Ans} Answers}

The key to answering our specific question, in the  `traditional total  microcanonical state' approach is the formula (\ref{microreduced})  which we already gave above for the reduced density operator, $\rho^{\mathrm{microc}}_{\mathrm{S}}$, on S.

We claim that the appropriate replacement for this formula in the `modern total-pure state approach' is
\begin{equation}
\label{purereduced}
\rho^{\mathrm{modapprox}}_{\mathrm{S}}=M^{-1}\left(\sum_{\epsilon=\Delta}^{E_c}
n_{\mathrm{B}}(E-\epsilon)\sum_{i=1}^{n_{\mathrm{S}}(\epsilon)}
 |\epsilon, i\rangle\langle \epsilon, i|+
\sum_{\epsilon=E_c+\Delta}^E
n_{\mathrm{S}}(\epsilon)\sum_{i=1}^{n_{\mathrm{B}}(E-\epsilon)}
|\widetilde{\epsilon, i}\rangle\langle
\widetilde{\epsilon, i}| \right ).
\end{equation}
On the right hand side of this equation, we continue to assume the spectrum to be slightly distorted in the way we explained before equation (\ref{microsysbath}), $n_{\mathrm{S}}$ and $n_{\mathrm{B}}$ to be defined as in (\ref{degdensS}) and (\ref{degdensB}) and the sums to be over integral multiples of $\Delta$, and we also continue to assume, as will be the case in our examples in Part~1, that $\sigma_{\mathrm{S}}$ and $\sigma_{\mathrm{B}}$ are monotonically increasing functions -- defining $E_c$ to be the energy value at which $\sigma_{\mathrm{S}}(E_c)=\sigma_{\mathrm{B}}(E-E_c)$. When $\epsilon > E_c$, 
the $|\widetilde{\epsilon, i}\rangle$ then denote the elements of an orthonormal basis of an $n_{\mathrm{B}}(E-\epsilon)$-dimensional subspace of the ($n_{\mathrm{S}}(\epsilon)$-dimensional) energy-$\epsilon$ subspace of ${\cal H}_{\mathrm{S}}$ which will depend on $\Psi$. As we shall see, this dependence on $\Psi$ will not matter for the developments in Part~1. We will postpone a full explanation of the way in which the subspace depends on $\Psi$ to Section \ref{Sect:Main} in Part~2.

It is important to notice that, as is easy to check, the constant, $M$, by which one needs to divide in order to normalize (\ref{purereduced}) has the same value, given by (\ref{sumnorm}) and (\ref{intnorm}) (and as explained after those equations, equal to the total number of states of the totem with energy in the interval $[E, E+\Delta]$) as the constant, $M$, by which one needs to divide in order to normalize (\ref{microreduced}). Moreover, while the states, $\rho^{\mathrm{microc}}_{\mathrm{S}}$ and $\rho^{\mathrm{modapprox}}_{\mathrm{S}}$, are clearly (usually) very different, both states share the same energy probability density, $P_{\mathrm{S}}(\epsilon)$ (\ref{enprobdens}). (There is of course a similar pair of equations to (\ref{microreduced}) and (\ref{purereduced}) with obvious reversals of the letters `S' and `B' and, in the case of
(\ref{purereduced}), with $E_c$ replaced by $E-E_c$.)

We now claim that the sense in which (\ref{purereduced}) is the appropriate replacement for (\ref{microreduced}) in the modern approach is then made clear by the following proposition, our argument for the correctness of which is given in (and is the main purpose of) Part~2:

\smallskip
\noindent
{\bf Proposition.} \cite{rigour} \textit{For a given, randomly chosen, pure state, $\Psi$, on the Hilbert space of our totem, with energy restricted to be in the range $[E, E+\Delta]$, the reduced density operator, $\rho^{\mathrm{modern}}_{\mathrm S}$ of the system may, as far as physical quantities of interest are concerned, with very high probability, be considered to be very close to the $\rho^{\mathrm{modapprox}}_{\mathrm{S}}$ of {\rm (\ref{purereduced})} for the appropriate (i.e.\ to the chosen vector $\Psi$) $n_{\mathrm{B}}(E-\epsilon)$-dimensional subspaces of ${\cal H}_{\mathrm S}$ spanned by the $|\widetilde{\epsilon, i}\rangle$ (see above and Part~2). (And a similar statement of course holds with system, {\rm S}, replaced by bath, {\rm B}.)}

\smallskip

What makes this proposition particularly useful is the fact that, while the $n_{\mathrm{B}}(E-\epsilon)$-dimensional subspaces (spanned by the $|\widetilde{\epsilon, i}\rangle$) of ${\cal H}_{\mathrm S}$ will depend on the choice of $\Psi$ (in a way which we shall explain in Section \ref{Sect:Main} in Part~2 where we point out, by the way, that they might themselves be said to be `random subspaces') as is easy to see and as we shall illustrate in Part~1, the values of physical quantities of interest, such as the mean energy and the von Neumann entropy of the system S (see (\ref{SvN}) below and Section \ref{Sect:Entropy}) calculated using $\rho^{\mathrm{modapprox}}_{\mathrm{S}}$, do not depend on which $n_{\mathrm{B}}(E-\epsilon)$-dimensional subspaces they are. Therefore we can conclude that, to the extent that the approximation of $\rho^{\mathrm{modern}}_{\mathrm{S}}$ by $\rho^{\mathrm{modapprox}}_{\mathrm{S}}$ is good (and we shall argue in Part~2 that, in our situations of interest, and when it is used for the purpose of calculating mean energy and entropy, it is very good) the actual values of these quantities must (with a very high probability) be largely independent of the choice of $\Psi$!   (Aside from mean energy, in fact we expect the entire energy probability density function, $P_{\mathrm{S}}(\epsilon)$, will most likely be close to that of $\rho^{\mathrm{modapprox}}_{\mathrm{S}}$ and hence also, similarly, for higher moments of the energy.)

Above, we recall that, for an arbitrary density operator, $\rho$, the von Neumann entropy is given by the formula
\begin{equation}
\label{SvN}
S(\rho)=-{\rm tr}(\rho\log\rho).
\end{equation}

We remark that it is easy to see from a comparison between (\ref{microreduced}) and (\ref{purereduced}) that the above proposition implies the `Canonical Typicality' result \cite{GoldsteinLebowitzetal} of Goldstein, Lebowitz et al., thus fulfilling our promise in Section \ref{Sect:Intro} to re-obtain the latter. For, in the relevant limit (see after Equation (\ref{sigmabpower})) $E_c$ in (\ref{purereduced}) will tend to $E$ and therefore $\rho^{\mathrm{modapprox}}_{\mathrm{S}}$ (\ref{purereduced}) will tend to $\rho^{\mathrm{microc}}_{\mathrm{S}}$ (\ref{microreduced}) which, in turn, will tend, by the traditional argument we reviewed in Section \ref{Sect:Intro}, to a Gibbs state (namely the $\rho^{\mathrm{Gibbs}}_{{\mathrm S},\beta}$ of (\ref{sGibbs}) for $\beta$ equal to the limiting value of $N_{\mathrm{B}}/E$).

To start now to address our specific question, we first observe that, whatever the densities of states, $\sigma_{\mathrm{S}}$ and $\sigma_{\mathrm{B}}$ (provided only they are monotonically increasing) as long as the total energy, $E$, of our totem is finite, then, of course neither of the density operators, $\rho^{\mathrm{microc}}_{\mathrm{S}}$ (\ref{microreduced}) and $\rho^{\mathrm{modapprox}}_{\mathrm{S}}$ (\ref{purereduced}), can be exactly thermal. To see this easily, it suffices to notice that the energy probability density, $P_{\mathrm{S}}(\epsilon)$ (\ref{enprobdens}), which these states share will obviously be zero for $\epsilon > E$, whereas, when $\sigma_{\mathrm{S}}(\epsilon)$ is sufficiently slowly growing for $\rho^{\mathrm{Gibbs}}_{{\mathrm S},\beta}$ (see (\ref{sGibbs})) to exist, the energy probability density for the Gibbs state (\ref{sGibbs}) will obviously take the form
\begin{equation}
\label{PsGibbs}
P^{\mathrm{Gibbs}}_{{\mathrm S}, \beta}(\epsilon)=Z_{S,\beta}^{-1}
\sigma_{\mathrm{S}}(\epsilon)\exp(-\beta\epsilon),
\end{equation}
where $Z_{S,\beta}$ is as in (\ref{Zs}), which (for a rising density of states $\sigma_{\mathrm{S}}$) will be non-zero for
all $\epsilon$.  However, one can ask whether $\rho^{\mathrm{microc}}_{\mathrm{S}}$ and/or $\rho^{\mathrm{modapprox}}_{\mathrm{S}}$ can be {\it approximately thermal}, say at sufficiently low energies.

We shall find that, for physically ordinary densities of states such as (cf.\ the discussion around (\ref{sigmabpower}))
\begin{equation}
\label{sigmasbpower}
\sigma_{\mathrm{S}}(\epsilon) =
A_{\mathrm{S}}\epsilon^{N_{\mathrm{S}}}, \quad \sigma_{\mathrm{B}}(\epsilon) =
A_{\mathrm{B}}\epsilon^{N_{\mathrm{B}}},
\end{equation}
then, when system, S, and energy bath, B, are large and of comparable size -- i.e.\ when $N_{\mathrm{S}}$ and $N_{\mathrm{B}}$ are both large, but comparably sized numbers -- then neither $\rho^{\mathrm{microc}}_{\mathrm{S}}$ nor $\rho^{\mathrm{modapprox}}_{\mathrm{S}}$ can even be approximately thermal. In particular, this is the case when both system and energy-bath densities of states are identical (i.e.\ when $A_{\mathrm{S}} =A_{\mathrm{B}}$ and $N_{\mathrm{S}}=N_{\mathrm{B}}$). Rather, we will show that, when system and energy bath are of comparable size (or identical) in the sense just explained, the energy probability density of both S and B will, instead of having the behaviour one would expect of a thermal state, deviate from the most likely distribution of energies between S and B according to a Gaussian probability distribution with width of the order of $E$ divided by the square root of $N_{\mathrm{S}}$ (equivalently $N_{\mathrm{B}}$).

On the other hand, we shall show that in certain well-defined senses, `approximately thermal' states are obtained for system, S, and energy bath, B, both on the traditional total microcanonical state approach and also on the modern total pure state approach if they both have identical densities of states which either rise exponentially with energy or rise as `quadratic exponentials' -- i.e.\ each as the exponential of a constant times the square of the energy -- the notion of `approximately thermal' depending both on the approach (i.e.\ the traditional total microcanonical state approach or the modern total pure state approach) and also on the behaviour of the densities of states (i.e.\ on whether they rise as the exponential of energy or of energy squared). See especially the notions of `$E$-approximately thermal' and `$E$-approximately semi-thermal' introduced in Section \ref{Sect:Exp} for the case of an exponentially rising density of states. (The extent to which these results generalize to non-identical densities of states is briefly discussed for the exponential case in Endnote \cite{Ftnt8} to Section \ref{Sect:Exp}.)

\subsection{\label{Sect:Entropy} Results on the origin of entropy}

Although it is not indicated in our title, besides our main question concerning the origin of thermality, we shall be greatly concerned throughout the paper, with the origin of entropy.  And we are particularly interested in understanding how the very large entropies of black holes come about.

To this end, we will obtain formulae (Equations  (\ref{Smicroexp}), (\ref{Spureexp}) in Section \ref{Sect:ExpForm} and Equations (\ref{Smicrocexpesq}) and (\ref{Smodernexpesq}) in Section \ref{Sect:ExpEsq}) for the entropy of our system, S, on both traditional and modern approaches, when system and energy bath both have either identical exponential or identical quadratic exponential densities of states.  (We will also obtain formulae for the mean energy of S and B.)   In the traditional approach, this is simply the mean entropy of the reduced density operator of the system when the totem is in a microcanonical state with given energy, $E$. In the modern approach, we remark, first, that, for every pure totem state, whether or not S and B have identical densities of states, the system entropy is necessarily always equal to the energy-bath entropy and both of these quantities are, in fact, identical \cite{Ftnt3} with the $\{$system$\}$-$\{$energy bath$\}$ entanglement entropy.  Second, the value of the entropy in the modern case is to be interpreted, in the light of our proposition, as the value that the system entropy ($=$ energy-bath entropy $=$ $\{$system$\}$-$\{$energy bath$\}$ entanglement entropy) of a randomly chosen totem pure state will, with very high probability, be very close to. One of the most significant of our overall conclusions, dependent on our proposition, which we argue for in Part~2, is the fact that there is such a value at all -- i.e.\ the fact that, with our basic general assumptions and for system and energy-bath densities of states of the sorts we discuss, the vast majority of totem states will have a system entropy close to one single value, namely $-{\rm tr} (\rho^{\mathrm{modapprox}}_{\mathrm{S}}\log(\rho^{\mathrm{modapprox}}_{\mathrm{S}}))$.
In terms of the language of Quantum Information Theory, this may be stated in the following way (below we temporarily suspend our terminological conventions, calling both S and B `systems' and our `totem' the `total system'):

\smallskip
\noindent
{\it Given two comparably-sized large systems, (}S {\it and} B{\it ), which are either uncoupled or weakly coupled, then (for physically reasonable densities of states and even some maybe physically unreasonable ones) if their total state is a random pure state, their degree of entanglement  (as measured by their entanglement entropy, $S$) will, with high probability, be close to the single value}
$-{\rm tr}(\rho^{\mathrm{modapprox}}_{\mathrm{S}}\log(\rho^{\mathrm{modapprox}}_{\mathrm{S}}))$.

\smallskip

\noindent
{\it (Similarly, we expect that the mean value of the energy of the system,} S {\it , will, with high probability, be close to the single value $-{\rm tr} (\rho^{\mathrm{modapprox}}_{\mathrm{S}}H_{\mathrm{S}})$.  Indeed we expect the full energy probability density function, $P_{\mathrm{S}}(\epsilon)$ [and hence also other moments of the energy], of} S {\it to, be, with high probability, close to that of $\rho^{\mathrm{modapprox}}_{\mathrm{S}}$ [and similarly with} S {\it replaced by} B{\it ].)}

Our results are that, for a totem with total energy $E$, for identical exponentially rising densities of states, $\sigma_{\mathrm{S}}(\epsilon)=\sigma_{\mathrm{S}}(\epsilon)=ce^{b\epsilon}$, on the traditional approach, the entropy, $S^{\mathrm{microc}}_{\mathrm{S}}$, will be $bE/2$ (up to a logarithmic correction) while, on the modern approach, the entropy (i.e.\ the single value as discussed in the previous paragraph) $S^{\mathrm{modapprox}}_{\mathrm{S}}$, will be $bE/4$ (up to a logarithmic correction). For identical quadratic exponential densities of states, $\sigma_{\mathrm{S}}(\epsilon) = \sigma_{\mathrm{B}}(\epsilon)=Ke^{q\epsilon^2}$, we find that $S^{\mathrm{microc}}_{\mathrm{S}}=qE^2/2$ (up to a correction of order 1 in $E$), while $S^{\mathrm{modapprox}}_{\mathrm{S}}$ will be tiny (i.e.\  a term of order 1 in $E$). (In both traditional and modern cases and with both equal exponential and equal quadratic exponential densities of states the mean energy of both system and energy bath will, of course be $E/2$ -- in the modern case, `mean energy' here meaning the value, $-{\rm tr} (\rho^{\mathrm{modapprox}}_{\mathrm{S}}H_{\mathrm{S}})$, that the mean energy of a random pure totem state will most likely be very close to.)

\subsection{\label{Sect:Out} Outline of the rest of the paper}

We shall give full details of the results outlined in Section \ref{Sect:Ans} in Part~1, the main sections of which comprise Section \ref{Sect:Power}, which discusses the case where the density of states of both system and energy bath goes as a power of the energy,  Sections \ref{Sect:Exp} and \ref{Sect:ExpForm}, which discuss the exponential case,  and Section \ref{Sect:ExpEsq}, which discusses the quadratic exponential case.  Section \ref{Sect:Gen}  develops the mathematical formalism to enable efficient computation of the expected energy and entropy of system, S, and energy bath, B, for the states $\rho^{\mathrm{microc}}_{\mathrm{S}}$ and $\rho^{\mathrm{modapprox}}_{\mathrm{S}}$ and this formalism is applied in Sections \ref{Sect:ExpForm} and \ref{Sect:ExpEsq} to obtain formulae for these quantities in the cases of exponential and quadratic exponential densities of states.  

Two further sections, \ref{Sect:Morentropy} and \ref{Sect:moretherm}, discuss some further related matters and can be skipped on a first reading.  Section \ref{Sect:Morentropy} discusses the special features of the entropy, in both modern and microcanonical cases, when the densities of states of system and energy bath are such that the energy probability density (\ref{enprobdens}) is sharply peaked (as is, for example, the case for our power law densities of states) and  derives some general formulae which enable us, e.g.\ to calculate the entropy for the states considered in Section \ref{Sect:Power}.  In passing, we clarify the relation with some traditional work on the microcanonical ensemble (where peaks are normally presupposed) and dispel some myths.  We also discuss the connection between the sum of the entropies of the partial states of system and energy bath with the totem entropy $\log(M)$.  In Section \ref{Sect:moretherm} we point out that if some of our basic assumptions are relaxed, then the prospects for systems to become hot become much less constrained and, in particular, there are ways in which a system can be hot while the totem is in a pure state which differ from the `modern' scenarios we discuss below.  In particular, we discuss the notion of `purification' (closely related to `thermofield dynamics').

The entropy formulae we obtain in Sections \ref{Sect:Exp},  \ref{Sect:ExpForm} and \ref{Sect:ExpEsq} (as outlined at the end of Section \ref{Sect:Entropy}) will play an important role in Section \ref{Sect:BHimplications} and in two companion papers \cite{Kaycompanion, Kayprefactor} where we discuss the application of the ideas and formulae of these sections  to the theory of quantum black holes.   In Section \ref{Sect:BHquad}, we point out an intriguing resemblance between our entropy and temperature formulae for quadratic exponential densities of states in the microcanonical strand of Section \ref{Sect:ExpEsq} with Hawking's energy and temperature formulae for (Schwarzschild) quantum black holes and point out an apparent lack of success for the modern strand of Section \ref{Sect:ExpEsq} in modelling black holes.  However, we argue that it is difficult to conclude anything decisive from these observations since (at least in a description in terms of a quantized Einsteinian metric) black holes presumably do not satisfy the basic assumptions underlying our results here -- in particular our assumption (see Equation (\ref{coupling})) of weak coupling.

What seems more promising is a connection between the formulae and results for entropy and temperature which we obtain in Sections \ref{Sect:Exp} and \ref{Sect:ExpForm} for exponentially growing densities of states and scenarios in which quantum black holes are viewed as strong string-coupling limits of certain states of weakly coupled strings.  In Section \ref{Sect:Betterstring} and in our two companion papers, \cite{Kaycompanion} and \cite{Kayprefactor},  we recall some of the existing work \cite{Susskind, HorowitzPolchinski, StromingerVafa, HorowitzReview} in this direction, and point out that, despite its great computational success, what is computed in this work is the {\it degeneracy} of certain black hole states; the fact that the resulting degeneracy formulae happen to agree with the previously known values of black hole entropy does not seem to have been explained hitherto.   We then go on to propose a modification of the existing string theory scenario, and in particular of the work of Susskind \cite{Susskind} and Horowitz and Polchinski \cite{HorowitzPolchinski, HorowitzReview}  based on the modern strand of the present paper and on our matter-gravity entanglement hypothesis and our entanglement picture of black hole equilibrium (see Section \ref{Sect:BlackHoles}) .  We argue that this modified scenario, which is based on an understanding of black hole equilibrium states as strong string-coupling limits of equilibria involving a long string coupled to a stringy atmosphere, does offer an explanation of black hole entropy and thereby also a satisfactory resolution to the Information Loss Puzzle.  The companion paper \cite{Kaycompanion} gives a brief announcement of the main results of the present paper with a focus on the main results and formalism, as well as discussing further our matter-gravity entanglement hypothesis and outlining the application of that, with the results of Sections \ref{Sect:Exp} and \ref{Sect:ExpForm}, to this string scenario.  The further companion paper \cite{Kayprefactor} develops the string scenario further.

Part~2, which comprises Sections \ref{Sect:Prelim},  \ref{Sect:Haar}, \ref{Sect:Main} and \ref{Sect:Further}, clarifies the meaning of Equation (\ref{purereduced}) and presents our arguments in favour of our proposition in Section \ref{Sect:Ans}.  A fuller description of the contents of Part~2 is given towards the end of Section \ref{Sect:Prelim}.

\section*{\large Part 1: Results for power law, exponential and quadratic exponential (equal) densities of states}

\bigskip

\section{\label{Sect:Power} Power-law densities of states}

If S and B have densities of states as in (\ref{sigmasbpower}) then,
by (\ref{intnorm}) and the remarks in the subsequent paragraph, we have
that $M$, i.e.\ the total number of totem states with energy in $[E, E+\Delta]$,
is given by
\begin{equation}
\label{Mpowerintegral}
M=A_{\mathrm{S}}A_{\mathrm{B}}\Delta\int_0^E
\epsilon^{N_{\mathrm{S}}}(E-\epsilon)^{N_{\mathrm{B}}}d\epsilon
\end{equation}
which can be rewritten
\begin{equation}
\label{Mpower}
M=A_{\mathrm{S}}A_{\mathrm{B}}\Delta E^{N_{\mathrm{S}}+N_{\mathrm{B}}+1}
B(N_{\mathrm{S}}+1, N_{\mathrm{B}}+1)
\end{equation}
where $B(x,y)$ is the usual beta function (see e.g.\ \cite{Gradshteyn}) --
related to the gamma and factorial functions by
\begin{equation}
\label{betagamma}
B(x+1, y+1)=\frac{\Gamma(x+1)\Gamma(y+1)}{\Gamma(x+y+2)}=\frac{x!y!}{(x+y)!(x+y+1)}.
\end{equation}
(For fractional arguments, we take $x!$ to mean $\Gamma(x+1)$.)
On the other hand, the number of such totem states with system energy in an
interval $[\epsilon, \epsilon+\delta]$ will, for suitable $\delta$,
be well-approximated by
\[
P_{\mathrm{S}}(\epsilon)\delta=
A_{\mathrm{S}}A_{\mathrm{B}}\delta\Delta M^{-1}\epsilon^{N_{\mathrm{S}}}(E-\epsilon)^{N_{\mathrm{B}}}
\]
\begin{equation}
\label{Psdeltapower}
=A_{\mathrm{S}}A_{\mathrm{B}}\delta\Delta M^{-1} E^{N_{\mathrm{S}}+N_{\mathrm{B}}}
\left(\frac{\epsilon}{E}\right )^{N_{\mathrm{S}}}
\left(1-\frac{\epsilon}{E}\right )^{N_{\mathrm{B}}}.
\end{equation}
Thus, combining (\ref{Mpower}),
(\ref{Psdeltapower}) and (\ref{betagamma}) we have that
\begin{equation}
\label{Pspower}
P_{\mathrm{S}}(\epsilon)=\frac{N_{\mathrm{S}}+N_{\mathrm{B}}+1}{E}
b(N_{\mathrm{S}};N_{\mathrm{S}}+N_{\mathrm{B}},\frac{\epsilon}{E})
\end{equation}
where (see e.g.\ \cite{Feller})
\begin{equation}
\label{binomial}
b(k;n,p)=\frac{n!}{k!(n-k)!}p^k(1-p)^{n-k}
\end{equation}
is, when $n$ and $k$ are integers, the binomial distribution function
which has the famous interpretation as the probability that $n$ `Bernoulli'
trials, each with probability $p$ for success and $q=1-p$ for failure,
result in $k$ successes and $n-k$ failures.  In order to take advantage of the insight
afforded by this connection with probability theory we shall (with
negligible error when $N_{\mathrm{S}}$ and $N_{\mathrm{B}}$ are large) assume from now on that $N_{\mathrm{S}}$ and $N_{\mathrm{B}}$, if not
already integers, are replaced by their nearest integers.

First we notice that we may use the well-known connection between the
binomial and the Poisson distribution to give an alternative derivation
of the fact that, in the limit as $E$ and $N_{\mathrm{B}}$ grow while $N_{\mathrm{S}}$ remains
constant and the ratio $N_{\mathrm{B}}/E$ converges to $\beta$, S's energy
probability density, $P_{\mathrm{S}}(\epsilon)$ (\ref{Pspower}), converges to the
Gibbs energy probability density
$P^{\mathrm{Gibbs}}_{{\mathrm S},\beta}(\epsilon)$ (see (\ref{PsGibbs}) and
(\ref{Zs})) with inverse temperature $\beta=N_{\mathrm{B}}/E$ for
$\sigma_{\mathrm{S}}(\epsilon)$ as in (\ref{sigmasbpower}) -- the latter Gibbs
energy probability density being given explicitly by
\begin{equation}
\label{PsGibbspower}
P^{\mathrm{Gibbs}}_{S,\beta}(\epsilon)=
\frac{\beta^{N_{\mathrm{S}}+1}\epsilon^{N_{\mathrm{S}}}e^{-\beta\epsilon}}{N_{\mathrm{S}}!}
\end{equation}
as one sees from (\ref{PsGibbs}) and (\ref{sigmasbpower}) after easily
checking from (\ref{Zs}) and (\ref{sigmasbpower}) that
\[
Z_{S,\beta}=A_{\mathrm{S}}^{-1}\frac{N_{\mathrm{S}}!}{\beta^{N_{\mathrm{S}}+1}}.
\]
This convergence result is of course a special case (i.e.\ the case where
S, as well as B, has a power-law density of states) of an easy corollary both of the
traditional thermality result (on the total microcanonical state
approach) and (bearing in mind the equality of the energy probability
density for both (\ref{microreduced}) and (\ref{purereduced})) of
the `Canonical Typicality' result of Goldstein Lebowitz et
al.\ (on the `modern' total pure state approach) which, as we discussed in Section
\ref{Sect:Intro}, both  hold in the same limit; we shall see shortly that the
alternative proof which we next give for this corollary easily implies
an alternative proof to the traditional thermality result itself and
thus also, by a remark we made in Section \ref{Sect:Ans} to
an alternative argument for `Canonical Typicality' when the
system and energy-bath densities of states both have power-law form.

As Feller puts it in \cite{Feller}, ``If $n$ is large and $p$ is small, whereas the product
$\lambda=np$ is of moderate magnitude'' then the binomial distribution
goes over to the Poisson distribution, i.e.\
\begin{equation}
\label{Feller}
b(k;n,p) \simeq \frac{\lambda^k}{k!}e^{-\lambda}.
\end{equation}
In particular (cf.\ e.g.\ \cite{Bauer}) for fixed $k$, the right hand side
of (\ref{Feller}) is the limit of $b(n;k,p)$ as $n\rightarrow\infty$
while $p\rightarrow 0$ in such a way that $np\rightarrow \lambda$.  From
this, and (\ref{Pspower}), we easily conclude that the limit, as
$E\rightarrow\infty$ while $N_{\mathrm{B}}/E\rightarrow\beta$ with
$N_{\mathrm{S}}$ fixed, of $P_{\mathrm{S}}(\epsilon)$ is  equal to
$P^{\mathrm{Gibbs}}_{{\mathrm S},\beta}(\epsilon)$ (\ref{PsGibbspower}).  So, to
summarize, in the appropriate limit of a large energy bath, the energy
probability density of S goes over to the energy probability density of
the appropriate Gibbs state;
\begin{equation}
\label{ProbDensPowerGibbs}
P_{\mathrm{S}}(\epsilon) \rightarrow P^{\mathrm{Gibbs}}_{{\mathrm S},\beta}(\epsilon)=
\frac{\beta^{N_{\mathrm{S}}+1}\epsilon^{N_{\mathrm{S}}}e^{-\beta\epsilon}}{N_{\mathrm{S}}!}.
\end{equation}
We remark that, by inspecting (\ref{PsGibbs}) and (\ref{Zs}) this is easily seen to be
equivalent to the statement that, in the same limit,
\[
M^{-1}n_{\mathrm{B}}(E-\epsilon)\rightarrow Z_{S,\beta}^{-1}e^{-\beta\epsilon}
\]
and, by inspecting (\ref{microreduced}) and (\ref{sGibbs}), one easily
sees that this implies that in the same limit
\[
\rho_{\mathrm{S}}^{\mathrm{microc}}\rightarrow
\rho^{\mathrm{Gibbs}}_{{\mathrm S},\beta}
\]
thus providing the alternative proof, which we promised
in Section \ref{Sect:Intro}, of the traditional result on the thermality of a small
system in contact with an energy bath in the traditional limit of a
large energy bath, in the case where both the energy bath, B, and
system, S, have a power-law density of states (and by our remark in Section \ref{Sect:Ans}
thus also providing
an alternative proof of `Canonical Typicality' for such S and B).

We will now demonstrate, however, that, when S and B are {\it of
comparable size}, then, if they both have power-law densities of states
as in  (\ref{sigmasbpower}), both the total microcanonical state
approach and the `modern' approach (i.e.\ with a total pure state)
predict that the reduced density operators of each of S and B will be
quite different from thermal!   We shall show this by showing that the
energy probability density of each of S and B (which we again recall
from the paragraph after (\ref{purereduced}) is the same in each
approach) will have a quite different form from the thermal form of
$P^{\mathrm{Gibbs}}_{{\mathrm S},\beta}(\epsilon)$.

First we notice that, when $k$ is a fixed fraction, $pn$, of $n$ (in
such a way that $0<p<1$ and also $pn$ is an integer) then, if $p$ and
$n$ are regarded as fixed, the binomial distribution function
(\ref{binomial}) $b(pn;n,p')$ is maximized when $p'=p$ and we easily
obtain the approximation \cite{Ftnt7} (now writing $p'=p+x$)
\begin{equation}
\label{binomapprox}
b(pn;n,p+x)\simeq \frac{1}{\sqrt{2\pi
np(1-p)}}\exp\left(-\frac{nx^2}{2p(1-p)}\right ).
\end{equation}
(\ref{binomapprox}) is obtained by expressing the left hand side in
terms of factorials and powers according to (\ref{binomial}).  We then
adopt Stirling's approximation, $N!\simeq
\sqrt{2\pi}N^{N+\frac{1}{2}}e^{-N}$ for each of the factorials and,
introducing $q=1-p$, write the term $(p+x)^{np}(1-p-x)^{n(1-p)}$ as
$p^{np}q^{nq}$ times $(1+x/p)^{np}(1-x/q)^{nq}$ and approximate the
latter by $\exp(-nx^2/2pq)$.  Clearly, as long as $n$ is extremely large
and $p$ is not extremely close to zero or 1, then this will be an
excellent approximation.

Combining (\ref{Pspower}) with the definition of $p$ before (\ref{binomapprox}) we see
that, if we identify $n$ with $N_{\mathrm{S}}+N_{\mathrm{B}}$, then
\[
P_{\mathrm{S}}(\epsilon)= \frac{n+1}{E}\,b\left (pn;n,\frac{\epsilon}{E}\right )
\]
where
\begin{equation}
\label{pCN}
p=\frac{N_{\mathrm{S}}}{N_{\mathrm{S}}+N_{\mathrm{B}}}
\end{equation}
and that, provided $n$ is extremely large and S and B are of `comparable
size', which of course, in view of (\ref{pCN}), corresponds exactly to
$p$ not being extremely close to zero or 1, then, by
(\ref{binomapprox}), to a high degree of accuracy, we will have the
approximation
\begin{equation}
\label{Pspowerapprox}
P_{\mathrm{S}}(\epsilon)\simeq
\frac{1}{E}\sqrt{\frac{\gamma}{\pi}}
\exp\left(-\gamma\frac{(\epsilon-\epsilon_0)^2}{E^2}\right )
\end{equation}
i.e.\ a Gaussian with a peak located (See Section \ref{Sect:peak} for an alternative perspective on Equation (\ref{antipopular})) at
\begin{equation}
\label{antipopular}
\epsilon_0=pE \quad\hbox{($p$ as in (\ref{pCN}))} \quad =\frac{N_{\mathrm{S}}}{N_{\mathrm{S}}+N_{\mathrm{B}}}E
\end{equation}
and
\begin{equation}
\label{gamma}
\gamma=\frac{n}{2p(1-p)}=\frac{(N_{\mathrm{S}}+N_{\mathrm{B}})^3}{2N_{\mathrm{S}}N_{\mathrm{B}}}
\end{equation}
and there will of course be an obvious counterpart formula for the energy probability density, $P_{\mathrm{B}}$, of B, similar to the above formula but with $p$ replaced by $1-p$.  (This of course changes the value of $\epsilon_0$ but not of $\gamma$.) So the energy of S will be in a Gaussian band around a most likely energy of $\epsilon_0$, the energy of B will be in a Gaussian band around a most likely energy of $E-\epsilon_0$, each having the same width which will be $E$ divided by a number (i.e.\ $\sqrt{2\gamma}$) which is of the order of the square root of either of the (comparable!) numbers $N_{\mathrm{S}}$, $N_{\mathrm{B}}$.   Moreover it is easy to see that, in both the traditional microcanonical and the modern total pure state approaches,  the two energy probability densities will be perfectly anticorrelated -- i.e.\ when S has energy in a small interval around $\epsilon$, then B with have energy in a similar small interval around $E-\epsilon$.

Above, by `width' we mean the standard deviation, $\mathfrak s$, from the mean of the energy probability density, i.e.
\[
{\mathfrak s}=(\overline{\epsilon^2} - {\bar\epsilon}^2)^{1/2}
\]
where
\begin{equation}
\label{moments}
\overline{\epsilon^n}=\int_0^E \epsilon^n P_{\mathrm S}(\epsilon)d\epsilon
\end{equation}
($={\rm tr}(\rho^{\mathrm{microc}}_{\mathrm{S}}H_{\mathrm{S}}^n) = {\rm tr}
(\rho^{\mathrm{modapprox}}_{\mathrm{S}}H_{\mathrm{S}}^n$) -- cf.\ Section \ref{Sect:Gen}).

\begin{figure}
\includegraphics[width = 300pt, height = 300pt]{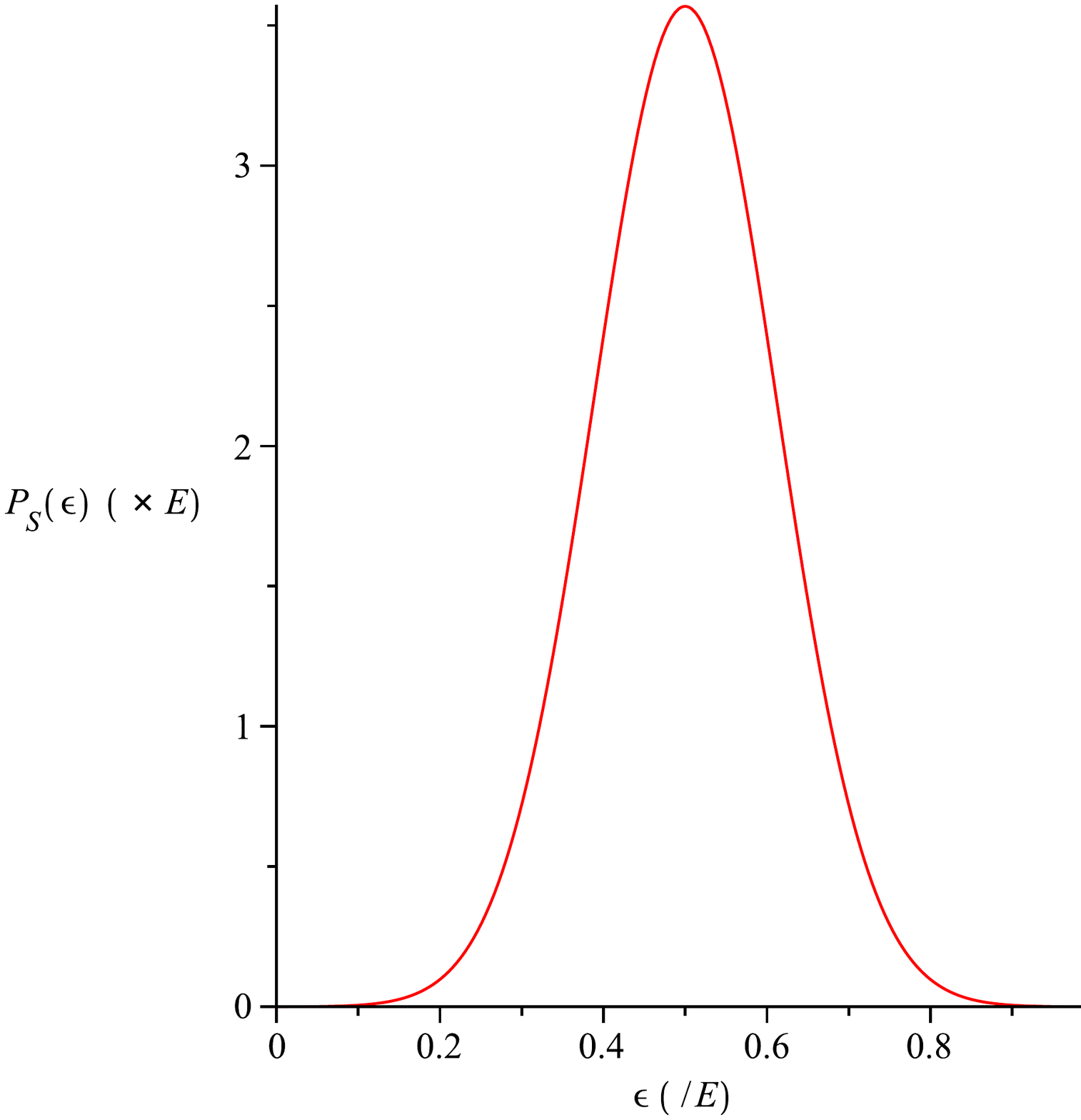}
\caption{\label{Fig1} Plot of the energy probability density (\ref{Pspower}), $P_S(\epsilon)$,
in the case S and B have the same density of states
$\sigma(\epsilon)=A\epsilon^{N}$ for the (`unusually' small) value $N=10$}
\end{figure}

\begin{figure}[ht]
\includegraphics[width = 300pt, height = 300pt]{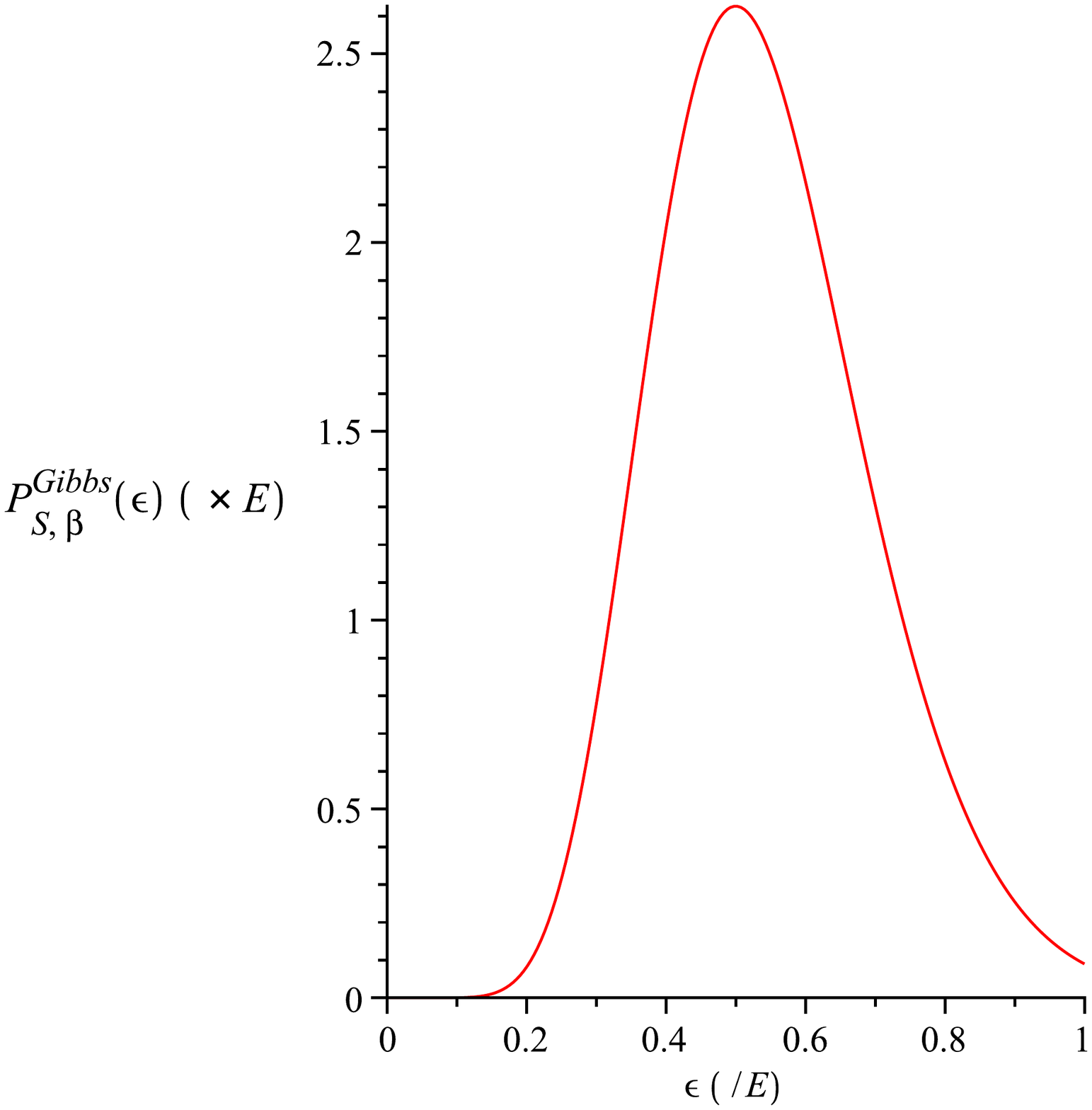}
\caption{\label{expGibbs} Plot of the energy probability density, $P^{\mathrm{Gibbs}}_{{\mathrm S},\beta}(\epsilon)$ for the thermal state at inverse temperature, $\beta$, on our system, S,  with density of states  $\sigma(\epsilon)=A\epsilon^N$, for the same (`unusually' small) value $N=10$ and for $\beta=22/E$ (i.e.\ the value of $\beta$ for which the mean energy is $E/2$). To be contrasted with the $P_S(\epsilon)$ of Figure \ref{Fig1}.}
\end{figure}

\noindent
(In the special case where $N_{\mathrm{S}}=N_{\mathrm{B}}=N$ say, one sees that $\gamma$ becomes $4N$ so the width, $\mathfrak s$, will be $E/2\sqrt{2N}$.) As we anticipated, this is a qualitatively very different behaviour from the energy probability density of thermal states and we conclude therefore, as promised, that, in both the traditional total microcanonical and the modern total pure state approaches, the reduced density operators of S and B must, when, S and B are of comparable size, be quite different in character from thermal density operators. To illustrate this point, we include a figure (Figure \ref{Fig1}) for the energy probability density, $P_S(\epsilon)$, in the case S and B have the same density of states $\sigma(\epsilon)=A\epsilon^N$ for the (unrealistically small) value $N=10$ and a comparison figure, Figure \ref{expGibbs}, showing the the energy probability density, $P^{\mathrm{Gibbs}}_{{\mathrm S},\beta}(\epsilon)$ for a thermal state at the inverse temperature, $\beta=2(N+1)/E$, chosen so that the mean energy takes the same value, $E/2$ -- again in the case $N=10$.

For the sake of a quantitative result, we note that, for general $N$, the width, $\mathfrak s$, of the energy probability density, $P^{\mathrm{Gibbs}}_{{\mathrm S},\beta}(\epsilon)$, of this comparison thermal state is (as is easily calculated) $E/2\sqrt{N+1}$ -- i.e.\ (to a very good approximation for large $N$) a factor of $\sqrt 2$ wider than the width of $P_S(\epsilon)$ while the height is (again by an easy calculation) a factor of $\sqrt 2$ smaller.

We shall postpone to Section \ref{Sect:Morentropy} a calculation of the (microcanonical and modern) entropies of S and B for general $N_{\mathrm S}$ and $N_{\mathrm B}$.  Suffice it to remark that, like the width, $\mathfrak s$, the microcanical entropy of S, differs, in general, from its value in the comparison thermal state at inverse temperature $\beta=2(N+1)/E$, albeit the difference is just a `small' constant (it is smaller by $\simeq\log 2/2$) independent of $\epsilon$.

Finally, we remark that, in this power-law density-of-states case, it
is clear from the developments in this section that the  `canonical' (i.e.\ thermal) behaviour
of $\rho^{\mathrm{microc}}_{\mathrm S}$ (or indeed of $\rho^{\mathrm{modapprox}}_{\mathrm
S}$) in the case that the system, S, is very much smaller than the
energy bath, B, may be reconciled with the above-discussed Gaussian
behaviour, when S and B are of comparable size, in that the relationship
between the two may be regarded as an instance of the well-known
relationship (see e.g.\ \cite{Feller} or \cite{Bauer})
between the Poisson and Gaussian distributions in probability theory.
(This obviously easily follows from the way we derived, above, both the
canonical behaviour and the Gaussian behaviour as limits of the binomial
distribution.)

\section{\label{Sect:Exp} Exponentially rising densities of states}

We now turn to discuss the quite different behaviour of the reduced
density operators $\rho^{\mathrm{microc}}_{\mathrm{S}}$ and
$\rho^{\mathrm{modapprox}}_{\mathrm{S}}$ when the densities of states of S and B increase
exponentially.  We shall confine our interest here to the case where
both densities of states, $\sigma_{\mathrm{S}}$ and $\sigma_{\mathrm{B}}$,
behave as $ce^{b\epsilon}$ with the same constants $c$ and $b$ in each expression:
\begin{equation}
\label{sigmasbexponential}
\sigma_{\mathrm{S}}(\epsilon) = ce^{b\epsilon}
, \quad \sigma_{\mathrm{B}}(\epsilon) = ce^{b\epsilon}.
\end{equation}
We remark, however, that, as may quite easily be checked, allowing
different values of $c$ (say $c_{\mathrm{S}}$ in the first formula and $c_{\mathrm{B}}$ in the second) will not essentially change our conclusions \cite{Ftnt8}.

The normalization constant $M$ is now easily seen -- either by using
(\ref{intnorm}) or, on recalling (\ref{degdensS}), by using (\ref{sumnorm}) -- to
be given by
\begin{equation}
\label{Mexp}
M=c^2e^{bE}E\Delta.
\end{equation}

We note that this will be large provided neither $c$ nor $\Delta$ are `too small' and provided also
\begin{equation}
\label{bEgg1}
bE\gg 1
\end{equation}
which will hold in cases of interest.

The formula, (\ref{microreduced}) for
$\rho^{\mathrm{microc}}_{\mathrm{S}}$ is then easily seen to coincide with the
formula, (\ref{sGibbs}), for a thermal density operator
$\rho^{\mathrm{Gibbs}}_{{\mathrm S},\beta}$, for the density of states
$\sigma_{\mathrm{S}}(\epsilon)$ as in (\ref{sigmasbexponential}) at inverse temperature
$\beta=b$, provided the latter formula is modified so that the sum over $\epsilon$
is truncated at the upper energy, $E$ and the partition function,
$Z_{S,\beta}$, is replaced by $cE$.  Of course, the un-truncated formula
(\ref{sGibbs}) will only make mathematical sense for $\beta > b$.
Nevertheless, the reduced density operator $\rho^{\mathrm{microc}}_{\mathrm{S}}$ (and similarly
also $\rho^{\mathrm{microc}}_{\mathrm{B}}$) clearly deserves to be called an
approximately thermal state at inverse temperature $b$.
(This will generalize from equal systems to
comparably sized systems if, by this, we mean systems with densities of
states with unequal $c_{\mathrm{S}}$ and $c_{\mathrm{B}}$ as discussed in
Endnote \cite{Ftnt8}).  We shall refer to the relevant notion of being
approximately thermal here as being \textit{$E$-approximately
thermal}.

Turning from the traditional total microcanonical state approach to the
modern total pure state approach, we see, on substituting
(\ref{sigmasbexponential}) into (\ref{purereduced}) and noting that
$E_c$ will obviously become $E/2$, that the $\epsilon$-summand in
(\ref{purereduced}) still agrees with the $\epsilon$-summand in
(\ref{sGibbs}) at inverse temperature $\beta=b$ up to energy $E/2$ and,
moreover, as always (cf.\ after Equation (\ref{purereduced})) the system
energy probability density of $\rho^{\mathrm{modapprox}}_{\mathrm{S}}$ is equal to
that of $\rho^{\mathrm{microc}}_{\mathrm{S}}$ and thus it agrees with the
energy probability density of a Gibbs state, for the same density of states, up to energy 
$E$.  We shall refer to the relevant notion of being approximately thermal here 
(i.e.\ agreement of the summand in the formula (\ref{purereduced}) with the summand in the
formula (\ref{sGibbs}) up to $\epsilon=E/2$ -- with a suitable change in
the value of $Z_{S,\beta}$ --  and agreement of the energy
probability density up to $E$) as being \textit{$E$-approximately
semi-thermal}.

\begin{figure}
\includegraphics[width = 300pt, height = 300pt]{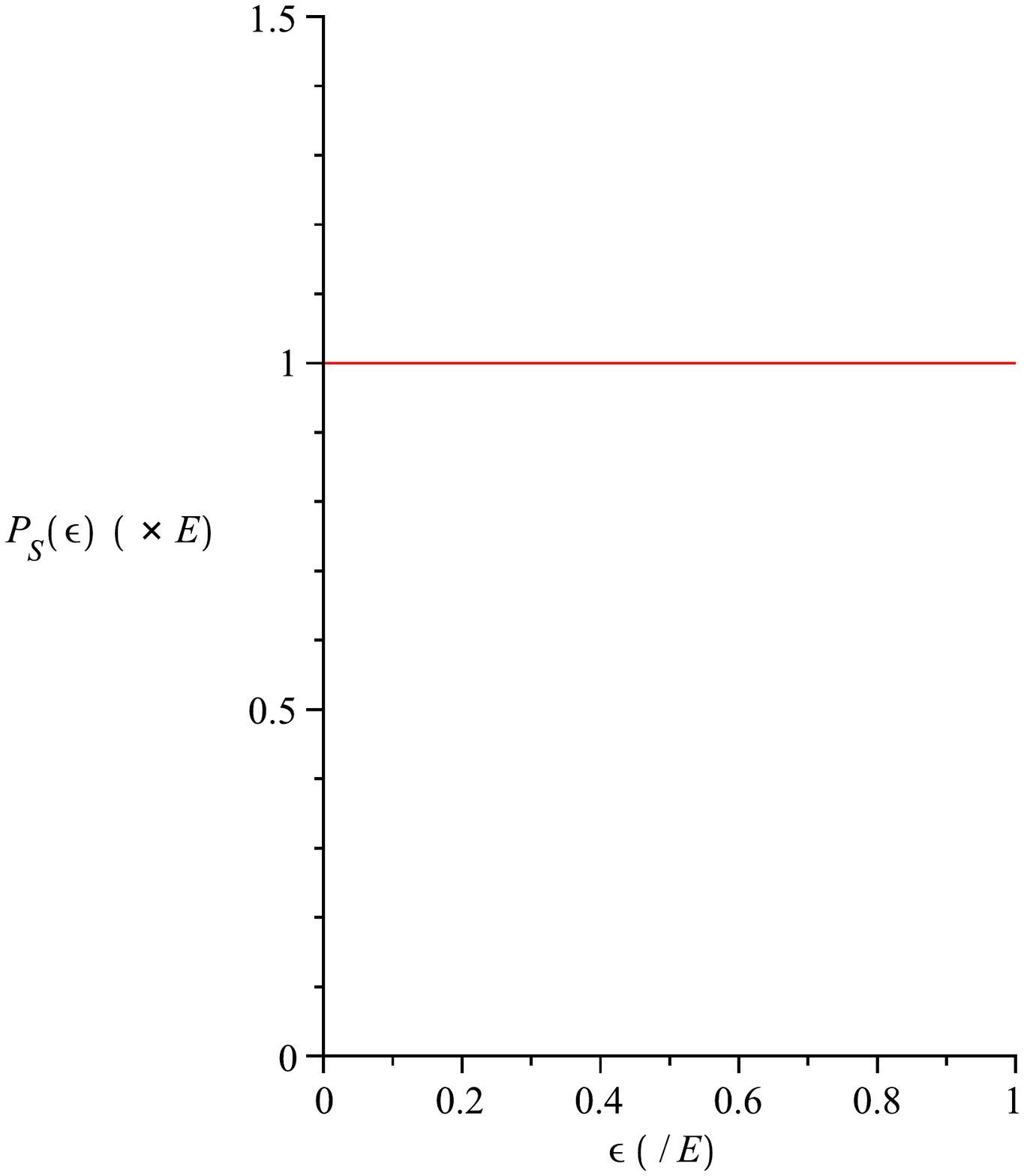}
\caption{\label{Fig2} Plot of the energy probability density (\ref{enprobdensexp}),
$P_{\mathrm S}(\epsilon)$, in the case S and B have the same density of states
$\sigma(\epsilon)=ce^{b\epsilon}$}
\end{figure}

We note here that, with the densities of states as in
(\ref{sigmasbexponential}), the energy probability density
$P_{\mathrm{S}}(\epsilon)$, which we recall by (\ref{enprobdens}) is
given in general by
\[
P_{\mathrm{S}}(\epsilon)
=\frac{\Delta}{M}\sigma_{\mathrm{S}}(\epsilon)\sigma_{\mathrm{B}}(E-\epsilon),
\]
will, with $M$ as in (\ref{Mexp}) and $\sigma_{\mathrm{S}}$ and
$\sigma_{\mathrm{B}}$ as in (\ref{sigmasbexponential}), reduce to
\begin{equation}
\label{enprobdensexp}
P_{\mathrm{S}}(\epsilon)=\frac{1}{E}.
\end{equation}
See Figure \ref{Fig2}.  Of course (cf.\ the paragraph after Equation (\ref{gamma})) the energies of S and B will, again, be perfectly anticorrelated.

Similar results to the the above results for S will obviously hold for
B.  We thus conclude, in fulfillment of our promise (cf.\ the start of
Section \ref{Sect:Spec}) that, with the appropriate meaning in
each case for the expression ``approximately thermal'', as above, when
the densities of states of S and B take the exponential form of
(\ref{sigmasbexponential}) then -- in contrast to the situation for
power-law densities of states discussed in Section \ref{Sect:Power}
--  on both the traditional total microcanonical state approach and also
on the modern total pure state approach, the reduced density operators
of both S and B will be approximately thermal (at inverse temperature
$\beta=b$) in an appropriate sense.

\section{\label{Sect:Gen} General formulae for mean energy and for entropy}

It is interesting (especially in preparation for the discussion in Section \ref{Sect:BHimplications} and in our companion papers \cite{Kaycompanion, Kayprefactor} about the connection with quantum black hole physics both of the results in Sections \ref{Sect:Exp} and \ref{Sect:ExpForm} concerning densities of states which grow exponentially with energy and, in Section \ref{Sect:ExpEsq}, concerning those which grow as quadratic exponentials in the energy) to calculate the mean
energy, $\bar\epsilon$, and also the von Neumman entropy, $S$, for each
of the density operators, $\rho^{\mathrm{microc}}_{\mathrm{S}}$,
$\rho^{\mathrm{modapprox}}_{\mathrm{S}}$ (and also for
$\rho^{\mathrm{microc}}_{\mathrm{B}}$,
$\rho^{\mathrm{modapprox}}_{\mathrm{B}}$). The former density operator will just give the usual mean energy and von Neumann entropy (defined as in (\ref{SvN})) of our system, S, when the totem is in the microcanonical ensemble.  The latter density operator will, according to our proposition in Section \ref{Sect:Ans}, give an energy value and an entropy value which will very probably be very close to the mean  value of the energy and the entropy of our system, S, when the totem is in a random pure state.

By (\ref{microreduced}), (\ref{purereduced}), and (\ref{sumnorm}), we have, in general,
that, with obvious notation,
\[
\bar\epsilon^{\mathrm{microc}}_{\mathrm{S}}:={\rm tr}
(\rho^{\mathrm{microc}}_{\mathrm{S}}H_{\mathrm{S}})=
M^{-1}\sum_{\epsilon=\Delta}^E \epsilon n_{\mathrm{S}}(\epsilon)n_{\mathrm{B}}(E-\epsilon).
\]
Similarly
\[
\bar\epsilon^{\mathrm{microc}}_{\mathrm{B}}:={\rm tr}
(\rho^{\mathrm{microc}}_{\mathrm{B}}H_{\mathrm{B}})=
M^{-1}\sum_{\epsilon=\Delta}^E \epsilon n_{\mathrm{B}}(\epsilon)n_{\mathrm{S}}(E-\epsilon)
\]
and one easily sees that necessarily,
$\bar\epsilon^{\mathrm{microc}}_{\mathrm{B}}=E-\bar\epsilon^{\mathrm{microc}}_{\mathrm{S}}$.
Moreover, we have
\begin{equation}
\label{puremeanenergy}
\bar\epsilon^{\mathrm{modapprox}}_{\mathrm{S}}
= {\rm tr}(\rho^{\mathrm{modapprox}}_{\mathrm{S}}H_{\mathrm{S}})
\end{equation}
\begin{equation}
= M^{-1}\left(\sum_{\epsilon=\Delta}^{E_c}
\epsilon n_{\mathrm{B}}(E-\epsilon)n_{\mathrm{S}}(\epsilon)
+ \sum_{\epsilon=E_c+\Delta}^E \epsilon n_{\mathrm{S}}
(\epsilon)n_{\mathrm{B}}(E-\epsilon) \right ),
\end{equation}
which is easily seen to be equal to
$\bar\epsilon^{\mathrm{microc}}_{\mathrm{S}}$.  Similarly,
$\bar\epsilon^{\mathrm{modapprox}}_{\mathrm{B}}=\bar\epsilon^{\mathrm{microc}}_{\mathrm{B}}$.

On the other hand, by (\ref{microreduced}) and (\ref{SvN}), we will have
\begin{equation}
\label{microentropy}
S^{\mathrm{microc}}_{\mathrm{S}}:=-{\rm tr}
(\rho^{\mathrm{microc}}_{\mathrm{S}}\log\rho^{\mathrm{microc}}_{\mathrm{S}})=
-M^{-1}\sum_{\epsilon=\Delta}^E n_{\mathrm{S}}(\epsilon)
n_{\mathrm{B}}(E-\epsilon)\log(M^{-1}n_{\mathrm{B}}(E-\epsilon))
\end{equation}
and by (\ref{purereduced}) and (\ref{SvN})
\[
S^{\mathrm{modapprox}}_{\mathrm{S}}:= -{\rm tr}
(\rho^{\mathrm{modapprox}}_{\mathrm{S}}\log\rho^{\mathrm{modapprox}}_{\mathrm{S}}))
\]
\begin{equation}
\label{pureentropy}
=-M^{-1}\left(\sum_{\epsilon=\Delta}^{E_c} n_{\mathrm{S}}
(\epsilon)n_{\mathrm{B}}(E-\epsilon)\log(M^{-1}n_{\mathrm{B}}(E-\epsilon))
+ \sum_{\epsilon=E_c+\Delta}^E n_{\mathrm{S}}
(\epsilon)n_{\mathrm{B}}(E-\epsilon)\log(M^{-1}n_{\mathrm{S}}(\epsilon)) \right )
\end{equation}
and similarly with S replaced by B.
We remark that it is not difficult to see from (\ref{purereduced}) and the counterpart equation for
$\rho^{\mathrm{modapprox}}_{\mathrm{B}}$ that $S^{\mathrm{modapprox}}_{\mathrm{S}}$ will necessarily equal
$S^{\mathrm{modapprox}}_{\mathrm{B}}$.  This is of course consistent with the fact that, by the general result recalled in Endnote \cite{Ftnt3}, for any pure totem state, $\Psi$, we necessarily have that the von Neumann entropies of the resulting reduced density operators $\rho^{\mathrm{modern}}_{\mathrm S}$ and $\rho^{\mathrm{modern}}_{\mathrm B}$
will necessarily be equal.  After all, as we claim in our Proposition in Section \ref{Sect:Ans} and argue in Part~2, for random $\Psi$,  $\rho^{\mathrm{modapprox}}_{\mathrm S}$ most probably gives a very good approximation of $\rho^{\mathrm{modern}}_{\mathrm S}$ and $\rho^{\mathrm{modapprox}}_{\mathrm B}$ of $\rho^{\mathrm{modern}}_{\mathrm B}$.

By referring to the second equality in (\ref{enprobdens}), it is easy to see that the formulae for $S^{\mathrm{microc}}_{\mathrm{S}}$ and $S^{\mathrm{modapprox}}_{\mathrm{S}}$ in  (\ref{microentropy}) and (\ref{pureentropy}) may be rearranged to give the following useful alternative expressions:
\begin{equation}
\label{altmicroentropy}
S^{\mathrm{microc}}_{\mathrm{S}}=\Delta
\sum_{\epsilon=\Delta}^E P_{\mathrm{S}}(\epsilon)\log\left (
\frac{n_{\mathrm{S}}(\epsilon)}{P_{\mathrm{S}}(\epsilon)\Delta}\right),
\end{equation}
\begin{equation}
\label{altpureentropy}
S^{\mathrm{modapprox}}_{\mathrm{S}}=
\Delta\left(\sum_{\epsilon=\Delta}^{E_c} P_{\mathrm{S}}(\epsilon)\log\left (
\frac{n_{\mathrm{S}}(\epsilon)}{P_{\mathrm{S}}(\epsilon)\Delta}\right)+\sum_{\epsilon=E_c+\Delta}^E P_{\mathrm{S}}(\epsilon)\log\left (
\frac{n_{\mathrm{B}}(E-\epsilon)}{P_{\mathrm{S}}(\epsilon)\Delta}\right)\right).
\end{equation}
Referring to (\ref{degdensS}) and making the replacement (\ref{continuum}) (and with the proviso made in the cautionary remark in \cite{spurious}) we see that (\ref{altmicroentropy}) and (\ref{altpureentropy}) have, as their continuum versions:
\begin{equation}
\label{contmicroentropy}
S^{\mathrm{microc}}_{\mathrm{S}}=
\int_0^E P_{\mathrm{S}}(\epsilon)\log\left (
\frac{\sigma_{\mathrm{S}}(\epsilon)}{P_{\mathrm{S}}(\epsilon)}\right)d\epsilon,
\end{equation}
\begin{equation}
\label{contpureentropy}
S^{\mathrm{modapprox}}_{\mathrm{S}}=
\int_0^{E_c} P_{\mathrm{S}}(\epsilon)\log\left (
\frac{\sigma_{\mathrm{S}}(\epsilon)}{P_{\mathrm{S}}(\epsilon)}\right)d\epsilon+\int_{E_c}^E P_{\mathrm{S}}(\epsilon)\log\left (
\frac{\sigma_{\mathrm{B}}(E-\epsilon)}{P_{\mathrm{S}}(\epsilon)}\right)d\epsilon.
\end{equation}
We notice, in passing, that the absence of the quantity $\Delta$ (or of any quantity that scales with $\Delta$) in the formulae $S^{\mathrm{microc}}_{\mathrm{S}}$ and $S^{\mathrm{modapprox}}_{\mathrm{S}}$ shows us the interesting fact that (for $\Delta$ in an appropriate not-too-large and not-too-small range, and to what, in typical applications will be an extremely good approximation) neither of these entropies depends on $\Delta$!

Finally, further useful insight concerning the form of Equations (\ref{altmicroentropy}) and (\ref{altpureentropy}) can be had by noticing that they can alternatively be derived as corollaries of the following easily proved Lemma, which we will also need to refer to in Section \ref{Sect:Further} in Part~2.

\noindent
{\bf Lemma:} \textit{Given density operators} $\rho_1, \rho_2, \dots$
\textit{on Hilbert spaces} ${\cal H}_1, {\cal H}_2, \dots $ \textit{respectively,
with von Neumann entropies} $S(\rho_1), S(\rho_2), \dots $ \textit{and given positive
real numbers} $\lambda_1, \lambda_2, \dots $ with $\sum_i \lambda_i =1$.
\textit{Then the density operator}
\[
\rho=\lambda_1\rho_1\oplus\lambda_2\rho_2\oplus\dots
\]
\textit{on the direct sum Hilbert space} ${\cal H}= {\cal H}_1\oplus {\cal H}_2
\dots $
\textit{will have an entropy,} $S,$ \textit{given by}
\begin{equation}
\label{entropylemma}
S=\sum_i \lambda_iS(\rho_i) -\sum_i\lambda_i\log\lambda_i.
\end{equation}

To apply this lemma to the calculation of
$S^{\mathrm{microc}}_{\mathrm{S}}$ and
$S^{\mathrm{modapprox}}_{\mathrm{S}}$ (for general densities of states) we first notice
that (\ref{microreduced}) and (\ref{purereduced}) can be rewritten as
\begin{equation}
\label{microreducedrewrite}
\rho^{\mathrm{micro}}_{\mathrm{S}}=\oplus_{\epsilon=\Delta}^E\,\lambda_\epsilon\rho_\epsilon
\end{equation}
and
\begin{equation}
\label{purereducedrewrite}
\rho^{\mathrm{modapprox}}_{\mathrm{S}}=
\oplus_{\epsilon=\Delta}^{E_c}\,\lambda_\epsilon\rho_\epsilon
+\oplus_{\epsilon=E_c+\Delta}^E\, \tilde\lambda_\epsilon\tilde\rho_\epsilon
\end{equation}
where
\begin{equation}
\label{rhoepsilon}
\rho_\epsilon =
n_{\mathrm{S}}(\epsilon)^{-1}\sum_{i=1}^{n_{\mathrm{S}}(\epsilon)}
|\epsilon, i\rangle\langle\epsilon, i|
\end{equation}
and
\begin{equation}
\label{tilderhoepsilon}
\tilde\rho_\epsilon=n_{\mathrm{B}}(E-\epsilon)^{-1}\sum_{i=1}^{n_{\mathrm{B}}(E-\epsilon)}
|\widetilde{\epsilon, i}\rangle\langle\widetilde{\epsilon, i}|
\end{equation}
where, $|\epsilon, i\rangle$ and $|\widetilde{\epsilon, i}\rangle$ are
as in (\ref{microreduced}) and (\ref{purereduced}), and where (recalling
that the sums in (\ref{microreduced}) and (\ref{purereduced}) are over
energies, $\epsilon$, which are integral multiples of $\Delta$)
\begin{equation}
\label{lambdaepsilon}
\lambda_\epsilon=\tilde\lambda_\epsilon=P_{\mathrm{S}}(\epsilon)\Delta
\end{equation}
where $P_{\mathrm{S}}(\epsilon)$ is the energy probability density
(\ref{enprobdens}).

We also easily see from (\ref{rhoepsilon}) and (\ref{tilderhoepsilon})
that, in general
\begin{equation}
\label{Srhoepsilon}
S(\rho_\epsilon)=\log(n_{\mathrm{S}}(\epsilon))\quad\hbox{and}\quad S(\tilde\rho_\epsilon)=
\log(n_{\mathrm{B}}(E-\epsilon)).
\end{equation}

Equations (\ref{altmicroentropy}) and (\ref{altpureentropy}) now follow by simple applications of the formula (\ref{entropylemma}) or of our above lemma to (\ref{microreducedrewrite}) and (\ref{purereducedrewrite}).

\section{\label{Sect:ExpForm} Formulae for mean energy and entropy for
exponentially rising densities of states}

Specialising to $n_{\mathrm{S}}$ and $n_{\mathrm{B}}$ given by
$n_{\mathrm{S}}(\epsilon)= \sigma_{\mathrm{S}}(\epsilon)\Delta$ and
$n_{\mathrm{B}}(\epsilon)=\sigma_{\mathrm{B}}(\epsilon)\Delta$ with
$\sigma_{\mathrm{S}}$ and $\sigma_{\mathrm{B}}$ as in
(\ref{sigmasbexponential}), we have that
\begin{equation}
\label{expenergy}
\bar\epsilon^{\mathrm{microc}}_{\mathrm{S}}=
\bar\epsilon^{\mathrm{modapprox}}_{\mathrm{S}}=E/2,
\end{equation}
and similarly with S replaced by B.  These results for the mean energies of S and B are of course
anyway obvious since the assumption of very weak coupling (see (\ref{coupling}) and the subsequent paragraph) implies that the mean energies of S and B will add to $E$, while (\ref{sigmasbexponential}) is symmetric under the replacement of S by B.   However, we remark that (\ref{expenergy}) turns out to remain exactly unchanged even when the densities of states are generalized so as to have different
pre-factors $c_{\mathrm{S}}$ and $c_{\mathrm{B}}$ (see Endnote \cite{Ftnt8}).  (Returning to the case of equal densities of states) we caution that the mean energies are just that, averages; they are not in any sense `most likely' energies.  In fact, as we saw in Section \ref{Sect:Exp}, the energy probability density (see (\ref{enprobdensexp}) and Figure \ref{Fig2}) is flat!

It is also straightforward to calculate, using the formulae of Section \ref{Sect:Gen}, that the entropies take the values
\begin{equation}
\label{Smicroexp}
S^{\mathrm{microc}}_{\mathrm{S}}=bE/2+\log(cE),
\end{equation}
\begin{equation}
\label{Spureexp}
S^{\mathrm{modapprox}}_{\mathrm{S}}=bE/4+\log(cE),
\end{equation}
and similarly with S replaced by B. (Again, see Endnote \cite{Ftnt8} for the
generalization to different prefactors, $c_{\mathrm{S}}$ and $c_{\mathrm{B}}$, in the first and second
formulae of (\ref{sigmasbexponential})).

In particular, inserting the formulae (\ref{sigmasbexponential}) and (\ref{enprobdensexp}) for $\sigma_{\mathrm{S}}$ and $P_{\mathrm{S}}$  in (\ref{altmicroentropy}) and (\ref{altpureentropy}) we obtain
\[
S^{\mathrm{microc}}_{\mathrm{S}}=\sum_{\epsilon=\Delta}^{E}\frac{\Delta}{E}(b\epsilon+\log(c\Delta)
 - \log(\Delta/E))
\]
\[
=\frac{\Delta}{E}\left
(b\Delta\frac{(E/\Delta)(E/\Delta+1)}{2}+(E/\Delta)\log(cE)\right )
\]
\begin{equation}
\label{Smicrocexpbis}
\simeq \frac{bE}{2}+\log(cE)
\end{equation}
while (assuming $E/\Delta$ is even)
\[
S^{\mathrm{modapprox}}_{\mathrm{S}}=\sum_{\epsilon=\Delta}^{E/2}\frac{\Delta}{E}(b\epsilon+\log(c\Delta)
 - \log(\Delta/E))+
\sum_{E/2+\Delta}^{E}\frac{\Delta}{E}(b(E-\epsilon)+\log(c\Delta)
 - \log(\Delta/E))
\]
\[
=2\sum_{\epsilon=\Delta}^{E/2}\frac{\Delta}{E}(b\epsilon+\log(c\Delta)
 - \log(\Delta/E))
\]
\[
=2\frac{\Delta}{E}\left
(b\Delta\frac{(E/2\Delta)(E/2\Delta+1)}{2}+(E/\Delta)\log(cE)\right )
\]
\begin{equation}
\label{Spureexpbis}
\simeq \frac{bE}{4}+\log(cE)
\end{equation}
which are the formulae (\ref{Smicroexp}) and (\ref{Spureexp}).
In the calculations above, we need to
recall that the sums in (\ref{microreduced}) and (\ref{purereduced}), and hence also in
the direct sums in (\ref{microreducedrewrite}) and
(\ref{purereducedrewrite}) and in the above equations, are over $\epsilon$ values which are
positive-integer multiples of $\Delta$.

We remark that the leading behaviour of $S^{\mathrm{microc}}_{\mathrm{S}}$ (\ref{Smicroexp})  (i.e.\ the term, $bE/2$, which remains when we ignore the logarithmic terms in (\ref{Smicroexp})) arises, in (say) the continuum version, (\ref{contmicroentropy}), of our general formula for $S^{\mathrm{microc}}_{\mathrm{S}}$ by replacing the logarithm in this formula by its `main part', by which we mean  the exponent, $b\epsilon$, in the formula,
(\ref{sigmasbexponential}) $\sigma_{\mathrm{S}}(\epsilon) = ce^{b\epsilon}$.  Similarly, the leading behaviour of $S^{\mathrm{modapprox}}_{\mathrm{S}}$ (i.e.\ the term $bE/4$ in (\ref{Spureexp})) arises by setting $E_c=E/2$ in  (\ref{contpureentropy}) and noticing that the main parts of the two logarithms in this formula are (in order) $b\epsilon$ and $b(E/2-\epsilon)$.

\section{\label{Sect:ExpEsq} Densities of states which grow as quadratic exponentials}

Next we discuss the behaviour of the reduced density operators,
$\rho^{\mathrm{microc}}_{\mathrm{S}}$ and
$\rho^{\mathrm{modapprox}}_{\mathrm{B}}$, when the densities of states of
S and B each increase as the exponential of a constant times the square of the energy.  We shall confine our interest to the case where both
densities of states, $\sigma_{\mathrm{S}}$ and $\sigma_{\mathrm{B}}$,
behave as $Ke^{q\epsilon^2}$ with the same constants, $K$ and $q$:
\begin{equation}
\label{sigmasbexpsquared}
\sigma_{\mathrm{S}}(\epsilon) = Ke^{q\epsilon^2}
, \quad \sigma_{\mathrm{B}}(\epsilon) = Ke^{q\epsilon^2}
\end{equation}
and shall just discuss the cases of $\rho^{\mathrm{microc}}_{\mathrm{S}}$ and
$\rho^{\mathrm{modapprox}}_{\mathrm{S}}$ -- those of $\rho^{\mathrm{microc}}_{\mathrm{B}}$ and
$\rho^{\mathrm{modapprox}}_{\mathrm{B}}$ obviously being similar.

\begin{figure}
\includegraphics[width = 300pt, height = 300pt]{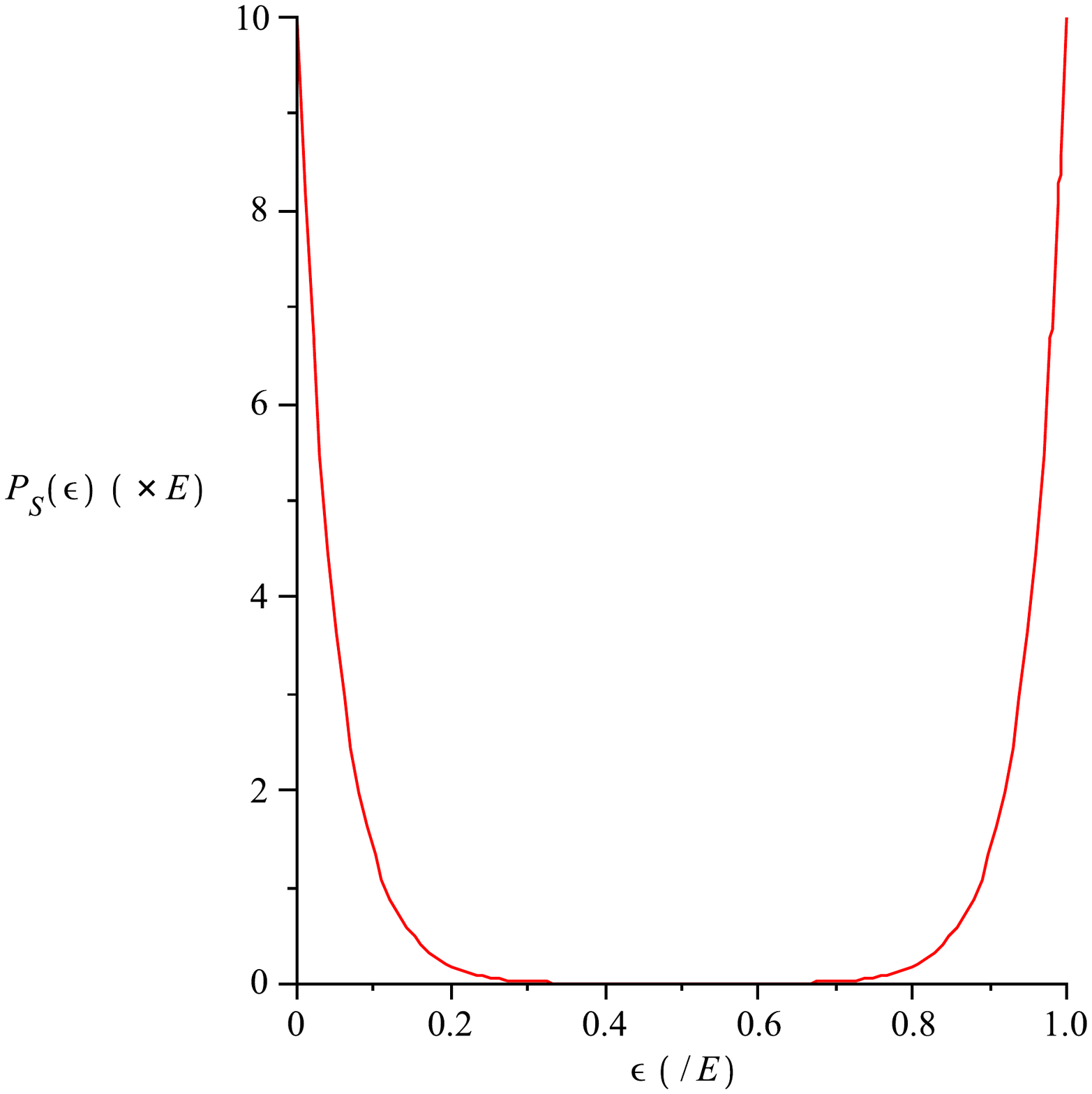}
\caption{\label{Fig3} Plot of the energy probability density (\ref{enprobdensexpsquared}),
$P_{\mathrm S}(\epsilon)$, in the case S and B have the same density of states
$\sigma(\epsilon)=Ke^{q\epsilon^2}$ for the (unrealistically small) value $qE^2=10$.}
\end{figure}

We shall assume that
\begin{equation}
\label{qEsquaredbig}
qE^2\gg 1.
\end{equation}
The energy probability density, $P_{\mathrm{S}}(\epsilon)$
(\ref{enprobdens}), now takes the form
\begin{equation}
\label{enprobdensexpsquared}
P_{\mathrm{S}}(\epsilon)=\frac{\Delta}{M}K^2e^{qE^2}e^{-2q\epsilon(E-\epsilon)}.
\end{equation}
and we sketch its graph in Figure \ref{Fig3}.

\begin{figure}
\includegraphics[width = 300pt, height = 300 pt]{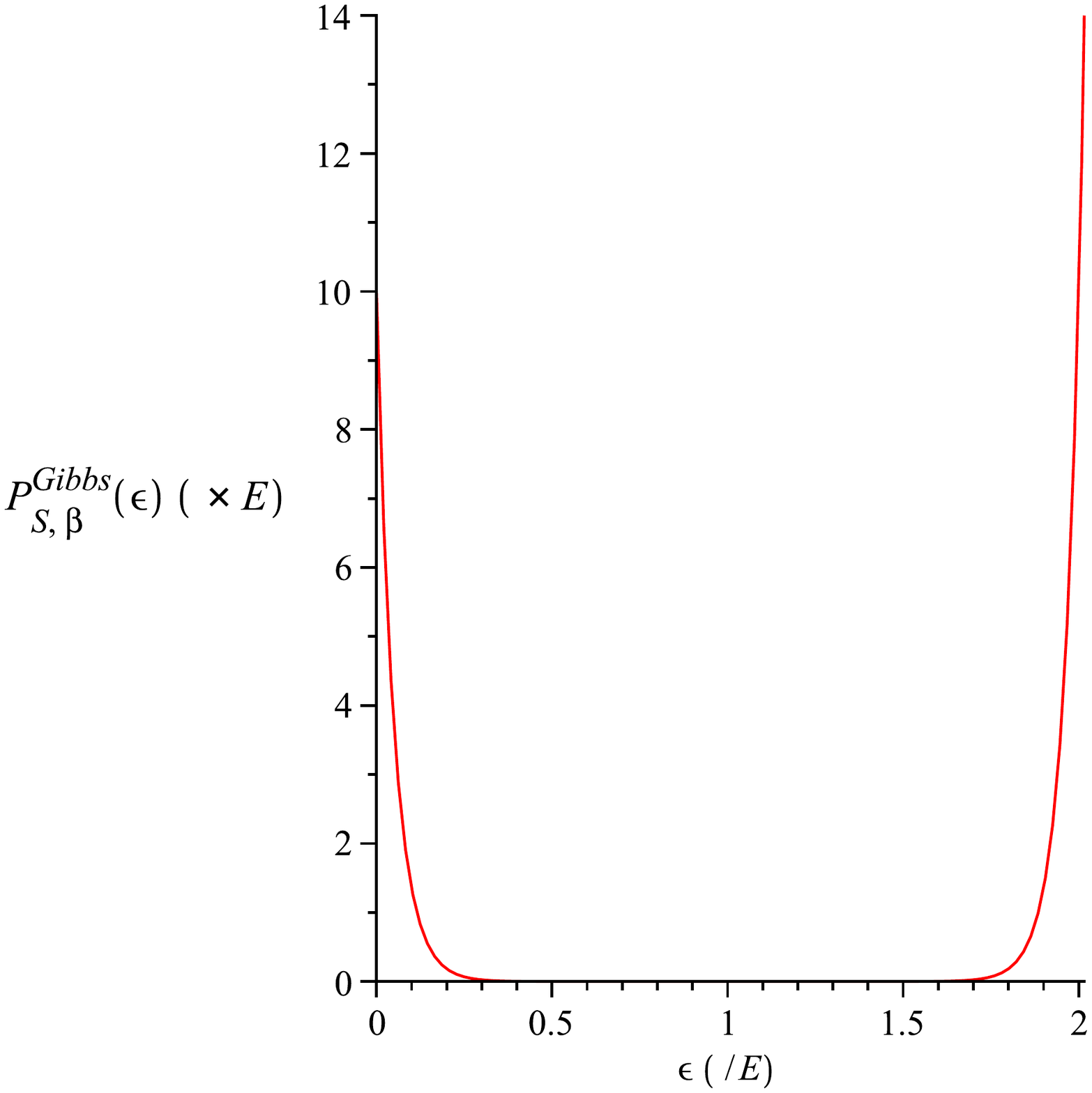}
\caption{\label{Fig4} Comparison figure for the (non-normalizable!) ``energy
probability density'' (\ref{PsGibbsexpsquared}) of the ``thermal state'', $P_{{\mathrm
S},\beta}^{\mathrm{Gibbs}}(\epsilon)$ at inverse temperature
$\beta=2qE$, again for the value $qE^2=10$}
\end{figure}

We note that it is symmetric about $E/2$ and also, in view of
(\ref{qEsquaredbig}), $P_{\mathrm{S}}(\epsilon)$
is very close to zero except when $\epsilon$ is close to $0$ or to $E$,
where it is well approximated by the exponentially decaying function
$\frac{\Delta}{M}K^2e^{qE^2}e^{-2qE\epsilon}$ (near $\epsilon=0$),
and by the exponentially rising function
$\frac{\Delta}{M}K^2e^{qE^2}e^{2qE(\epsilon-E)}$ (near $\epsilon=E$).
Approximating the integral from $0$ to $E$ of $P_{\mathrm{S}}(\epsilon)$
by the sum of the (equal) integrals (from $0$ to $\infty$ and from $-\infty$ to $E$)
of these exponential approximations, and demanding that the result must
equal $1$ we see that $P_{\mathrm{S}}(\epsilon)$ will be well approximated on
its domain $[0, E]$ by
\begin{equation}
\label{enprobdensexpsquaredapprox}
P_{\mathrm{S}}(\epsilon)\simeq qE(e^{-2qE\epsilon}+e^{2qE(\epsilon-E)})
\end{equation}
from which we infer that the normalization constant, $M$, will be well
approximated \cite{Ftnt9} by
\begin{equation}
\label{Mexpenergysquared}
M\simeq K^2 \frac{e^{qE^2}}{qE}\Delta.
\end{equation}

We pause here to record that (cf.\ (\ref{expenergy})) either by calculation or by the above-noted symmetry of $P_{\mathrm{S}}(\epsilon)$, we will obviously have that the mean energy of both S and E will be $E/2$:
\begin{equation}
\label{expesqenergy}
\bar\epsilon^{\mathrm{microc}}_{\mathrm{S}}=
\bar\epsilon^{\mathrm{modapprox}}_{\mathrm{S}}=E/2,
\end{equation}

However, (cf.\ our remark after Equation (\ref{expenergy})) even more emphatically than in the case of exponentially rising densities of states , these are {\it not} `most likely energies'.  In fact, the energy probability density, $P_{\mathrm{S}}(\epsilon)$, tells us that the energy (say of S) will be highly likely, and with equal likelikhoods, either to be close to $0$ or to be close to $E$ -- and highly unlikely to be close to $E/2$ (and similarly for B).  In addition of course (cf.\ after Equation (\ref{gamma}) and Equation (\ref{enprobdensexp})) the energy of S and the energy of B will be perfectly anticorrelated.  So when the energy of S is near $0$, the energy of B will be near $E$, and when the energy of S is near $E$, the energy of B will be near $0$.

In analogy to what we did in Sections \ref{Sect:Exp}, \ref{Sect:ExpForm}, we next wish
to compare the formulae (\ref{enprobdensexpsquared}) and
(\ref{enprobdensexpsquaredapprox}) for $P_{\mathrm{S}}(\epsilon)$ with
\begin{equation}
\label{PsGibbsexpsquared}
P^{\mathrm{Gibbs}}_{{\mathrm S}, \beta}(\epsilon)= C e^{-\beta\epsilon}Ke^{q\epsilon^2},
\end{equation}
where $C$ is a suitable constant, which, but for the fact that it is not
integrable on the interval $(0, \infty)$ (for any value of $\beta$!)
would deserve to be called `the energy probability density of a thermal state at
inverse temperature $\beta$' (cf.\ (\ref{PsGibbs})) for the same density of states
(\ref{sigmasbexpsquared}).
If $\beta E \gg 1$, then, on the interval $[0, \beta/q]$,
$P^{\mathrm{Gibbs}}_{{\mathrm S}, \beta}(\epsilon)$
will be very close to zero except when $\epsilon$ is close to $0$ or to $\beta/q$,
where it will be well approximated by the exponential decay
$Ce^{-\beta\epsilon}$ (near $0$),
and by the exponentially rising function
$Ce^{\beta(\epsilon-\beta/q)}$ (near $\epsilon=\beta/q$). Put otherwise, on the
interval $[0,\beta/q]$, $P^{\mathrm{Gibbs}}_{{\mathrm S}, \beta}(\epsilon)$ will take the
approximate form
\begin{equation}
\label{enprobdensGibbsexpsquaredapprox}
P^{\mathrm{Gibbs}}_{{\mathrm S}, \beta}(\epsilon)\simeq CK(e^{-\beta\epsilon}+
e^{\beta(\epsilon-\beta/q)}).
\end{equation}
Beyond $\epsilon=\beta/q$,
$P^{\mathrm{Gibbs}}_{{\mathrm S}, \beta}(\epsilon)$ will, of course, grow rapidly.
If we now choose to make the identification
\begin{equation}
\label{betaequals2qE}
\beta=2qE,
\end{equation}
we see that (\ref{enprobdensGibbsexpsquaredapprox}) can be written
\begin{equation}
\label{enprobdensGibbsexpsquaredapproxE}
P^{\mathrm{Gibbs}}_{{\mathrm S}, \beta}(\epsilon)
\simeq CK(e^{-\beta\epsilon}+e^{\beta(\epsilon-2E)})
\end{equation}
while (\ref{enprobdensexpsquaredapprox}) can be written
\begin{equation}
\label{enprobdensexpsquaredapproxbeta}
P_{\mathrm{S}}(\epsilon)\simeq qE(e^{-\beta\epsilon}+e^{\beta(\epsilon-E)})
\end{equation}
and comparison of (\ref{enprobdensGibbsexpsquaredapproxE}) and
(\ref{enprobdensexpsquaredapproxbeta}) or a glance at Figures
\ref{Fig3} and \ref{Fig4}, immediately shows that
these resemble one-another closely -- each having an equally-rapidly
exponentially decaying peak located near $\epsilon=0$ and each having an
equally rapidly exponentially rising, equal-sized second peak -- the
only discrepancy being that, in the case of $P_{\mathrm{S}}(\epsilon)$,
the second peak is located near $\epsilon=E$ while, in the case of
$P^{\mathrm{Gibbs}}_{{\mathrm S}, \beta}(\epsilon)$, it is located near
$\epsilon=2E$. Thus, at low energies -- and indeed at all energies,
$\epsilon$, up to a little below $E$ -- the energy probability density,
$P_{\mathrm{S}}(\epsilon)$, is closely approximated by the thermal
energy probability density for $\beta=2qE$, while there is a qualitative
resemblance between $P_{\mathrm{S}}(\epsilon)$ on its full interval
$[0,E]$ and $P^{\mathrm{Gibbs}}_{{\mathrm S}, \beta}(\epsilon)$ on the
interval $[0,2E]$ (with the  above-mentioned quantitative discrepancy
that the second peak in $P_{\mathrm{S}}(\epsilon)$ occurs near $E$ while
the second peak in $P^{\mathrm{Gibbs}}_{{\mathrm S}, \beta}(\epsilon)$
occurs near $2E$).

Moreover, one may easily check that, except for discrepancies
corresponding to the above discrepancy for the energy probability
densities, the reduced density  operators,
$\rho^{\mathrm{microc}}_{\mathrm{S}}$ (defined as in
(\ref{microreduced})) and $\rho^{\mathrm{modapprox}}_{\mathrm{S}}$
(defined as in (\ref{purereduced})), respectively, will stand in
relation to $\rho^{\mathrm{Gibbs}}_{{\mathrm S}, \beta}$ (defined as in
(\ref{sGibbs})) for $\beta=2qE$, in a similar way to the relationships
which we termed `$E$-approximately thermal'  and `$E$-approximately
semi-thermal' in Section \ref{Sect:Exp}.

In conclusion, except for the discrepancy pointed out above, we may say
that, in contrast again to the situation for power-law densities of
states and with many similarities (but also a few differences) to what
we found in Sections \ref{Sect:Exp}, \ref{Sect:ExpForm}, for densities of states which 
grow exponentially with energy, also densities of states which grow, (\ref{sigmasbexpsquared}),  
as quadratic exponentials lead to reduced density operators on S and B which are, in the 
sense we have explained above, approximately thermal.

Next we turn to calculate the von Neumann entropies of
$\rho^{\mathrm{microc}}_{\mathrm{S}}$ and $\rho^{\mathrm{modapprox}}_{\mathrm{S}}$
when the densities of states are as in (\ref{sigmasbexpsquared}).

In the spirit of the last paragraph of Section \ref{Sect:ExpForm} we expect the leading term in $S^{\mathrm{microc}}_{\mathrm{S}}$ to be given by
\begin{equation}
\label{Smicrocexpesq}
S^{\mathrm{microc}}_{\mathrm{S}}\simeq
\int_0^E  P_{\mathrm S}(\epsilon) \, q\epsilon^2 d\epsilon
\simeq\frac{qE^2}{2}
\end{equation}
where, for the first approximate equality, we have replaced the logarithm in (\ref{contmicroentropy}) by its `main part' -- i.e.\ by the exponent, $K\epsilon^2$, in the first equation in (\ref{sigmasbexpsquared}),
and, for the second approximate equality, we have used the fact that the energy probability density,
$P_{\mathrm S}(\epsilon)$ (see (\ref{enprobdensexpsquaredapprox}) and Figure \ref{Fig3}) consists of two sharp peaks, each of area $1/2$, one located at $\epsilon=0$ and one at $\epsilon=E$.

Proceeding similarly for $S^{\mathrm{modapprox}}_{\mathrm{S}}$, we similarly expect the leading term to be given by approximating
$(\ref{contpureentropy})$ by
\begin{equation}
\label{Smodernexpesq}
S^{\mathrm{modapprox}}_{\mathrm{S}}\simeq
\int_0^{E/2}  P_{\mathrm S}(\epsilon)\, q\epsilon^2 d\epsilon+\int_{E/2}^E  P_{\mathrm S}(\epsilon)\, q(E-\epsilon)^2 d\epsilon
\simeq 0.
\end{equation}

It is straightforward to check that the error in both (\ref{Smicrocexpesq})
and (\ref{Smodernexpesq}) is only of order 1 in $E$; one needs only to be
careful to realize that this is one situation in which (cf.\  Endnote \cite{spurious}) it is important to work with the discrete sum versions, (\ref{microentropy}) and (\ref{pureentropy}) or alternatively
(\ref{altmicroentropy}) and (\ref{altpureentropy}), of our entropy formulae; if one were to work unthinkingly with (\ref{contmicroentropy}) and (\ref{contpureentropy}), one might (wrongly)
conclude there is a (for some values of $K$, $q$ and $E$, negative!) correction to both equations of form
$\log(K/qE) + O(1)$ -- the problem being caused by the steeply rising
behaviour of $P_{\mathrm{S}}(\epsilon)$ near $\epsilon=0$ and
$\epsilon=E$.

Thus, for our densities of states which grow as quadratic exponentials, there is an even more dramatic difference between the value of $S^{\mathrm{microc}}_{\mathrm{S}}$ and the value of
$S^{\mathrm{modapprox}}_{\mathrm{S}}$ than we found, in Section
\ref{Sect:ExpForm}, for densities of states which grow exponentially with energy (where
they differed by a factor of 2).

\section{\label{Sect:Morentropy} More about Entropy}

\noindent
{\it Note}: The reader may wish to skip this, and the next, section on a first reading and go directly to Section \ref{Sect:BHimplications}.

\subsection{\label{Sect:peak} Special facts about the entropy when the energy probability density is sharply peaked}

In the case of our power-law density of states example, an alternative way of arguing that the location of the peak of the energy probability density, $P_{\mathrm{S}}(\epsilon)$, of the system, S, is given by the formula (\ref{antipopular}) is to assume foreknowledge of the existence of a (single) peak in the energy probability density, $P_{\mathrm S}(\epsilon)$, in the interior of the energy-interval
$[0, E]$ and then to note that, by (\ref{enprobdens}), this must occur at an energy, $\epsilon$ for which
\begin{equation}
\label{popsigma}
\frac{d\log\sigma_{\mathrm S}(\epsilon)}{d\epsilon}+\frac{d\log\sigma_{\mathrm B}(E-\epsilon)}{d\epsilon}=0
\end{equation}
which easily implies (\ref{antipopular}).

In a popular approach (cf.\ also \cite{Hawking:1976de}) to such problems involving the microcanonical ensemble of a pair of weakly coupled systems with such a (say unique, interior) peak, such a calculation often appears in the following guise:

One writes $\epsilon=\epsilon_1$ and $E-\epsilon=\epsilon_2$.  One calls $\log\sigma_{\mathrm S}(\epsilon)$ ``$S_1(\epsilon_1)$'', and one calls $\log\sigma_{\mathrm B}(E-\epsilon)$ ``$S_2(\epsilon_2)$''.  Then one
writes the equations
\[
\epsilon_1+\epsilon_2=E,
\]
\[
\frac{\partial S_1}{\partial \epsilon_1}-\frac{\partial S_2}{\epsilon_2}=0,
\]
(equivalent to (\ref{popsigma})
and
\begin{equation}
\label{popstable}
\frac{\partial^2 S_1}{\partial\epsilon_1^2}+\frac{\partial^2 S_2}{\partial\epsilon_2^2} < 0
\end{equation}
(expressing the fact that it is a peak and not a trough).

It is often then assumed, or, at least, tacitly implied, that $S(\epsilon_1)$ and $S(\epsilon_2)$ are the ``entropies'' of Systems 1 and 2 (our systems S and `energy bath' B) and that $\partial S_1/\partial \epsilon_1$ and $\partial S_2/\partial \epsilon_2$ are the ``temperatures'' of Systems 1 and 2.   Finally,  the equation (\ref{popstable}) is interpreted as telling us that Systems 1 and 2 are in ``stable equilibrium''.

Concerning this popular approach, we would remark and emphasize:

\smallskip

(a) $S(\epsilon_1)$ and $S(\epsilon_2)$ (our $\log\sigma_{{\mathrm S}}(\epsilon)$ and  $\log\sigma_{{\mathrm B}}(E-\epsilon)$) are {\it not} entropies (they are logarithms of densities of states).  To make sense of the logarithms one would, at least, need to multiply
$\sigma_{{\mathrm S}}(\epsilon)$ and  $\sigma_{{\mathrm B}}(E-\epsilon)$ by ``constants'' with the dimensions of energy, first, to make the overall arguments of the logarithms dimensionless.  This may not matter if the resulting logarithm is anyway destined to be differentiated with respect to $\epsilon$ to define a `temperature' (see Paragraph (b) below).  However it {\it will} matter if one wishes to talk meaningfully about the logarithms themselves (evaluated at the peak values of $\epsilon$ and $E-\epsilon$) as `entropies'.  One could, of course, insert, in each logarithm, an arbitrary constant with the dimensions of energy, and try to argue that it doesn't make much difference, in practice, what is the precise value of this constant provided it is of a ``reasonable'' order of magnitude.  But, even if it were the case that all that was at stake was such a ``constant'', one would expect, in a fundamental understanding of the origin of entropy, its value to be determined in terms of the physical parameters of the problem.   In fact, as we shall see below, what actually needs to be inserted is not a constant, but rather (for given system and energy-bath densities of states) a quantity (which we call $Q$ below) with the dimensions of energy which (like the peak values of $\epsilon$ and $E-\epsilon$ themselves) depends on the totem energy, $E$.

\smallskip

(b) It is true that one can think of $\partial S_1/\partial \epsilon_1$ and $\partial S_2/\partial\epsilon_2$ as `energy-dependent temperatures' in the sense (cf.\ Section \ref{Sect:Background} and Endnote \cite{fictTemp}) that, if System~1 had energy $\hat\epsilon_1$ and were uncoupled to System~2, but, rather, coupled to another and much smaller system, then that smaller system would likely get itself into a thermal state at the temperature $\partial S_1/\partial \epsilon_1$ evaluated at $\hat\epsilon_1$ (and similarly for System~2).  However, in the `equilibrium' in question, where System~1 and System~2 are coupled to one-another and neither can be regarded as `small', {\it neither} System~1 nor System~2 is in a thermal state (as we have shown in Section \ref{Sect:Power} for our power-law case)!

\smallskip

(c) Finally, this `popular' point of view is only of value in cases where the energy probability density (say of System 1) has a peak.  Whereas, we emphasize, as explained in this paper, one still predicts definite energy probability density functions when System~1 and System~2 have densities of states (such as our equal exponential and equal quadratic exponential cases discussed in Sections \ref{Sect:Exp} and \ref{Sect:ExpEsq})  which do not lead to an interior peak.  (In the equal exponential case, we find an energy probability density which is flat, and, in the quadratic exponential case, it is concave with peaks at the extremities of the range $[0,E]$ which are not `maxima' in the sense of the above equations for $S_1$ and $S_2$.)  In such latter cases, the significant quantity of interest is not the location of a peak (there may even be no peak) but rather the full energy probability density function itself.

\smallskip

We turn next to the value of the entropy of the system in cases where, for given system and energy-bath densities of states, and given total energy, $E$,  the energy probability density,
$P_{\mathrm{S}}(\epsilon)$, has a single peak, say at $\epsilon=\epsilon_0$. We are, of course, interested in both of the entropies, $S^{\mathrm{microc}}_{\mathrm{S}}$ and $S^{\mathrm{modapprox}}_{\mathrm{S}}$.   As we anticipated in Point (a) above, when the totem is in the microcanonical ensemble with energy in our interval $[E, E+\Delta]$, $S^{\mathrm{microc}}_{\mathrm{S}}$ will be $\log(Q\sigma_{\mathrm{S}}(\epsilon_0))$ for a suitable quantity, $Q$, with the dimensions of energy determined by the parameters of the problem, although we emphasize again that $Q$ is not a ``constant''; rather (for given system and energy-bath densities of states) it is a certain function of totem energy, $E$, which has the dimensions of energy and which is determined by the detailed shape of the peak in $P_{\mathrm{S}}(\epsilon)$.  In fact, by the general formula (\ref{contmicroentropy}) (one can see that the issues mentioned in Endnote \cite{spurious} will not be relevant for a sharp peak which is well inside the interior of the interval $[0,E]$) we will have
\[
S^{\mathrm{microc}}_{\mathrm{S}}=
\int_0^E P_{\mathrm{S}}(\epsilon)\log(L\sigma_{\mathrm{S}}(\epsilon))d\epsilon
- \int_0^E P_{\mathrm{S}}(\epsilon)\log(LP_{\mathrm{S}}(\epsilon))d\epsilon
\]
where we have {\it temporarily} introduced an arbitrary non-zero constant, $L$, with the dimensions of energy, which will of course cancel out in the final result.

Since $P_{\mathrm{S}}(\epsilon)$ is, by assumption, sharply peaked at $\epsilon=\epsilon_0$, and assuming $\sigma(\epsilon)$ is relatively slowly varying (as will be true in typical examples such as the power-law density of states case treated below) the value of the first integral will be very well approximated by $\log(L\sigma_{\mathrm{S}}(\epsilon_0))$.   The second integral will obviously take the form $\log (L/
Q)$ where $Q$ is a quantity with the dimensions of energy which can in principle be computed in terms of our system and energy-bath densities of states and the value of $E$.
So we will have
\begin{equation}
\label{Smicrocpower}
S^{\mathrm{microc}}_{\mathrm{S}}=\log(Q\sigma_{\mathrm{S}}(\epsilon_0)).
\end{equation}
On the other hand, if we consider the totem to be in a pure state, randomly chosen amongst all states with energy in the range $[E, E+\Delta]$, then, by (\ref{contpureentropy}), we expect the entropy to most probably be very close to
$S^{\mathrm{modapprox}}_{\mathrm{S}}$ given by
\[
S^{\mathrm{modapprox}}_{\mathrm{S}}=
\int_0^{E_c} P_{\mathrm{S}}(\epsilon)\log(L\sigma_{\mathrm{S}}(\epsilon))d\epsilon+
\int_{E_c}^E P_{\mathrm{S}}(\epsilon)\log(L\sigma_{\mathrm{B}}(E-\epsilon))d\epsilon
- \int_0^E P_{\mathrm{S}}(\epsilon)\log(LP_{\mathrm{S}}(\epsilon))d\epsilon
\]
which, by a similar argument, and in view of the definition of $E_c$ (see after Equation (\ref{purereduced})), will be very well approximated by
\[
S^{\mathrm{modapprox}}_{\mathrm{S}}=\mathrm{min}(\log(Q\sigma_{\mathrm{S}}(\epsilon_0)), \,
\log(Q\sigma_{\mathrm{B}}(E-\epsilon_0))
\]
for the same value of $Q$.

We next illustrate the computation of $Q$ in the case where both of system, S, and energy bath, B, have the power law densities of states (\ref{sigmasbpower}) which we discussed in Section \ref{Sect:Power}.  We have
\[
\log(L/Q)=\int_0^E P_{\mathrm{S}}(\epsilon)\log(LP_{\mathrm{S}}(\epsilon))d\epsilon
\]
where $P_{\mathrm{S}}(\epsilon)$ is given by (\ref{Pspowerapprox}) with $\gamma$ as in
(\ref{gamma}).  As long as $N_{\mathrm{S}}$ and $N_{\mathrm{B}}$ are comparable in size, the location, (\ref{antipopular}), of the peak, $\epsilon_0$, will be far from the extremities of the interval $[0, E]$ and we may clearly replace the limits of the above integral by $-\infty$ and $\infty$ with very little error.

So
\[
\log(L/Q) \simeq \frac{1}{E}\sqrt{\frac{\gamma}{\pi}}\int_{-\infty}^{\infty}\log\left(\frac{L}{E}\sqrt{\frac{\gamma}{\pi}}\exp\left(-\gamma\frac{(\epsilon-\epsilon_0)^2}{E^2}\right ) \right)
\exp\left(-\gamma\frac{(\epsilon-\epsilon_0)^2}{E^2}\right ) d\epsilon
\]
\[
=-\log\left(\frac{L}{E}\sqrt{\frac{\gamma}{\pi}}\right) +\frac{\gamma}{E^3}\sqrt{\frac{\gamma}{\pi}}\int_{-\infty}^\infty (\epsilon-\epsilon_0)^2\exp\left(-\gamma\frac{(\epsilon-\epsilon_0)^2}{E^2}\right ) d\epsilon.
\]
The second term above may easily be calculated by writing it as\hfil\break
$-(\gamma^{3/2}/E\pi^{1/2})\partial/\partial\gamma\int_\infty^\infty\exp\left(-\gamma\frac{(\epsilon-\epsilon_0)^2}{E^2}\right ) d\epsilon$.  This is equal to $(\gamma^{3/2}/E\pi^{1/2})\partial/\partial\gamma(E\pi^{1/2}\gamma^{-1/2})=1/2$. So
\[
\log(L/Q) \simeq -\log\left(\frac{L}{E}\sqrt{\frac{\gamma}{e\pi}}\right)
\]
or, equivalently,
\begin{equation}
\label{Qpowerpower}
Q=\sqrt{\frac{e\pi}{\gamma}}E.
\end{equation}

For the purposes of the comparison we make in Section \ref{Sect:Power}, we also calculate $Q$ for a {\it thermal} state at some given inverse temperature, $\beta$, of a system, S, with a power law density of states $\sigma(\epsilon)=A\epsilon^N$.  By (\ref{ProbDensPowerGibbs}) we will now have
\[
\log(L/Q)=\frac{\beta^{N+1}}{N!}\int_0^\infty\log\left(L\frac{\beta^{N+1}}{N!}\epsilon^Ne^{-\beta\epsilon}\right)\epsilon^Ne^{-\beta\epsilon}d\epsilon
\]
\[
=-\log\left(\frac{L\beta}{N!}\right)-\frac{\beta^{N+1}}{N!}\int_0^\infty(N\log(\beta\epsilon)-\beta\epsilon)\epsilon^Ne^{-\beta\epsilon}d\epsilon.
\]
We may do the integral here by noticing that $\int_0^\infty x^n\log x e^{-bx}dx$ $=d/d\alpha\int_0^\infty x^\alpha e^{-bx}dx|_{\alpha=n}$ $=d/d\alpha(b^{-(\alpha+1)} \Gamma(\alpha+1))|_{\alpha=n}$
$=-(\log b) b^{-(\alpha+1)}\Gamma(\alpha+1)+b^{-(\alpha+1)}d/d\alpha\Gamma(\alpha+1))|_{\alpha=n}$ $=-(\log b) b^{-(n+1)}n!+b^{-(n+1)}\Gamma(n+1)\psi(n+1)$ where $\Gamma$ denotes the gamma function and $\psi(n+1)$ the digamma function (see e.g.\ \cite{Gradshteyn}) of $n+1$ which (see again \cite{Gradshteyn}) is equal to $\sum_{k=1}^n 1/k-C$ where $C$ is Euler's constant ($=0.5772\dots$).  Using this, we conclude that $\log(L/Q)=$
\[
-\log(L\beta)+\log N! -N\psi(N+1)+N+1
\]
which, using Stirling's approximation (which tells us that $\log n!=(n+1/2)\log n - n +(1/2)\log(2\pi)+1/(12n) +O(1/n^2))$ and the asymptotic expansion of $\psi(n+1)$ ($=\log n + 1/(2n) + O(1/n^2)$) is equal to
\begin{equation}
\label{Qpowerthermal}
-\log(L\beta)+\frac{1}{2}\log N+\frac{1}{2}\log(2\pi)+\frac{1}{2} + O(1/N)
\end{equation}
from which we conclude that
\begin{equation}
\label{Qpowerthermal}
Q\simeq \frac{\sqrt{2e\pi N}}{\beta}.
\end{equation}
We note that if we identify $N$ here with $N_{\mathrm S}$ and if $N_{\mathrm B} \gg N_{\mathrm S}$, then, by (\ref{gamma}), $\gamma \simeq N_{\mathrm B}^2/2N_{\mathrm S}$.  If, additionally, we take $\beta$ in (\ref{Qpowerthermal}) to be $d\log\sigma_{\mathrm B}(\epsilon)/d\epsilon|_{\epsilon=E}$ which (with  $\sigma_{\mathrm B}(\epsilon)=A\epsilon^{N_{\mathrm B}}$) equals $N_{\mathrm B}/E$, then the  $Q$ of (\ref{Qpowerpower}) and (\ref{Qpowerthermal}) both take the same value
$\sqrt{2e\pi N_{\mathrm S}}E/N_{\mathrm B}$.
This agreement is to be expected since, as we discussed in Sections \ref{Sect:Intro} and \ref{Sect:Power}, in this situation, where S is small and in the case of a total microcanonical state, the reduced density operator of S will be close to that of a thermal state at
inverse temperature $d\log\sigma_{\mathrm B}(\epsilon)/d\epsilon|_{\epsilon=E}$.  So the agreement of the two $Q$ in this regime serves as a check on the correctness of our two formulae (\ref{Qpowerpower}) and (\ref{Qpowerthermal}).

However, in Section \ref{Sect:Power} we were interested in comparing the properties of the (as we show there) {\it non-thermal} reduced state of S when $N_{\mathrm S}$ and $N_{\mathrm B}$ are of comparable size and the totem is in a microcanonical state with the properties of a thermal state of S with the same expected energy.  Treating, for simplicity, the case where $N_{\mathrm S}=N_{\mathrm B} = N$, say, the relevant inverse temperature, $\beta$, is $2(N+1)/E$ ($\simeq 2N/E$ for large $N$) and $\gamma$ (\ref{gamma}) is $4N$.  We then find that the `thermal' $Q$ (Equation (\ref{Qpowerthermal})) becomes $\sqrt{e\pi/2N}E$ whereas the `microcanonical' $Q$ (Equation (\ref{Qpowerpower})) becomes $(1/2)\sqrt{e\pi/N}E$.  Thus the entropy of the comparison thermal state of S is bigger than the entropy, $S^{\mathrm{microc}}_{\mathrm{S}}$, of the reduced state of S  by $\log2/2$.  While this is a small difference it is conceptually significant that it is not zero.

\subsection{\label{totement} On the entropy of the totem and more about $\Delta$}

Our general framework involves a system, S, and an energy bath, B, comprising a totem.  A natural question is:  What is the relationship between the entropy of S, the entropy of B, and the entropy of the totem? The answer to this question depends, first of all, on whether we are contemplating the, traditional, microcanonical, scenario, or the modern scenario in which the state of the totem is pure -- albeit chosen at random amongst the set of states in our energy range.  In the latter, modern, scenario, there is only one entropy:  As we mentioned in Sections \ref{Sect:Entropy} and \ref{Sect:Gen}, the (von Neumann) entropy of S is equal to the (von Neumann) entropy of B and both are equal to the $\{$system$\}$-$\{$energy bath$\}$ entanglement entropy of the totem; the von Neumann entropy of the totem is of course zero.

In the microcanonical scenario, one might, naively, expect the entropy of the totem to be the sum of the entropy of S and the entropy of B.  But, as we shall see, this is not true.  One way to see that it cannot be true is to notice (see again below) that the entropy of the totem (which will obviously have the value
$\log M$, $M$ as in (\ref{micro}) and (\ref{sumnorm})) -- below we shall call this $S^{\mathrm{microc}}_{\mathrm{totem}}$  -- depends on the width, $\Delta$, of our energy band $[E, E+\Delta]$ whereas, as we saw in Section \ref{Sect:Gen} (for a suitable range of $\Delta$) the entropies of S and B do not.   What is of course always true (and applies to both modern and microcanonical scenarios) is the property of {\it subadditivity} \cite{Wehrl} which guarantees that the the entropy of the totem,  must be less than or equal to the sum of the entropies of S and B.

Actually, it turns out, in all three cases we have studied here (i.e.\ with system and energy bath densities of states of power-law form and [equal] exponential or quadratic exponential form) that the sum of the entropies of S and B is {\it close to} the entropy of the totem.  We have not attempted to formulate any general precise statement of what we mean by this, nor have we attempted to offer any general explanation as to why this should be the case but content ourselves simply with making the content of this statement manifest for each of our three density-of-states models:

For our (equal) exponential densities of states (\ref{sigmasbexponential}) we notice that, by (\ref{Mexp})
\[
S^{\mathrm{microc}}_{\mathrm{totem}}=\log M = bE + \log(c^2E\Delta)
\]
whereas, by (\ref{Smicroexp})
\[
S^{\mathrm{microc}}_{\mathrm{S}}+S^{\mathrm{microc}}_{\mathrm{B}}=bE + \log(c^2E^2).
\]

For our (equal) quadratic exponential densities of states (\ref{sigmasbexpsquared}), by (\ref{Mexpenergysquared})
\[
S^{\mathrm{microc}}_{\mathrm{totem}}=\log M = qE^2 + \log\left(\frac{K^2\Delta}{2E}\right)
\]
whereas (see (\ref{Smicrocexpesq}) and the paragraph after (\ref{Smodernexpesq}))
\[
S^{\mathrm{microc}}_{\mathrm{S}}+S^{\mathrm{microc}}_{\mathrm{B}}=qE^2 + O(1).
\]

For our power-law densities of states (\ref{sigmasbpower}), we first obtain a good approximate formula for $M$ by noting that the value of the integral in  (\ref{Mpowerintegral}) is equal to the ratio of the maximum value of
its integrand, $\epsilon_0^{N_{\mathrm{S}}}(E-\epsilon_0)^{N_{\mathrm{B}}}$ ($\epsilon_0$ as in (\ref{antipopular})) to its maximum value when normalized, which, by our Gaussian approximation,
(\ref{Pspowerapprox}), is well-approximated by $(1/E)\sqrt{\gamma/E}$, $\gamma$ as in (\ref{gamma}).  Thus we have (to a very good approximation)
\[
S^{\mathrm{microc}}_{\mathrm{totem}}=\log M = \log\left(A_{\mathrm S}A_{\mathrm B}E\Delta
\sqrt{\frac{\pi}{\gamma}}\epsilon_0^{N_{\mathrm S}}(E-\epsilon_0)^{\mathrm N_{\mathrm B}}\right)
\]
whereas, by (\ref{Smicrocpower}) for S and its obvious counterpart for B and (\ref{Qpowerpower}), 
\[
S^{\mathrm{microc}}_{\mathrm{S}}+S^{\mathrm{microc}}_{\mathrm{B}}= 
\log\left(A_{\mathrm S}A_{\mathrm B}E^2\frac{e\pi}{\gamma}\epsilon_0^{N_{\mathrm S}}(E-\epsilon_0)^{\mathrm N_{\mathrm B}}\right).
\]

We see that subadditivity in the exponential case entails $\Delta < E$.  In the quadratic exponential case, (and neglecting the $O(1)$ term) it entails $\Delta < 2E/K^2$, and in the power-law case, it entails $\Delta < e\sqrt{\pi/\gamma}E$.  When $N_{\mathrm S}=N_{\mathrm B} = N$, say, $\gamma = 4N$ and this latter inequality amounts to $\Delta < (e\sqrt{\pi}/2)(E/\sqrt{N})$.  The first of these inequalities ($\Delta < E$) is obviously consistent with almost any sort of smallness assumption on $\Delta$.  The other two inequalities indicate a need to be more precise than we were, in our rather sketchy remarks in Section \ref{Sect:Intro} and in our subsequent derivations, about what is the appropriate range of $\Delta$ for any given pair of densities of states, $\sigma_{\mathrm S}$ and $\sigma_{\mathrm B}$, in order for our arguments and approximations to be valid.   We shall, however, not pursue this issue further in the present paper except to deduce that the above inequalities must be necessary conditions on the value of $\Delta$.

\section{\label{Sect:moretherm} More about thermality: Purification}

Throughout the preceding sections we have assumed (see the paragraph after Equation (\ref{coupling}))  that both our system, S, and our energy bath, B, have densities of states which are positively supported (i.e.\ the Hamiltonians $H_{\mathrm{S}}$ and $H_{\mathrm{B}}$ are positive) and monotonically increasing and we have been concerned exclusively with totem states which are (close to) stationary states for a totem Hamiltonian, (\ref{coupling}), which weakly couples S and B.   In this short section, we point out that the prospects for the thermality of either S or B become much less constrained if we relax some of these assumptions.   In particular, and in the spirit of the `modern' approach, given {\it any} system density of states, $\sigma_{\mathrm{S}}(\epsilon)$, whatsoever (provided only it grows sufficiently slowly for the desired thermal state to exist) one can always find an energy bath density of states and a pure totem state such that the reduced state of the system is an exactly thermal state at any given temperature.  

In fact, there is a well-known procedure, in the spirit of the modern approach,  
known as `purification' (see e.g.\ \cite{KayPuri} and references therein; see also the papers on `thermofield
dynamics' \cite{ThermoField} and \cite{PureThermal} which are based on the same idea -- we note that the term `purification' seems to be due to Powers and St$\o$rmer \cite{Purification}) by which a system, S, with any density of states whatsoever (but we shall assume it to be positively supported and monotonically increasing)  and in any non-pure state one wishes to prescribe (but we are interested
in a thermal state at some inverse temperature $\beta$) may be provided with a notional energy bath, B, such that there is a choice of pure state on the resulting totem for which the reduced density operator on S is equal to the the prescribed state.

The essential idea of purification is based on the fact that any density operator, $\rho$, on a Hilbert space, $\cal H$, takes the form
\[
\rho=\sum_i \rho_i|\psi_i\rangle\langle\psi_i|,
\]
the $\rho_i$ being positive numbers which sum to 1 and $\psi_1,\psi_2, \dots$ being an orthonormal basis for $\cal H$, and on the observation that this can be viewed as arising as the partial trace over the second
copy of $\cal H$, in the tensor product ${\cal H}\otimes {\cal H}$, of the pure density operator, 
$|\Psi\rangle\langle\Psi|$, where 
\[
\Psi=\sum_i \rho_i^{1/2}\psi_i\otimes\psi_i.
\]
The easiest way to see this is to notice that, for any linear operator, $A$, on $\cal H$,
\[
\langle\Psi|(A\otimes I) \Psi\rangle_{{\cal H}\otimes{\cal H}}={\rm tr}(\rho
A)_{\cal H}.
\]
If we now specialize to the thermal situation where the $\psi_i$ are the energy eigenstates of a Hamiltonian, $H$, on $\cal H$ with energy eigenvalues say $e_i$ and $\rho_i=Z^{-1}e^{-\beta e_i}$ and identify $\cal H$ both with the system Hilbert space, ${\cal H}_{\mathrm{S}}$, and also with the Hilbert space, ${\cal H}_{\mathrm{B}}$, of our notional energy bath, then, if we also take the energy bath Hamiltonian to equal $H$ (and therefore the energy levels of the energy bath to be the same as the energy levels of the system)  then $\Psi$ provides us with a pure totem state with the property that the reduced state of the system (and also the reduced state of the energy bath) is exactly thermal.  With the notation of Section \ref{Sect:Background}, we would write ${\cal H}_{\mathrm{B}}={\cal H}_{\mathrm{S}}$ and 
take $\Psi$ on ${\cal H}_{\mathrm{S}}\otimes {\cal H}_{\mathrm{B}}$ to be given by
\begin{equation}
\label{pure}
\Psi=Z_{S,\beta}^{-1/2}
\sum_{\epsilon=\Delta}^\infty e^{-\beta\epsilon/2} \sum_{i=1}^{n_{\mathrm{S}}(\epsilon)}
|\epsilon,i\rangle\otimes|\epsilon,i\rangle.
\end{equation}
Then the partial trace of $|\Psi\rangle\langle\Psi|$ over ${\cal H}_{\mathrm{S}}$ will equal the $\rho^{\mathrm{Gibbs}}_{{\mathrm S},\beta}$ of Equation (\ref{sGibbs}).  

This achievement of exact thermality contrasts with the situation discussed in Section \ref{Sect:Ans} (see in and after the paragraph containing Equation (\ref{PsGibbs})) where (on the `modern approach') the totem state is assumed to be randomly chosen from amongst totem states in a narrow range of energies for a weakly coupled total Hamiltonian.  As we saw there and in the rest of Part 1, with that assumption, and for systems of comparable size, thermality can never be achieved exactly and can only be approximately achieved for certain special densities of states -- such as, in particular, the exponential and quadratic exponential cases discussed in Sections \ref{Sect:Exp} and \ref{Sect:ExpEsq}.  However, in the purification mechanism described here, the state, $\Psi$, of the totem is -- say if we regard the totem Hamiltonian to be given by Equation (\ref{coupling}) with $H_{\mathrm{B }}=H_{\mathrm{S}}$ (and, say, with no coupling term at all) -- clearly not even close to an energy eigenstate; in other words, the totem is in a highly non-stationary state.  We remark that this purification mechanism does not seem to play much of a role in everyday physics as a mechanism by which a system can get to be hot, although, interestingly, essentially this mechanism has been made use of in the laboratory \cite{YurkePotasek} to produce thermal states of photons.  (See also the remark about the Unruh effect in the next paragraph and the remarks about quantum black holes in Section \ref{Sect:BHimplications}.)

We further remark that there is an alternative reinterpretation of Equation (\ref{pure}) in which one ascribes to the `energy bath', B, the Hamiltonian $H_{\mathrm{B }}=-H_{\mathrm{S}}$ (and substitutes these Hamiltonians into the totem Hamiltonain formula (\ref{coupling})).  With this interpretation, the state of the totem is an eigenstate of totem energy (with totem energy eigenvalue zero!) and so a stationary state.  But now the density of states of the energy bath is negatively supported!  This latter interpretation can be said to be realized in the Unruh effect (see e.g.\ \cite{KayDoubleWedge} and reference therein) whereby the vacuum state of a relativistic quantum field theory in Minkowski space, restricted to a right-Rindler wedge, is a thermal state with respect to Lorentz boosts;  the left-Rindler wedge plays the role of our energy bath, B, and this can be thought of as a copy of the right-wedge quantum system but with a Hamiltonian which is the negative of the right-wedge Rindler Hamiltonian.

\section{\label{Sect:BHimplications} Implications for quantum black holes}

\subsection{\label{Sect:BHquad}The (problematic) connection between our results on quadratic exponential densities of states in Section \ref{Sect:ExpEsq} and black hole thermodynamics}

There is a striking, at least superficial, resemblance between  Equations (\ref{betaequals2qE}) and 
(\ref{Smicrocexpesq}) in Section \ref{Sect:ExpEsq} -- i.e.\ $\beta=2qE$ and 
$S^{\mathrm{microc}}_{\mathrm{S}} (= S^{\mathrm{microc}}_{\mathrm{B}})= qE^2/2$ (up to an $O(1)$ correction) -- for the inverse temperature and the traditional microcanonical entropy  of a weakly coupled system, S, and energy bath, B,  with equal quadratic exponential densities of states
(\ref{sigmasbexpsquared}) and the equations (see Section \ref{Sect:BlackHoles}) $\beta=8\pi G{\cal M}$ and $S=4\pi G{\cal M}^2$ for the Hawking inverse temperature and entropy of a Schwarzschild black hole.  In fact, if we identify, say (see below), the mean energy of B -- i.e.\ (see (\ref{expesqenergy})) half the totem mean energy, $E/2$ -- with the black hole mass, $\cal M$, and identify $q$  with $2\pi G$, the entropy,  $S^{\mathrm{microc}}_{\mathrm{B}}$  matches the Hawking entropy and the (inverse) temperatures match too.

This might seem to suggest that, if one identifies the system, S, with `matter' and the energy bath, B, with `gravity', then the traditional microcanonical strand of Section  \ref{Sect:ExpEsq} may provide a good model for a black hole in contact with its thermal atmosphere in a box and a good explanation for the microscopic origin of the entropy of this system.    (Of course, we must bear in mind that, in this model, the state is only thermal in the approximate sense explained in Section \ref{Sect:ExpEsq}.)  And on the other hand, our result, (\ref{Smodernexpesq}), that, with the densities of states (\ref{sigmasbexpsquared}), the `modern' totem pure state entropy, $S^{\mathrm{modapprox}}_{\mathrm{S}}$ vanishes (up to a term of order 1 in $E$) might seem to be at odds with our matter-gravity entanglement hypothesis described in \cite{Kay1, Kay2, KayAbyaneh} and in Section \ref{Sect:BlackHoles} -- which entails that the total matter-gravity state of a black hole is a pure state.   However, we need to realize that the results of Section \ref{Sect:ExpEsq} assume that the dynamics of the totem is governed by a totem Hamiltonian  of the schematic form (\ref{coupling}) with both S and B (now to be interpreted as `matter' and `gravity') Hamiltonians positive and weakly coupled.  Yet, notoriously, it seems unlikely to be possible to have a quantum theory of gravity within the scope of these basic assumptions (albeit these assumptions do seem to apply to the weak-string coupling limit if the fundamental degrees of freedom are taken to be those of a string rather than of the gravitational field itself -- see Section \ref{Sect:Betterstring}).   Already classical general relativity is nonlinear and, unlike the situation for (\ref{coupling}), energy (mass) is not additive.  So the fact that the mean energies of system and energy bath (modelling matter and gravity) are equal in our model seems strange.  (It is also unclear whether we should identify the entropy of the black hole with  $S^{\mathrm{microc}}_{\mathrm{B}}$ as we did above, or with $S^{\mathrm{microc}}_{\mathrm{B}}+S^{\mathrm{microc}}_{\mathrm{S}}$, which is twice as big, or with
$\log M$ -- see also Section \ref{totement}.)  Furthermore (see the remarks after Equation (\ref{expesqenergy})) in this model, the mean energy, $E/2$, of each of matter and of gravity is anyway just the mean of an energy probability density (the $P_{\mathrm S}(\epsilon)$  of (\ref{enprobdensexpsquaredapproxbeta}) and of Figure \ref{Fig3})  which is peaked at the extremes, $\epsilon=0$ and $\epsilon=E$, while of course (cf.\ after Equation (\ref{expesqenergy}))  the energy probability densities of matter and gravity are perfectly anticorrelated.  So the model predicts large statistical fluctuations, with (to the extent that it makes sense to talk about the energy of subsystems in general relativity) probability distributions for matter and gravity being such that, approximately, with probability one half, gravity has all the energy and matter none, and with probability one half, matter has all the energy and gravity none.  The latter case (where there is presumably no black hole) is then a particular problem because (see the next paragraph) presumably the quadratic exponential form of the density of states presupposed in the model for matter becomes invalid when a black hole is not present.

In fact, turning to more specifically quantum aspects, aside from all the usual problems of quantum gravity (non-renormalizability etc.)\ it would seem to be incorrect to assume that one can ascribe a single density of states to each of gravity and matter throughout changes of state which include the formation of black holes.  Rather it would seem that one has to assign, in some sense, a `state-dependent density of states' to matter; in the absence of black holes, the densities of states of common forms of matter (including photons) grow much more slowly than quadratic exponentials, while, plausibly, when a black hole is present, they do grow as quadratic exponentials (with subleading corrections).  (Evidence for this latter statement is provided by the `brick wall' approach \cite{tHooftBrick, MukohyamaIsrael, KayOrtiz} which suggests that the matter entropy is comparable to the gravitational entropy when a black hole is present.  We shall also argue in \cite{Kayprefactor} that the string scenario we advocate in Section \ref{Sect:Betterstring} and discuss further in \cite{Kaycompanion, Kayprefactor} leads, plausibly, to just such a state-dependent density of states for matter.)  Moreover, the situation is further complicated by the very different status of the concept of `time' in general relativity from that presupposed in traditional formulations of quantum theory.

In the light of all these problems and difficulties, and of our current lack of knowledge as to how to resolve them (other than to assume that a black hole is an [ill-understood] strong string-coupling limit of a certain [better understood] state of string theory at weak coupling -- see Section \ref{Sect:Betterstring}) it seems to us still reasonable to cling to our matter-gravity entanglement hypothesis and our entanglement picture of black hole equilibrium (see Section \ref{Sect:BlackHoles}).   Indeed, there would seem just as much reason to believe in a model along the following `modern' lines (inspired by the idea of `purification' outlined in Section \ref{Sect:moretherm}) as to believe in the above model within the traditional microcanonical strand of Section  \ref{Sect:ExpEsq}:  

\smallskip

{\it A tentative possible `modern' model with the correct Hawking value for the entropy:} `Matter' is modelled as a `system', `S', and gravity as an `energy bath', `B', which each have a density of states as in (\ref{sigmasbexpsquared}), and  the total state of the matter-gravity totem which corresponds to a black hole of mass ${\cal M}=E/2$ is modelled by the pure density operator 
$|\Psi\rangle\langle\Psi|$ where $\Psi$ is 
\begin{equation}
\label{modernBHmodel}
\frac{1}{\sqrt{M}}\sum_{\epsilon=\Delta}^E 
\sqrt{n_{\mathrm{B}}(E-\epsilon)}\sum_{i=0}^{n_{\mathrm{S}}(\epsilon)}
|\epsilon, i\rangle_{\mathrm{S}}|\epsilon, i\rangle_{\mathrm{B}}
\end{equation}
where (cf.\ (\ref{degdensS}), (\ref{degdensB}) and (\ref{sigmasbexpsquared})) 
$n_{\mathrm{B}}(\epsilon)=\exp(q\epsilon^2)\Delta$ and $q=2\pi G$.

\smallskip

This state is easily seen to have partial traces over S and B identical to those, i.e.\ the $\rho^{\mathrm{microc}}_{\mathrm{B}}$ and
$\rho^{\mathrm{microc}}_{\mathrm{S}}$ of (\ref{microreduced}), of the microcanonical state (\ref{microsysbath}) $\rho_{\mathrm{microc}}$ for the same densities of states (i.e.\ (\ref{sigmasbexpsquared})) and thus, obviously, it will equally well predict an inverse temperature of $2qE$ and a system entropy ($=$ energy bath entropy) of $qE^2/2$ (plus the same $O(1)$ correction).

While we have argued that this latter `modern' model is no less justified than our above microcanonical model, in view of the difficulties and problems mentioned above, it is still quite unclear what status should be assigned to it or how seriously it should be taken.   There is also an apparent flaw in this tentative model in that the pure state of the totem is far from being an energy-eigenstate.  It could possibly be that this is the best one can do when one attempts to force a strong-coupling situation into a weak-coupling mould, or maybe the model should be modified along the lines of the alternative reinterpretation of Equation (\ref{pure}) in Section \ref{Sect:moretherm} so that the totem state is modelled as an energy eigenstate, at the expense of having densities of states which are not monotonically increasing and/or not positively supported.   Finally, there is the same strange feature that we raised above for our microcanonical model, that both the gravity and the matter are modelled as having, on average, exactly half of the mass (i.e.\ $E/2$) of the totem.   Also, as in the microcanonical model, the energy probability densities of matter and gravity are predicted to each have 
the same energy probability density  (the $P_{\mathrm S}(\epsilon)$  of (\ref{enprobdensexpsquaredapproxbeta}) and of Figure \ref{Fig3}) with equal-sized peaks at $\epsilon=0$ and  $\epsilon=E$.  Albeit, interestingly (and related to the fact that the totem state is far from being an energy eigenstate) this model differs from the microcanonical model in that the two energy probability densities are now no longer anticorrelated but, instead, perfectly correlated:  One sees immediately from (\ref{modernBHmodel}) that when gravity has energy near 0, so will matter, and when gravity has energy near $E$, so will matter.   So at least one of the problems we mentioned above (the one we referred to as a ``particular problem'') for the microcanoncial model seems to be alleviated in the above proposed `modern' model.  Another problem which is alleviated with this tentative modern model is that it is clear, in this modern model that the entropy should be identified with $S^{\mathrm{modern}}_{\mathrm{B}}$, whereas, as we remarked above, in the microcanonical model it was not clear whether it should be identified with $S^{\mathrm{microc}}_{\mathrm{B}}$ or with (approximately -- see Section \ref{totement}) twice this value; in the modern model, there {\it is} only one entropy -- i.e.\ the S-B entanglement entropy ($=S^{\mathrm{modern}}_{\mathrm{B}}=S^{\mathrm{modern}}_{\mathrm{S}}\simeq S^{\mathrm{modapprox}}_{\mathrm{B}}=S^{\mathrm{modapprox}}_{\mathrm{S}}$)!

One would also wish to be able to relate our discussion, in Section \ref{Sect:peak}, of cases where the probability density is sharply peaked, to the results, \cite{Hawking:1976de}, of Hawking on his microcanonical approach to quantum black holes.  These latter results of Hawking do seem to form a physically compelling and coherent picture and one would like to understand whether and, if so, how, they can be reconciled with the modern strand of results in Section \ref{Sect:peak} even though they seem, superficially, to be more easy to understand in terms of the microcanonical strand of work there and seem, superficially, to be at odds with the modern strand.  We hope to address this question elsewhere.  Suffice it to to say here that, again, the difficulties and problems mentioned above are of at least equal relevance also to this issue and thus it seems  difficult, also for this issue, to reach a fully convincing conclusion either way \cite{micro}.

\subsection{\label{Sect:Betterstring}Towards a better understanding of black hole entropy in terms of string entropy}

Where we have been able to make a, we think, persuasive, case for the relevance, of the `modern' strand of ideas in the present paper -- used in combination with the (see Section \ref{Sect:BlackHoles}) matter-gravity entanglement hypothesis of \cite{Kay1, Kay2, KayAbyaneh} -- to the understanding of black hole entropy is with a model which relates the work in Sections \ref{Sect:Exp} and \ref{Sect:ExpForm} here, concerning densities of states which grow exponentially with energy, to an understanding of black hole entropy based on the idea that black holes are strong string-coupling limits of states of weakly coupled string theory.   This application of our work to quantum black holes seems to be more well-founded because, unlike in the situations discussed above, we {\it would} expect the general assumptions we made at the outset here (positive Hamiltonians, weak coupling etc.)\ to be applicable to the weak-coupling regime of string theory.

In 1993, Susskind \cite{Susskind} proposed, and in 1997, Horowitz and Polchinski \cite{HorowitzPolchinski, HorowitzReview}, gave further evidence, of a semi-qualitative nature, for, the hypothesis that a (say 4-dimensional, Schwarzschild) black hole can be interpreted as the strong string-coupling limit of a certain state of string theory at weak coupling consisting of a (single) long string.  These authors argued that one obtains, with this interpretation, an explanation of black hole entropy in terms of known formulae, based on the `counting of states', in string theory at low string coupling.  Moreover, in related work on extremal, and near extremal, black holes (an important early paper was Strominger and Vafa \cite{StromingerVafa}), full quantitative agreement was found between the results of such a string theory approach to black hole entropy (and other related quantities) and the previously established Hawking formulae.  It was then claimed that this work, not only gave a microscopic explanation for black hole entropy but also, in view of the fact that string theory is a standard quantum mechanical theory with a unitary time evolution,  that it resolved  the Information Loss Puzzle (see Section \ref{Sect:BlackHoles}).  We next wish to argue that, while we agree all this work seems to provide us with an important clue towards the microscopic understanding of black-hole entropy which, plausibly, may well turn out to be consistent with a resolution to the Information Loss Puzzle, it cannot, by itself, be regarded as a complete explanation of these things.   (We shall expand on these arguments in two companion papers, \cite{Kaycompanion} and \cite{Kayprefactor}.)  Our point is simply that, what is actually calculated in the cited work (for example in \cite{StromingerVafa}) is not the {\it entropy} of a particular black hole state, but rather the (logarithm of the) {\it degeneracy} of a given black hole energy level.  No explanation was given in the cited work as to why the logarithm of this degeneracy should be interpreted physically as an entropy.  After all the $n$'th energy level of the textbook non-relativistic Hydrogen-atom Hamiltonian has a degeneracy of $n^2$ but we would not predict from this that a Hydrogen atom has an entropy of $\log n^2$!   Of course it is true that the logarithm of the degeneracy of an energy level is the same thing as the von Neumann entropy of the microcanonical density operator $(1/d)\sum_{i=1}^d|i\rangle\langle i|$ where we denote by $|i\rangle$, ($i=1\dots d$) the elements of a basis of states with the given energy.    But if we were to attempt to interpret e.g.\ the Strominger Vafa results as meaning that a black hole should be modelled by such an (impure!) microcanoncial state,  then the Information Loss Puzzle would surely return:  How, in a string theoretic description of a dynamical process of black hole formation, can a presumably pure initial string theory state evolve into such an (impure) microcanonical state?  (Such a microcanonical state also wouldn't fit with our picture of black holes as thermal states.)  

Actually, the Horowitz-Polchinski work is couched in terms, not of the degeneracy of a particular energy level of string theory, but rather of the (averaged out) density of states of a long string. The problem is then compounded by the fact that a density of states is not a dimensionless quantity, so it is not physically meaningful to take its logarithm \cite{guess}.  (In fact similar remarks apply to those we make in Paragraph (a) of Section \ref{Sect:peak}.)

Focusing on this Horowitz-Polchinski work, we shall next propose a modified version of their scenario, based on the modern strand of Sections \ref{Sect:Exp} and \ref{Sect:ExpForm} of the present paper, which seems to overcome the above difficulties and to offer the promise of a fully satisfactory explanation of black hole entropy in terms of string theory, consistent with unitarity and consistent with a resolution to the Information Loss Puzzle -- namely with the resolution to the Information Loss Puzzle we proposed in \cite{Kay1, Kay2, KayAbyaneh} based on our matter-gravity entanglement hypothesis (see above and Section \ref{Sect:BlackHoles}).  This will be further discussed in \cite{Kaycompanion} and further developed in \cite{Kayprefactor}, whose content we indicate very briefly at the end of this subsection. 

We begin by briefly recalling the basic argument of Susskind, Horowitz and Polchinski \cite{Susskind, HorowitzPolchinski} as expounded in \cite{HorowitzReview}.   Their basic hypothesis is that, as one scales the string length scale, $\ell$, up and the string coupling constant, $g$, down from their physical values, keeping Newton's gravitational constant, $G=g^2\ell^2$, fixed, a (4-dimensional)  Schwarzschild black hole of mass ${\cal M}$ will turn into a long string with roughly the same energy,  $\epsilon={\cal M}$.  The density of states of such a long string, in the limit of weak coupling, is known, very roughly (i.e.\ omitting an approximately inverse-power prefactor -- see below) to take the exponential form
\begin{equation}
\label{longstringapproxdens}
\sigma_{\mathrm{long\ string}}(\epsilon) = C_{\mathrm{ls}}e^{\ell \epsilon}
\end{equation}
($C_{\mathrm{ls}}$ a constant with the dimensions of inverse energy of the same order of magnitude as $\ell$).

The gist of the argument is that the `logarithm' of this is approximately given by 
\begin{equation}
\label{Horo1}
S_{\mathrm{long\ string}}=\ell\epsilon
\end{equation}
and they refer to this quantity as (approximately) the `entropy of the long string at energy $\epsilon$'.  They then argue that this should be equated with the entropy of a (Schwarzschild) black hole provided that one does the equating (i.e.\ during the process of scaling $\ell$ described above) when, to within an order of magnitude or so,
\begin{equation}
\label{Horo2}
\ell=G{\cal M} 
\end{equation}
which is roughly the `size' of the black-hole.  (Cf.\ the fact that the Schwarzschild radius is $2G{\cal M}$.)

Combining (\ref{Horo1}) and (\ref{Horo2}) (and replacing $\epsilon$ by ${\cal M}$) they thus claim to predict that the entropy of the black hole will be within an order of magnitude or so of a constant times $G{\cal M}^2$ which agrees, up to an undetermined value for the constant, with the Hawking value, $4\pi G{\cal M}^2$, for the entropy of a black hole.  

In our view, what one is actually entitled to say, instead, is that, the number of energy eigenstates of a black hole in a band of width $\Delta$ around energy $\epsilon$ will, by (\ref{longstringapproxdens}) 
be $\ell\epsilon + \log(C_{\mathrm{ls}}\Delta)$, which, for a `reasonable-sized' $\Delta$ will be approximately the same as $\ell\epsilon$.  The argument in the previous two paragraphs then tells us that the number of energy eigenstates of a black hole in a band of width $\Delta$ around energy (i.e.\ mass) $\cal M$ will be within an undetermined constant, say $C$, of the order of 1, times $G{\cal M}^2$ (and thus, by the way, that the density of states of a Schwarzschild black hole should behave roughly as a constant times $\exp(CG\epsilon^2)$.) However, in our view, it remains a challenge to explain why the logarithm of this formula for the density of states of a black hole should equal (to within an unknown constant, $C$ of the order of 1) the Hawking formula for black hole entropy.

In our attempt to meet this challenge, we first posit that the scenario in which a black hole goes over to a single long string should be replaced by a scenario in which an equilibrium state (i.e.\ energy eigenstate) consisting of a black hole in contact with its matter atmosphere in a suitable box (on our view described by a pure total state -- see our `entanglement picture of black hole equilibrium' in Section \ref{Sect:BlackHoles}) with approximate total energy, $E$, goes over (again, as one scales the string length scale, $\ell$, up and the string coupling  constant, $g$, down from their physical values, keeping Newton's gravitational constant, $G=g^2\ell^2$ fixed) to a (pure) equilibrium state, with a similar total energy, consisting of a single long string in contact with an atmosphere of small strings in a suitably rescaled box.  

We now assume that the density of states of the long string takes (to the same rough approximation as above) the form of (\ref{longstringapproxdens}) and that the density of states of the stringy atmosphere, $\sigma_{\mathrm{string\ atmosphere}}(\epsilon)$,
takes (again, to a rough approximation) the similar, exponential, form
\begin{equation}
\label{stringatmosapproxdens}
\sigma_{\mathrm{string\ atmosphere}}(\epsilon) = C_{\mathrm{sa}}e^{\ell \epsilon}.
\end{equation}

If we now regard (most of) the stringy atmosphere as corresponding to `matter' and as playing the role of our `system', S, and the long string as corresponding to (most of) `gravity' and as playing the role of our energy-bath, `B', then it is plausible that these may be described by Hilbert spaces and Hamiltonians which, since we are at weak string coupling, should fall within the scope of the present paper, and in particular the dynamics should be described by a totem Hamiltonian of form (\ref{coupling}).  In view of the exponential growth of the densities of states, (\ref{longstringapproxdens}) and (\ref{stringatmosapproxdens}), we may therefore apply the formalism of Sections \ref{Sect:Exp} and \ref{Sect:ExpForm} (modified as explained in Endnote \cite{Ftnt8} to take into account the different prefactors, $C_{\mathrm{sa}}$ and $C_{\mathrm{ls}}$).  In particular, the modern strand of these sections tells us that a typical pure equilibrium state of our $\{$string atmosphere$\}$-$\{$long string$\}$ totem with energy around $E$ will, with a high probability, have an entropy very close to that given by Equation (\ref{Spureexp}) with $b=\ell$  (with the modification to the logarithmic term given in Endnote \cite{Ftnt8}).  I.e.\ ignoring the logarithmic term, by
\[
S=\ell E/4,
\]
while the (expected) energy of the long string, $\bar\epsilon_{\mathrm{long\ string}}$ will (see again Endnote \cite{Ftnt8}) be given by
\[
\bar\epsilon_{\mathrm{long\ string}}=E/2
\]
(and, of course the mean energy of the stringy atmosphere will also be $E/2$ in this model).

In parallel to the philosophy of \cite{Susskind, HorowitzPolchinski, HorowitzReview}, we now assume that, when we scale $g$ back up and $\ell$ down, keeping $G=g^2\ell^2$ constant and keeping $\Psi$ the `same', we can equate $\bar\epsilon_{\mathrm{long\ string}}$ with the mass, $\cal M$, of the black hole when $\ell=XG{\cal M}$, say, where $X$ is an adjustable parameter of the order of 1. We thereby obtain the prediction $S=XGM^2/2$, as the value for the entanglement entropy of black hole and thermal atmosphere in the `same' (i.e.\ after rescaling) state $\Psi$.  But it is also plausible (as indicated above -- see the relevant Endnote in \cite{Kayprefactor} for further discussion) that this is approximately the same as the entanglement entropy between gravity and matter which, according to the matter-gravity entanglement hypothesis of \cite{Kay1, Kay2, KayAbyaneh} and Section \ref{Sect:BlackHoles} is the physical entropy of the black hole.  We thus predict that the physical entropy of our black hole is (approximately) $XGM^2/2$.   This agrees with the Hawking entropy of $4\pi{\cal M}^2$ if we take $X=8\pi$.

Furthermore, we showed, in Section \ref{Sect:Exp}, that both S and B will be `$E$-approximately semi-thermal',  in the sense explained there, at inverse temperature $\beta= \ell$.  Equating this with the inverse black hole temperature when $\ell=XG{\cal M}$ predicts a black hole inverse temperature of $XG{\cal M}$ which, intriguingly, agrees with the inverse Hawking temperature for the same value of $X$ (i.e.\ $8\pi$).  We remark that we would not have correctly predicted both Hawking temperature and Hawking entropy for a single value of $X$ had we followed the traditional microcanonical, instead of the modern strand, of Section \ref{Sect:Exp} nor if we had adopted the approach of \cite{Susskind, HorowitzPolchinski, HorowitzReview} and defined the inverse temperature by $\beta=d(\log(\sigma_{\mathrm{long\ string}}(\epsilon))/d\epsilon$.  (In each case, the necessary values of $X$ for fitting the Hawking entropy and the Hawking temperature would have differed by a factor of 2.)  However we caution that it is not clear whether this nice feature of our modern model with exponential densities of states survives when (see next paragraph) we improve the model to include the appropriate approximately inverse-power prefactors.  We discuss this further in \cite{Kayprefactor}.  Nevertheless, our main point is that a `modern' model for black hole entropy, based on our matter-gravity entanglement hypothesis seems able to predict a temperature of the order of the Hawking temperature and an entropy of the order of the Hawking entropy.

The main deficiency in the above scenario is the adoption of the equations (\ref{longstringapproxdens}) and (\ref{stringatmosapproxdens}) for the approximate forms of the long-string and stringy-atmosphere densities of states.  These formulae omit important (dimension-dependent) approximately inverse-power prefactors,  and when one takes these into proper account, it turns out (with some, seemingly reasonable, assumptions) that the account of the origin of black hole entropy above and in \cite{Kaycompanion} needs significant changes and is even, in certain respects, misleading, although one arrives at similar final conclusions.  The prefactors are also needed to explain why an equilibrium weakly coupled string state in a box consists of a single long string surrounded by an atmosphere of short strings as we posited above.  Also, the statistical spread in energy of the string (and hence the predicted statistical spread in energy of the black hole)  around the mean energy $E/2$ will be altered with the correct prefactors.  All these matters will be discussed in our second companion paper \cite{Kayprefactor}.

\bigskip

\section*{\large Part 2: Full explanation of the formula (\ref{purereduced}) and arguments for the validity of the proposition in Section \ref{Sect:Ans}}

\bigskip

\section{\label{Sect:Prelim} The work of Lubkin and Page and other preliminaries and outline of the remainder of Part~2}

In Section \ref{Sect:Intro}, we mentioned the work of Lubkin, \cite{Lubkin},  where it is shown that a randomly
chosen pure density operator, $\rho^{mn}=|\Psi\rangle\langle\Psi|$,
on the tensor-product Hilbert space, ${\cal H}_m\otimes {\cal H}_n$, of a pair
of quantum systems -- ${\cal H}_m$ being $m$-dimensional and ${\cal H}_n$ being
$n$-dimensional -- will, for fixed $m$ and $n\gg m$, have, with high probability, a
reduced density operator, $\rho^{mn}_m$, on ${\cal H}_m$, which is close
to the maximally mixed density operator -- with components, in any
Hilbert space basis,  ${\rm diag}(1/m, \dots, 1/m)$.   We first need to recall some more details about this work as well as some further related developments which will be relevant throughout Part~2.

Lubkin justified the above statement and made it precise by obtaining a result which is (easily
seen to be) equivalent to the following exact formula for the mean value
(i.e.\ over Haar measure on the set of unit vectors $\Psi$) $\langle
{\rm tr}((\rho^{mn}_m)^2)\rangle$, of $(\rho^{mn}_m)^2$:  In our
notation
\begin{equation}
\label{Lubkinmeanrhosquared}
\langle {\rm tr}((\rho^{mn}_m)^2)\rangle=\frac{m+n}{mn+1}.
\end{equation}
We shall re-derive this result of Lubkin with a somewhat different
method in the next section (Section \ref{Sect:Haar}) since our full explanatioin of the formula in Equation
(\ref{purereduced}) and our arguments for the validity of our proposition of Section \ref{Sect:Ans}
will be closely based on it.  Lubkin then gave a simple general argument which amounts to the statement that, for any density operator, $\rho_m$, on an $m$-dimensional Hilbert space,
whenever  $m\langle{\rm tr}(\rho_m^2)\rangle-1\ll 1$, then the mean
value,  $\langle S(\rho_m)\rangle$, of the von Neumann entropy,
$S(\rho_m)$, of $\rho_m$ will be well-approximated by
\begin{equation}
\label{Lubkingeneralentropy}
\langle S(\rho_m)\rangle\simeq\log m-\frac{1}{2}\left(m\langle{\rm
tr}(\rho_m^2)\rangle-1\right ).
\end{equation}
Applying this result to $\rho^{mn}_m$,  (\ref{Lubkinmeanrhosquared}) implies that
whenever $m\ll n$,
\begin{equation}
\label{Lubkinmeanentropy}
\langle S(\rho^{mn}_m)\rangle\simeq \log m -\frac{m^2-1}{2(mn+1)}
\end{equation}
which may also be regarded as an alternative quantitative expression of the
qualitative property that, when $m \ll n$, most
$\rho^{mn}_m$ must be close to maximally mixed.

In (\ref{Lubkingeneralentropy}) and (\ref{Lubkinmeanentropy}) above, the von Neumann
entropy is defined in the usual way as in Equation (\ref{SvN}).

Around 15 years later, Page \cite{Page} showed that the formula
\begin{equation}
\label{Pagemeanentropy}
\langle S(\rho^{mn}_m)\rangle\simeq \log m -\frac{m}{2n}
\end{equation}
is a good approximation (with error term of order $1/mn$)
whenever $1\ll m\le n$, and noted that this agrees with
(\ref{Lubkinmeanentropy}) on their common domain of validity \cite{Ftnt2}. We note
here, in passing that, combining
the two estimates (\ref{Lubkinmeanentropy}) and (\ref{Pagemeanentropy}),
we can clearly write, simply, that whenever $m\le n$,
$\langle S(\rho^{mn}_m)\rangle = \log m -m/2n + O(1/mn)$ (since, when
$m$ and $n$ are both of order 1, the entropy can anyway only be of order 1) and
hence, combining this result with $m$ and $n$ interchanged with the
equality of $S(\rho^{mn}_m)$ and
$S(\rho^{mn}_{\ \ n})$ \cite{Ftnt3} (by which we mean the reduced density operator of $\rho^{mn}$ on
${\cal H}_n$ -- see Section \ref{Sect:Haar})
\begin{equation}
\label{LubkinPagemeanentropy}
\langle S(\rho^{mn}_m)\rangle = \log ({\rm min}(m,n)) -\frac{{\rm
min}(m,n)}{2{\rm max}(m,n)}+O\left (\frac{1}{mn}\right ).
\end{equation}

In the remainder of Part~2, we first introduce, in Section \ref{Sect:Haar}, a
useful coordinatization for unit vectors in an $N$-dimensional Hilbert
space in terms of which the  `Haar' measure of Section \ref{Sect:Intro} takes a
particularly convenient form.  In order to prepare the ground for our
subsequent generalization (see below) we then use this coordinatization
to obtain an alternative derivation of Lubkin's result
(\ref{Lubkinmeanrhosquared}). We also discuss in more detail the
qualitative consequence of Lubkin's result concerning the `almost
maximal mixedness' of the density operator $\rho^{mn}_m$ when $m \ll n$
(as previously pointed out by Lubkin, as mentioned in Section
\ref{Sect:Intro}) and we also point out a related important second
qualitative consequence concerning the nature of $\rho^{mn}_m$ in the
`opposite' situation when $m \gg n$.   Then, in Section \ref{Sect:Main}, we use a generalization of our alternative derivation of Equation
(\ref{Lubkinmeanrhosquared}) as well as a suitable generalization of our
argument for its two qualitative consequences to give the full statement of Equation (\ref{purereduced}) including an explanation of how the $n_{\mathrm{B}}(E-\epsilon)$-dimensional
subspace of the ($n_{\mathrm{S}}(\epsilon)$-dimensional)
energy-$\epsilon$ subspace of ${\cal H}_{\mathrm{S}}$ spanned by the $|\widetilde{\epsilon, i}\rangle$ depend on $\Psi$ and also to give our argument for the validity of our proposition (stated in full in Section \ref{Sect:Ans}) that  $\rho^{\mathrm{modern}}_{\mathrm S}$ (see the discussion after
(\ref{purereduced})) is well approximated by the $\rho^{\mathrm{modapprox}}_{\mathrm S}$ of Equation (\ref{purereduced}).  We end, in Section \ref{Sect:Further}, with two calculations which provide confirmatory evidence of the goodness of our approximation in situations such as those we discuss in Part~1.

\section{\label{Sect:Haar} A useful representation of Haar measure and
details on, and further consequences of, Lubkin's result}

Let $\cal H$ be an $N$-dimensional Hilbert space and let $\lbrace
E_1\dots E_N\rbrace$ be an arbitrary orthonormal basis.  Then, as usual,
we coordinatize an arbitrary vector, $\psi\in {\cal H}$, by the $N$-tuple
of complex numbers $(z_1, \dots, z_N)$ where $\psi = \sum_{a=1}^N z_a
E_a$. $\psi$ is, of course, then a unit vector if and only if $\sum_{a=1}^N |z_a|^2=1$.
So the set of normalized vectors in our Hilbert space is coordinatized
as the unit sphere in ${\mathbb C}^N$. (Writing $z_a=x_a + iy_a$ etc.\ we
see that this is obviously the `same thing' as the real unit
(2$N$-1)-sphere.)  Next we change to polar coordinates in each copy of
${\mathbb C}$ by setting $z_a=r_ae^{i\theta_a}$, whereupon the usual
volume element $dz_1\dots dz_N$ on ${\mathbb C}^N$  takes the form
$r_1\dots r_N dr_1\dots dr_N d\theta_1\dots d\theta_N$. Changing
coordinates further from $(r_1,\dots, r_N; \theta_1,\dots,\theta_N)$ to
$(r_1,\dots, r_{N-1};R;\theta_1,\dots,\theta_N)$, where
\begin{equation}
\label{unitsphere}
R^2=\sum_{a=1}^N r_a^2,
\end{equation}
this volume element is easily seen to become $r_1\dots r_{N-1}Rdr_1\dots
dr_{N-1}dR d\theta_1\dots d\theta_N$.  Next we note that, in these
latter coordinates, the unit sphere in ${\mathbb C}^N$ is defined by the
condition $R=1$.  Thus the usual area element, $dA$, on our unit sphere
is obtained by setting $R=1$ and removing the term $dR$ from this
formula.  i.e.
\[
dA=r_1\dots r_{N-1}dr_1\dots dr_{N-1}d\theta_1\dots d\theta_N.
\]
It is now convenient to replace the coordinates $r_a$ $(a=1,\dots, N-1)$ by
$w_a$ $(a=1,\dots, N-1)$ where
$w_a=r_a^2$, whereupon clearly
\[
dA=2^{N-1}dw_1\dots dw_{N-1}d\theta_1\dots d\theta_N.
\]
In view of the relation between the $w_a$ and the first $N-1$ of the
$r_a$ and the fact that the $r_a$ satisfy (\ref{unitsphere}) with $R=1$,
the ($N$-1)-tuple $(w_1,\dots, w_{N-1})$ clearly takes values which range
over the simplex defined by the inequalities $0\le w_a$ for each
$a$-value from 1 to $N-1$, together with the inequality  $\sum_{a=1}^{N-1}
w_a \le 1$. (We shall call this the {\it standard ($N$-1)-simplex}.) We
remark that we can think of the quantity $1-\sum_{a=1}^{N-1} w_a$ as
`$w_N$' for we will then have $w_N^2=r_N^2$.  So the latter inequality
can then be expressed as $0\le w_N$.  The $\theta_a$ values each, of
course, range over $[0,2\pi)$.  So the area of our unit sphere is the
integral of $dA$ over the above ranges for our variables which is
$2^{N-1}(2\pi)^N$ times the volume of our simplex.  But the latter is
easily seen to be $1/(N-1)!$.  So the area of our unit sphere is
$2\pi^N/(N-1)!$ which, of course, is the well-known value for the
surface area of the real (2$N$-1)-sphere.   We want our Haar measure to be
a probability measure, so we need to normalize it by dividing by this
surface area.  In conclusion, (up to an irrelevant set of measure zero)
we have coordinatized the set of normalized vectors in our $N$-dimensional
Hilbert space by products of ($N$-1)-tuples $(w_1,\dots, w_{N-1})$ whose
values range over our standard ($N$-1)-simplex, with $N$-tuples
$(\theta_1,\dots, \theta_N)$ whose values range over the standard (i.e.\
with all periods equal to $2\pi$) $N$-torus, and, with this
coordinatization, (normalized) Haar measure is simply the product
\begin{equation}
\label{Haar}
d\hbox{Haar}=d(\hbox{Simplex})\times d(\hbox{Torus})
\end{equation}
where
\begin{equation}
\label{Simplex}
d\hbox{Simplex}=(N-1)!dw_1\dots dw_{N-1}
\end{equation}
and
\begin{equation}
\label{Torus}
d\hbox{Torus}=(2\pi)^{-N}d\theta_1\dots d\theta_N.
\end{equation}

For later convenience, we next define, and record, the easy-to-check values of, certain
integrals of certain products of w's over our simplex w.r.t. $d\hbox{Simplex}$:

\begin{equation}
\label{J1}
J_1:=\int w_1 d\hbox{Simplex}=1/N,
\end{equation}
\begin{equation}
\label{J11}
J_{11}:=\int w_1^2 d\hbox{Simplex}=2/N(N+1),
\end{equation}
\begin{equation}
\label{J12}
J_{12}:=\int w_1w_2 d\hbox{Simplex}=1/N(N+1).
\end{equation}
Obviously we assume here, for $J_1$ and $J_{11}$, that $N$ is at least 2, and for
$J_{12}$, that $N$ is at least 3.  We note that $J_p$
(defined as $J_1$ but with $w_1$ replaced by $w_p$) will equal $J_1$ for
any other value of $p$ between 1 and N. Similarly (and with an obvious
corresponding notation) $J_{pp}=J_{11}$ for any other $p$ between 1 and
$N$, and $J_{qp}=J_{12}$ for any pair of {\it different} $q$ and $p$
between 1 and $N$. We reiterate that all this holds even if $q$ or $p$
is equal to $N$, in which case, as we remarked above, $w_N$ is taken to
mean $1-w_1- \dots -w_{N-1}$.

We next use this coordinatization to compute the average,
$\langle\rho^{mn}_m\rangle$, of $\rho^{mn}_m$ (see the paragraph before
Equation (\ref{Lubkinmeanrhosquared}) in Section \ref{Sect:Prelim}) over Haar
measure (with the result (\ref{meanrhomnm}) below) and also to
(re-)derive Lubkin's formula (\ref{Lubkinmeanrhosquared}) for
$\langle{\rm tr}((\rho^{mn}_m)^2)\rangle$.  (Here, as in Section \ref{Sect:Prelim},
we indicate averages with respect to Haar measure with angle-brackets
$\langle\quad\rangle$.) Let $\Psi$ be an arbitrary unit vector in ${\cal
H}_m\otimes {\cal H}_n$ and choose (arbitrary) bases, $\lbrace e_1, \dots,
e_m\rbrace$ for ${\cal H}_m$ and $\lbrace f_1, \dots, f_n \rbrace$ for
${\cal H}_n$.  Then we may write
\begin{equation}
\label{Psimn}
\Psi=c_{ak}e_a\otimes f_k \quad\hbox{(summed over $a$ and $k$)}.
\end{equation}
(where $c_{ak}c_{ak}^*$ [summed over $a$ and $k$] $=1$) and the reduced density
operator (see Sections \ref{Sect:Intro} and \ref{Sect:Prelim}) $\check\rho^{mn}_m$ on ${\cal H}_m$ takes the form
$(\check\rho^{mn}_m)_{a\hat a} |e_a\rangle\langle e_{\hat
a}|$ (summed over $a$ and $\hat a$ from 1 to $m$) where
\begin{equation}
\label{rhomnm}
(\check\rho^{mn}_m)_{a\hat a}=c_{ak}c_{\hat ak}^* \quad\hbox{(summed over k)}.
\end{equation}
(The reason for the `check' `$\check\quad$' is that we will also want,
below, to talk about the ($m\times m$) matrix whose components are
$(\check\rho^{mn}_m)_{a\hat a}$.  And we call this `$\check\rho^{mn}_m$'
to distinguish it from the operator $\rho^{mn}_m$ on ${\cal H}_m$.) We
want to average this over the unit sphere in ${\mathbb C}^N$ for $N=mn$
where each factor of $\mathbb C$ accommodates one of the $N=mn$
components of $c_{ak}$.  So we replace $c_{ak}$ in (\ref{rhomnm}) by
$r_{ak}e^{i\theta_{ak}}$ and then by $w_{ak}^{1/2}e^{i\theta_{ak}}$ and
similarly for $c_{\hat ak}$, obtaining
\[
(\check\rho^{mn}_m)_{a\hat a}=w_{ak}^{1/2}w_{\hat
ak}^{1/2}e^{i(\theta_{ak}-\theta_{\hat ak})} \quad\hbox{(summed over k)}
\]
and we integrate this over Haar measure (\ref{Haar}). Integrating
over the $\theta$s first (with $d\hbox{Torus}$ (\ref{Torus})) will obviously give a factor
of $\delta_{a\hat a}$ for each $k$ in the sum (from 1 to $n$) over $k$.  We are
thus left with a sum (over $k$) of $n$ integrals,
\[
\int w_{ak} d\hbox{Simplex}
\]
for each $a$, each of which takes
the form of $J_1$ (\ref{J1}) for $N=mn$.  So we conclude that
\[
\langle(\check\rho^{mn}_m)_{a\hat a}\rangle=\frac{n\delta_{a\hat
a}}{mn}=\frac{1}{m}\delta_{a\hat a}
\]
and hence, obviously,
\begin{equation}
\label{meanrhomnm}
\langle\rho^{mn}_m\rangle=\frac{1}{m}I_m
\end{equation}
where $I_m$ denotes the identity operator on ${\cal H}_m$. Similarly,
denoting, by $\rho^{mn}_{\ \ n}$ the reduced density operator on ${\cal H}_n$
we will have
\begin{equation}
\label{meanrhomnn}
\langle\rho^{mn}_{\ \ n}\rangle=\frac{1}{n}I_n.
\end{equation}
We remark that we don't strictly need this result for the qualitative
consequences of Lubkin's results we discuss below, but it is anyway
interesting and also serves as a useful preliminary to the recalculation
of Lubkin's result to which we will turn next -- see especially the
remark after equation (\ref{rhonggmnsim}).  More importantly, we will
need the counterpart to this result in our argument, below, for the
closeness of $\rho^{\mathrm{modern}}_{\mathrm S}$ and
$\rho^{\mathrm{modapprox}}_{\mathrm S}$.

Proceeding similarly, it is straightforward to see that
\[
{\rm tr}((\rho^{mn}_m)^2)=c_{ak}c_{\hat ak}^*c_{\hat al}c_{al}^*
\quad\hbox{(summed over $k$, $l$, $a$ and $\hat a$)}
\]
\[
= w_{ak}^{1/2}w_{\hat ak}^{1/2}w_{\hat
al}^{1/2}w_{al}^{1/2}e^{i(\theta_{ak}-\theta_{\hat ak})}e^{-i(\theta_{al}-\theta_{\hat
al})}\quad\hbox{(summed over $k$, $l$, $a$ and $\hat a$)}.
\]
We integrate this over $d\hbox{Haar}$, again doing the
$\theta$-integrals first.  Clearly the latter will vanish unless either
$a=\hat a$ or $k=l$ (or both) whereupon, for fixed values of $a$, $\hat
a$, $k$ and $l$, the complex exponential will integrate (with $d\hbox{Torus}$ (\ref{Torus}))
to 1.  Moreover, (i)  If $a=\hat a$ and $k=l$,
then the w-integral over the simplex will equal $J_{11}$ (\ref{J11}) for $N=mn$ -- and
there are $mn$ such cases; (ii) If $a \ne \hat a$ and $k=l$, then the
w-integral over the simplex will equal $J_{12}$ (\ref{J12}) for $N=mn$ -- and there
are $nm(m-1)$ such cases; and finally (iii) If $a=\hat a$ and $k\ne l$,
then the w-integral over the simplex will equal $J_{12}$ for $N=mn$ again,
and there are $n(n-1)m$ such cases.  Thus we conclude that
\begin{equation}
\label{reLubkin}
\langle {\rm tr}((\rho^{mn}_m)^2)\rangle=\frac{2mn+mn(m-1)+n(n-1)m}{mn(mn+1)}
=\frac{m+n}{mn+1}
\end{equation}
in agreement with (\ref{Lubkinmeanrhosquared}).

Lubkin's result (\ref{Lubkinmeanrhosquared})/(\ref{reLubkin}) is
important for us because of two qualitative
consequences:  First, as we mentioned in Section \ref{Sect:Prelim} and as Lubkin
himself essentially argued, if $n\gg m$ then
$\langle {\rm tr}((\rho^{mn}_m)^2)\rangle$ will be close to $1/m$.  The only way
this can happen is if {\it most} (in the sense we clarify below) totem
states have reduced system density operators $\rho^{mn}_m$ close to the maximally
mixed density operator $(1/m)I_m$.
To see this, notice that (adopting the convention of counting each
eigenvalue, $\lambda_a$, $\nu$ times when $\nu$ is its multiplicity)
amongst density operators, $\rho$, on an $m$-dimensional Hilbert space,
the eigenvalues of $\rho$ have to satisfy both $\sum_{a=1}^m
\lambda_a=1$ (since every density operator has unit trace) as well as
$\sum_{a=1}^m\lambda_a^2={\rm tr}(\rho^2)$ and one easily sees from
these two conditions that the minimum  value of ${\rm tr}(\rho^2)$ is
$1/m$ and that this minimum value is attained only when each of the
$\lambda_a$ equals $1/m$.  If we next consider the set of such $\rho$
for which ${\rm tr}(\rho^2)$ is equal to $1/m + \eta$ where
$\eta$ denotes a (small) positive number, then one easily sees
(again by considering the sum of the eigenvalues and the sum of their
squares) that each of the $\lambda_a$ must take the form $1/m +
\delta_a$ where $\sum_{a=1}^m\delta_a^2=\eta$.  Applying this result
to each of our reduced density operators $\rho^{mn}_m$, now writing the
eigenvalues of each of these in the form $1/m + \delta_a$, then we
immediately see that if $\langle {\rm tr}((\rho^{mn}_m)^2)\rangle =
1/m+\eta$ (which will hold with
$\eta=(m+n)/(mn+1)-1/m=(m^2-1)/(mn+1)$ which will be small if $n\gg
m$) then the statement in words:
\begin{equation}
\label{rhonggmmsim}
\hbox{For $n\gg m$,}\ \rho^{mn}_m\ \hbox{will probably be close to}\ \frac{1}{m}I_m
\end{equation}
will hold in the sense that $\sum_{a=1}^m\langle\delta_a^2\rangle = \eta$.

Similar results will obviously hold for $\rho^{mn}_{\ \ n}$:
\begin{equation}
\label{rhomggnnsim}
\hbox{For m $\gg n$,}\ \rho^{mn}_{\ \ n}\ \hbox{will probably be
close to}\ \frac{1}{n}I_n
\end{equation}
in a similar sense to above.

Our second qualitative consequence of Lubkin's result for $\rho^{mn}_m$
arises as a corollary to the above statement about $\rho^{mn}_{\ \ n}$:
To explain what it is, note first that, just as we had the formula
(\ref{rhomnm}) for the components, $(\check\rho^{mn}_m)_{a\hat a}$, of
the $m\times m$ matrix $\check\rho^{mn}_m$, so we clearly have that
$\rho^{mn}_{\ \ n} =(\check\rho^{mn}_{\ \ n})_{k\hat
k}|f_k\rangle\langle f_{\hat k}|$ where, in the notation of
(\ref{Psimn}), the $n\times n$ matrix $\check\rho^{mn}_{\ \ n}$ is given
by
\begin{equation}
\label{rhomnn}
(\check\rho^{mn}_{\ \ n})_{k\hat k}=c_{ak}{c_{a\hat k}}^*\  \hbox{(summed over $a$)}.
\end{equation}
So, denoting by $C$ the ($m \times n$) matrix whose components are the $c_{ak}$ and by $C^+$ its ($n\times m$) adjoint matrix, we clearly have
\begin{equation}
\label{matrices}
\check\rho^{mn}_m=CC^+ \ \hbox{and}\ \check{\rho^{mn}_{\ \ n}}^*\ =C^+C.
\end{equation}
It easily follows from  (\ref{matrices}) that $x\in {\mathbb C}^m$ can
be an eigenvector of $\check\rho^{mn}_m$ with a non-zero (positive) eigenvalue,
$\lambda$, if and only if $y=\lambda^{-1/2}(C^+x)^*$ ($\in {\mathbb C}^n$)  is an eigenvector of $\check\rho^{mn}_{\ \ n}$ with the same eigenvalue.  (The factor of $\lambda^{-1/2}$ is easily seen to be needed if we want to ensure that $y$ is normalized whenever $x$ is normalized.)
Moreover we note for future reference (in our digression on the Schmidt
decomposition below) that we then have $Cy^*=\lambda^{1/2} x$ -- i.e.
\begin{equation}
\label{cyequalslambdax}
c_{ak}{y^k}^*=\lambda^{1/2} x^a
\end{equation}
-- the left hand side being summed over $k$.
We conclude (continuing to adopt the convention of counting any eigenvalue  $\nu$ times if
it has multiplicity $\nu$) that, if $m>n$ and if $\check\rho^{mn}_{\ \ n}$ has eigenvalues
$\lambda_1, \dots, \lambda_n$, then $\check\rho^{mn}_m$ will have this
same set of eigenvalues together with $m-n$ more, all of which will,
however, be zero!   Moreover (cf.\ the discussion after equation (\ref{rhonggmmsim})), since
$\langle{\rm tr}(\rho^{mn}_{\ \ n})^2\rangle=1/n+\eta$ for $\eta$ now
equal to $(m+n)/(mn+1)-1/n=(n^2-1)/(mn+1)$ -- which will be small if $m\gg n$ -- we will have
$\lambda_k=1/n + \delta_k$ where
$\sum_{k=1}^n\langle\delta_k^2\rangle=\eta$.  So, we may say that
\begin{equation}
\label{rhomggnmsim}
\hbox{For $m\gg n$,}\  \rho^{mn}_m\ \hbox{will probably be close to}\
\frac{1}{n}\sum_{k=1}^n|\tilde e_a\rangle\langle \tilde e_a|
\end{equation}
where $\lbrace\tilde e_1, \dots, \tilde e_n\rbrace$ is a basis for an
$n$-dimensional subspace of ${\cal H}_m$ (which will depend on $\Psi$).
This is the second qualitative
consequence of Lubkin's result  we promised to arrive at at the outset.  As far as we are aware,
it does not appear to have been pointed out before.  But, for our purposes, it will be of
equal importance to the first consequence.

Similarly, of course:
\begin{equation}
\label{rhonggmnsim}
\hbox{For $n\gg m$,}\  \rho^{mn}_{\ \ n}\ \hbox{will probably be close to}\
\frac{1}{m}\sum_{a=1}^m|\tilde f_a\rangle\langle \tilde f_a|
\end{equation}
where $\lbrace\tilde f_1, \dots, \tilde f_m\rbrace$ is a basis for an
$m$-dimensional subspace of ${\cal H}_n$ (which will again depend on $\Psi$).

We remark that, in preparation for the argument we give below for the
claim that $\rho^{\mathrm{modern}}_{\mathrm S}$ is well-approximated by
$\rho^{\mathrm{modapprox}}_{\mathrm S}$, it is useful to observe
that/how (\ref{rhomggnmsim}) and (\ref{rhomggnnsim}) are consistent with
(\ref{meanrhomnm}) and (\ref{meanrhomnn}).

Further insight into the origin of (\ref{rhonggmmsim}), (\ref{rhomggnnsim}),
(\ref{rhomggnmsim}) and
(\ref{rhonggmnsim}) can be had by recalling that a given vector $\Psi\in
{\cal H}_m\otimes {\cal H}_n$ -- which we have written so far in the
form (\ref{Psimn}) -- can also be written as a single sum
\begin{equation}
\label{Schmidt}
\Psi=\sum_{i=1}^{{\mathrm{min}}(m,n)}\lambda_i^{1/2}\tilde e_i\otimes\tilde f_i
\end{equation}
for suitable choices of basis $\lbrace\tilde e_1,\dots, \tilde e_m\rbrace$
on ${\cal H}_m$ and $\lbrace\tilde f_1,\dots, \tilde f_n\rbrace$ on ${\cal
H}_n$.  This is the well-known Schmidt decomposition (cf.\ e.g.\
\cite{BengtZyc} and/or the next paragraph) and the $\lambda_i$ are the same as those
discussed above.  (\ref{rhonggmmsim}) and (\ref{rhonggmnsim}) may then
be viewed as (easy) consequences of the fact that, when $n\gg m$, the
$\lambda_i$ in (\ref{Schmidt}) are probably close to $1/m$ for $i=1, \dots, m$, while
they are zero for $i>m$.  Similarly (\ref{rhomggnnsim}) and
(\ref{rhomggnmsim}) may be viewed as consequences of the fact that,
when $m\gg n$, the $\lambda_i$ in (\ref{Schmidt}) are probably close to $1/n$ for
$i=1, \dots, n$, while they are zero for $i>n$.

The Schmidt decomposition in the form (\ref{Schmidt}) can actually be
derived easily from the results following Equation (\ref{matrices}).  In
this paragraph, we digress to point out how, treating the cases where
$n\ge m$: Denote by $\lbrace x_1,\dots, x_m\rbrace$ a complete set of
orthonormal eigenvectors of $\check\rho^{mn}_m$, and denote by $x^a_i$ the
$a$th component of $x_i$. In view of the sentence following Equation
(\ref{matrices}), we can clearly find a complete set, $\lbrace
y_1^*,\dots,y_n^*\rbrace$, of orthonormal eigenvectors of $\check\rho^{mn}_{\ \
n}$ so that, denoting by ${y^k_j}^*$ the $k$th component of $y_j^*$, we have,
by (\ref{cyequalslambdax}),  $c_{ak}{y^k_j}^*=\lambda_j^{1/2}x^a_j$ (the left hand
side being summed over $k$).  (We only need to make sure that the
$i$-value of every $x_i$ belonging to each non-zero eigenvalue of
$\check\rho^{mn}_m$ matches the $j$-value of a $y_j^*$ belonging to an equal
non-zero eigenvalue of $\check\rho^{mn}_{\ \ n}$ -- there being necessarily an
equal number of each; for any other $y_j$ [and of course there have to
be others whenever $n>m$] the right hand side will anyway vanish.)  Also
introduce a new basis $\lbrace \tilde e_1, \dots, \tilde e_m\rbrace$ for
${\cal H}_m$ such that  $e_a=x^a_i\tilde e_i$ (summed over $i$) and,
similarly, introduce a new basis $\lbrace \tilde f_1 \dots \tilde
f_n\rbrace$ for ${\cal H}_n$ such that $f_k={y^k_j}^*\tilde f_j$ (summed over
$j$).  Then we have (see (\ref{Psimn}))
\[
\Psi=c_{ak}e_a\otimes f_k \  \hbox{(summed over $a$ and $k$)}=c_{ak}x^a_i{y^k_j}^*\tilde
e_i\otimes \tilde f_j \  \hbox{(summed over $i$, $j$, $a$ and $k$)}
\]
\[
= \ \ \hbox{by (\ref{cyequalslambdax})}
\ \ \lambda_j^{1/2}x^a_ix^a_j\tilde e_i\otimes \tilde f_j \  \hbox{(summed over
$i$, $j$ and $a$)} = \lambda_j^{1/2}\delta_{ij}\tilde e_i\otimes \tilde f_j
\  \hbox{(summed over $i$ and $j$)}
\]
So we have
\[
\Psi=\lambda_i^{1/2} \tilde e_i\otimes\tilde f_i \ \hbox{(summed over $i$)}
\]
thus establishing (\ref{Schmidt}) in cases where $n\ge m$.  The cases
where $m\ge n$ are obviously similar.  This ends our digression.

\section{\label{Sect:Main} Main argument for the validity of the approximation (\ref{purereduced}) of
$\rho^{\mathrm{modern}}_{\mathrm S}$ by $\rho^{\mathrm{modapprox}}_{\mathrm S}$}

Let us now turn to consider the set of unit vectors, $\Psi$, in the Hilbert space
of a totem as specified in Section \ref{Sect:Spec} -- i.e.\ in the
subspace of states with total energies in the range $[E,E+\Delta]$ of
${\cal H}_{\mathrm S}\otimes {\cal H}_{\mathrm B}$.  Allowing ourselves
to make the slight distortion explained before equations
(\ref{degdensS}) and (\ref{degdensB}), we may assume ${\cal H}_{\mathrm
S}$ has an orthonormal basis consisting of vectors $|\epsilon_{\mathrm
S}, i\rangle$, where $\epsilon_{\mathrm S}$ ranges from $\Delta$ to $E$ in
steps of $\Delta$, while, for each $\epsilon_{\mathrm S}$, the integer, $i$,
ranges from 1 to $n_{\mathrm S}(\epsilon_{\mathrm S})$; and similarly ${\cal
H}_{\mathrm B}$ has an orthonormal basis consisting of vectors
$|\epsilon_{\mathrm B}, j\rangle$, where $\epsilon_{\mathrm B}$ ranges
from $\Delta$ to $E$ in steps of $\Delta$, while, for each $\epsilon_{\mathrm
B}$, $j$ ranges from 1 to $n_{\mathrm B}(\epsilon_{\mathrm B})$.
(Below, and as in Section \ref{Sect:Intro}, we shall sometimes drop the S and B
subscripts on the $\epsilon_{\mathrm S}$ when no ambiguity can arise.)  Then, we
are interested in the set of unit vectors,  $\Psi$, in the subspace,
which we shall call below ${\cal H}_M$, of ${\cal H}_{\mathrm S}\otimes {\cal H}_{\mathrm B}$
with total energy exactly $E$.  The reason for
the name ${\cal H}_M$ is that ${\cal H}_M$ will clearly have dimension
$M$, where $M$ is as in (\ref{sumnorm}).

Each such $\Psi\in {\cal H}_M$ is writeable in the form
\begin{equation}
\label{Psibigtotem}
\Psi=\sum_{\epsilon=\Delta}^E
\sum_{i=1}^{n_{\mathrm S}(\epsilon)}\sum_{j=1}^{n_{\mathrm
B}(E-\epsilon)}
c^{ij}_\epsilon |\epsilon,
i\rangle\otimes|E-\epsilon, j\rangle
\end{equation}
where we recall (see above and cf.\ before Equation (\ref{microsysbath})) that the sum
over $\epsilon$ goes up in integer multiples of $\Delta$.  We also note that
since $\Psi$ is a unit vector, the sum (with the above indicated ranges)
over $\epsilon$, $i$ and $j$ of $|c^{ij}_\epsilon|^2$ equals 1.
For such a $\Psi$, the partial trace of $|\Psi\rangle\langle\Psi|$ over
${\cal H}_{\mathrm B}$, i.e.\ the reduced density operator,
$\rho^{\mathrm{modern}}_{\mathrm S}$, on ${\cal
H}_{\mathrm S}$, will then clearly be given by
\begin{equation}
\label{modexact}
\rho^{\mathrm{modern}}_{\mathrm S}=\sum_{\epsilon=\Delta}^E
\sum_{i=1}^{n_{\mathrm S}(\epsilon)}\sum_{\hat i=1}^{n_{\mathrm S}(\epsilon)}
{\check r}^{\mathrm S^{i\hat i}}_\epsilon|\epsilon, i\rangle\langle \epsilon, \hat i|,
\end{equation}
where
\begin{equation}
\label{reducedrhomodexact}
{\check r}^{\mathrm S^{i\hat i}}_\epsilon=\sum_{j=1}^{n_{\mathrm B}(E-\epsilon)}
c^{ij}_\epsilon {c^{\hat ij}_\epsilon}^*.
\end{equation}

We shall find it useful sometimes to think of ${\cal H}_{\mathrm S}$ as a direct
sum $\oplus_{\epsilon=0}^E {\cal H}^{\mathrm S}_\epsilon$ (and similarly for ${\cal
H}_{\mathrm B})$ where ${\cal H}^{\mathrm S}_\epsilon$
is spanned by the $|\epsilon, i\rangle$ for fixed $\epsilon$ as $i$
varies from 1 to $n_{\mathrm S}(\epsilon)$ and we shall call the
restriction of $\rho^{\mathrm{modern}}_{\mathrm S}$ to
${\cal H}^{\mathrm S}_\epsilon$ simply $r^{\mathrm S}_\epsilon$ -- its
$i\hat i$ components in the basis consisting of the $|\epsilon,
i\rangle$ being obviously the ${\check r}^{\mathrm S^{i\hat i}}_\epsilon$ introduced above.

Our aim is to give an argument in favour of the claimed correctness of
the proposition, which we state in Section \ref{Sect:Ans}, that, in situations
of interest, $\rho^{\mathrm{modern}}_{\mathrm S}$ will be
well-approximated by the $\rho^{\mathrm{modapprox}}_{\mathrm S}$ of
(\ref{purereduced}) and, in the course of giving this argument, to make clear how the $n_{\mathrm{B}}(E-\epsilon)$-dimensional
subspace of the ($n_{\mathrm{S}}(\epsilon)$-dimensional)
energy-$\epsilon$ subspace of ${\cal H}_{\mathrm{S}}$ spanned by the $|\widetilde{\epsilon, i}\rangle$ depend on $\Psi$.  (We should amplify on this statement by explaining
that, when we say that $\rho^{\mathrm{modern}}_{\mathrm S}$ is
well-approximated by $\rho^{\mathrm{modapprox}}_{\mathrm S}$, what we
mean is that the values of physical quantities of interest, such as the
mean energy and the entropy of the system S, calculated using
$\rho^{\mathrm{modern}}_{\mathrm{S}}$ will be close to the values of
the same quantities calculated using
$\rho^{\mathrm{modapprox}}_{\mathrm{S}}$.)

The main ingredients in our argument concern the average, $\langle
r^{\mathrm S}_\epsilon\rangle$, of $r^{\mathrm S}_\epsilon$ and also the
average, $\langle {\mathrm tr}((r^{\mathrm S}_\epsilon)^2)\rangle$, of
${\mathrm tr}((r^{\mathrm S}_\epsilon)^2)$, where both averages are
taken as $\Psi$ ranges over the whole of ${\cal H}_M$ (with respect to
Haar measure on ${\cal H}_M$). The calculations of these quantities are
closely similar to the preliminary calculations we carried out above for
the average of the density operator $\rho^{mn}_m$ and the average of the
trace of its square; the difference being that we are now averaging over
all unit $\Psi$ in our full $M$-dimensional Hilbert space ${\cal H}_M$
(with $M$ as in (\ref{sumnorm})) even though what we are averaging is
only the restriction, $r^{\mathrm S}_\epsilon$, (and the trace of the square of the restriction)
of $\rho^{\mathrm{modern}}_{\mathrm S}$ to
${\cal H}^{\mathrm S}_\epsilon$ for  fixed $\epsilon$.  As a result,
while the counterpart of the product, $mn$, in the denominator in our
preliminary calculation would just be the single product  $n_{\mathrm
S}(\epsilon)n_{\mathrm B}(E-\epsilon)$ were our average only to be over
${\cal H}^{\mathrm S}_\epsilon\otimes {\cal H}^{\mathrm
B}_{E-\epsilon}$, since we average over the unit vectors of the full
Hilbert space ${\cal H}_M$, the counterpart will be turn out to be $M$.
Aside from this difference, to calculate  $\langle r^{\mathrm
S}_\epsilon\rangle$ one proceeds very similarly to the  passage, above,
between equations (\ref{rhomnm}) and (\ref{meanrhomnm}); the reader can
easily supply the details simply by replacing (\ref{rhomnm}) by
(\ref{reducedrhomodexact}) etc.\  One clearly obtains (instead of
(\ref{meanrhomnm}))
\begin{equation}
\label{meanrhoepsilon}
\langle r^{\mathrm S}_\epsilon\rangle=\frac{n_{\mathrm
B}(E-\epsilon)}{M}I^{\mathrm S}_\epsilon
\end{equation}
where $I^{\mathrm S}_\epsilon$ denotes the identity on ${\cal
H}^{\mathrm S}_\epsilon$.
Similarly, proceeding as in the passage between equations
(\ref{meanrhomnm}) and (\ref{reLubkin}) (but again it will turn out that one
needs to replace $mn$ in the denominator by $M$) we easily find:
\[
\langle {\rm tr}((r^{\mathrm S}_\epsilon)^2)\rangle=
\frac{2n_{\mathrm B}(E-\epsilon)n_{\mathrm S}(\epsilon)+
n_{\mathrm B}(E-\epsilon)n_{\mathrm S}(\epsilon)n_{\mathrm S}(\epsilon)-1)+n_{\mathrm B}(E-\epsilon)(n_{\mathrm B}(E-\epsilon)-1)n_{\mathrm S}(\epsilon)}{M(M+1)}
\]
\begin{equation}
\label{meantracerhoepsilonsquared}
=\frac{n_{\mathrm B}(E-\epsilon)n_{\mathrm S}(\epsilon)
(n_{\mathrm B}(E-\epsilon)+n_{\mathrm S}(\epsilon))}{M(M+1)}.
\end{equation}

Now, rather as in our arguments for the two qualitative consequences of
Lubkin's result, (but now our arguments will involve both the
counterpart, (\ref{meanrhoepsilon}), to (\ref{meanrhomnm}) as well as
the counterpart, (\ref{meantracerhoepsilonsquared}), to
(\ref{reLubkin})) we observe from (\ref{meantracerhoepsilonsquared})
that,  whenever $n_{\mathrm S}(\epsilon)\gg n_{\mathrm B}(E-\epsilon)$,
$\langle{\rm tr}(r^{\mathrm S}_\epsilon)^2\rangle$ will be very close
to  $M^{-2}(n_{\mathrm B}(E-\epsilon))^2$, which, \textit{in the
presence of} (\ref{meanrhoepsilon}), easily implies (i.e.\ by similar
reasoning to that used above in our derivation of (\ref{rhonggmmsim})
and (\ref{rhomggnnsim})) that $r^{\mathrm S}_\epsilon$ must be very
close to $M^{-1}n_{\mathrm B}(E-\epsilon)\sum_{i=1}^{n_{\mathrm
S}(\epsilon)}|\epsilon, i\rangle\langle \epsilon, i|$. Moreover,
whenever $n_{\mathrm B}(E-\epsilon)\gg n_{\mathrm S}(\epsilon)$,
$\langle {\rm tr}((r^{\mathrm S}_\epsilon)^2)\rangle$ will be very close to
$M^{-2}(n_{\mathrm S}(\epsilon))^2$, which, again in the presence of
{\rm (\ref{meanrhoepsilon})}, easily implies (i.e.\ by similar reasoning to
that used above in our derivation of (\ref{rhomggnmsim}) and
(\ref{rhonggmnsim})) that $r^{\mathrm S}_\epsilon$ will be very close
to $M^{-1}n_{\mathrm{S}}(\epsilon)\sum_{i=1}^{n_{\mathrm{B}}(E-\epsilon)}
|\widetilde{\epsilon, i}\rangle\langle \widetilde{\epsilon, i}|$ where
$|\widetilde{\epsilon, i}\rangle$ denote the elements of an orthonormal basis for an
$n_{\mathrm{B}}(E-\epsilon)$-dimensional subspace of the
($n_{\mathrm{S}}(\epsilon)$-dimensional) energy-$\epsilon$ subspace,
${\cal H}^{\mathrm S}_\epsilon$ of ${\cal H}_{\mathrm{S}}$ which will
depend on $\Psi$.  Comparing these conclusions with the form of Equation
(\ref{purereduced}) we immediately see that, if it were the case that
for all $\epsilon$, either $n_{\mathrm S}(\epsilon)\gg n_{\mathrm
B}(E-\epsilon)$ or $n_{\mathrm B}(E-\epsilon)\gg n_{\mathrm
S}(\epsilon)$, then (\ref{purereduced}) would obviously be a
good approximation (at least each term in the sum over $\epsilon$ will be) for all $\epsilon$.
However, of course, in typical situations of interest, there will be a region of $\epsilon$
values around the value $E_c$ -- see the definitions of terms
immediately after Equation (\ref{purereduced}) -- where neither of these
statements will hold and  $n_{\mathrm S}(\epsilon)$  and  $n_{\mathrm
B}(E-\epsilon)$  will be of comparable size. (We remark in passing, though, that,
typically, [one will be able to choose $\Delta$ so that] each of these
quantities will be much greater than 1 for all or very nearly all $\epsilon$
[which are multiples of $\Delta$ and] in the range $[0,E]$.)

Nevertheless for the sort of situations of interest to us -- and, in
particular, for the densities of states which increase according to the
power law, (\ref{sigmasbpower}), as considered in Section \ref{Sect:Power}
or which increase exponentially, (\ref{sigmasbexponential}), as
considered in Sections \ref{Sect:Exp} and \ref{Sect:ExpForm},
or which increase as quadratic exponentials, (\ref{sigmasbexpsquared}),  as considered in Section
\ref{Sect:ExpEsq} -- and assuming the totem energy $E$ and our choice
of energy-increment, $\Delta$ (see Section \ref{Sect:Intro}) are such
that $M\gg 1$ (to ensure that the system [and bath] has access to a very
large number of states) -- one can check that the region of
$\epsilon$-values around $E_c$ where $n_{\mathrm B}(E-\epsilon)$ and
$n_{\mathrm S}(\epsilon)$ are of comparable size will always be very
small in size compared to $E$, while the sum over this region of
$n_{\mathrm S}(\epsilon)n_{\mathrm B}(E-\epsilon)$ will be very small
compared to $M$. (In other words, the integral over this energy-region
of the energy-probability density $P_{\mathrm{S}}(\epsilon)$ [see
(\ref{enprobdens})] will be very much less than 1.)  Moreover, as
$\epsilon$ decreases towards zero, or increases towards $E$ from $E_c$,
then for all three densities of states, (\ref{sigmasbpower}),
(\ref{sigmasbexponential}), (\ref{sigmasbexpsquared}), one may check
that the ratio  $n_{\mathrm S}(\epsilon)/n_{\mathrm B}(E-\epsilon)$,
respectively  $n_{\mathrm B}(E-\epsilon)/n_{\mathrm S}(\epsilon)$, and
hence the counterparts (i.e.\ with $n_{\mathrm S}(\epsilon)$ replacing
$m$  and $n_{\mathrm B}(E-\epsilon)$ replacing $n$) to the quantities
which we called $\eta$ before Equation (\ref{rhonggmmsim}), respectively
Equation (\ref{rhomggnmsim}), will get rapidly smaller and hence the
relevant notion of closeness (i.e.\ as in  (\ref{rhonggmmsim}),
(\ref{rhomggnmsim})) will get rapidly  stronger. It is then
straightforward to argue from these statements that quantities of
interest such as (cf.\ (\ref{puremeanenergy}))
$\bar\epsilon^{\mathrm{modern}}_{\mathrm{S}}
=:{\rm tr}(\rho^{\mathrm{modern}}_{\mathrm{S}}H_{\mathrm{S}})$, ${\rm tr}
((\rho^{\mathrm{modern}}_{\mathrm{S}})^2)$ itself, and  (cf.\
(\ref{pureentropy})) $S^{\mathrm{modern}}_{\mathrm{S}}=:-{\rm tr}
(\rho^{\mathrm{modern}}_{\mathrm{S}}\log\rho^{\mathrm{modern}}_{\mathrm{S}})$
will be closely approximated (respectively) by
$\bar\epsilon^{\mathrm{modapprox}}_{\mathrm{S}}
=:{\rm tr}(\rho^{\mathrm{modapprox}}_{\mathrm{S}}H_{\mathrm{S}})$, ${\rm
tr} ((\rho^{\mathrm{modapprox}}_{\mathrm{S}})^2)$, and  (cf.\
(\ref{pureentropy})) $S^{\mathrm{modapprox}}_{\mathrm{S}}=:-{\rm tr}
(\rho^{\mathrm{modapprox}}_{\mathrm{S}}\log\rho^{\mathrm{modapprox}}_{\mathrm{S}})$.

Concerning the latter two quantities -- i.e.\ the trace of the square of
the reduced density operator of S and its von Neumann entropy -- there are reasons to
expect the approximation of  $S^{\mathrm{modern}}_{\mathrm{S}}$ by
$S^{\mathrm{modapprox}}_{\mathrm{S}}$ to be even better than the
approximation of ${\rm tr} ((\rho^{\mathrm{modern}}_{\mathrm{S}})^2)$ by
${\rm tr} ((\rho^{\mathrm{modapprox}}_{\mathrm{S}})^2)$.

\section{\label{Sect:Further} Further checks and details on the validity of the Approximation (\ref{purereduced})}

As a partial check of various aspects of all of the above argument, and
in justification of our latter remark, it is instructive first to
consider the case where, for all $\epsilon$ ($=0, \Delta, 2\Delta,
\dots$) in the range $[0,E]$, we have $n_{\mathrm
S}(\epsilon)=1=n_{\mathrm B}(E-\epsilon)$ where, of course it is {\it
never} true that $n_{\mathrm S}(\epsilon)\gg n_{\mathrm B}(E-\epsilon)$
or that $n_{\mathrm S}(\epsilon)\ll n_{\mathrm B}(E-\epsilon)$ (nor that
each of these quantities is very much greater than 1!) so we can think
of this as one sort of `worst case scenario'.  Of course this is not an
example that interests us in Part~1, but it would apply e.g.\ to a
totem consisting of a pair of weakly coupled quantum harmonic
oscillators (with equal spring constants) for a total energy much
greater than the level spacing (and a choice of $\Delta$ equal to the
level spacing). For this model, we may clearly write
\begin{equation}
\label{Psisho}
\Psi=\sum_{\epsilon=\Delta}^E c_\epsilon|\epsilon\rangle|E-\epsilon\rangle
\end{equation}
so that the reduced density operator of the system, S, will be
\[
\rho_{\mathrm S}^{\mathrm{modern}}=\sum_{\epsilon=\Delta}^E
|c_\epsilon|^2|\epsilon\rangle\langle\epsilon|,
\]
while the formula, (\ref{purereduced}), for $\rho_{\mathrm S}^{\mathrm{modapprox}}$
(\ref{purereduced}) becomes simply
\[
\rho_{\mathrm S}^{\mathrm{modapprox}}=\frac{1}{M}\sum_{\epsilon=\Delta}^E
|\epsilon\rangle\langle\epsilon|,
\]
where (cf.\ (\ref{sumnorm})) $M=E/\Delta$. Clearly, (cf.\ (\ref{puremeanenergy})) the
approximate mean energy,
\[
\bar\epsilon^{\mathrm{modapprox}}_{\mathrm{S}}
:={\rm tr}(\rho^{\mathrm{modapprox}}_{\mathrm{S}}H_{\mathrm{S}})=
\frac{1}{M}\sum_{\epsilon=\Delta}^E \epsilon
\]
\[
=\frac{1}{2M}E\left (\frac{E}{\Delta}+1\right )=\frac{E}{2}.
\]
(In calculating the value of the above sum, we need of course to recall that the
sum is over values of $\epsilon$ which are integer multiples of $\Delta$.)
Therefore, since this doesn't depend on the $c_\epsilon$, its average over Haar measure
(indicated with ``$\langle\quad\rangle$'') takes the same value:
\[
\langle\bar\epsilon^{\mathrm{modapprox}}_{\mathrm{S}}\rangle=\frac{E}{2}.
\]
On the other hand, the average over Haar measure of the exact mean
energy, $\bar\epsilon^{\mathrm{modern}}_{\mathrm{S}}$,
may be calculated as follows:
\[
\langle\bar\epsilon^{\mathrm{modern}}_{\mathrm{S}}\rangle = \langle
\sum_{\epsilon=\Delta}^E \epsilon |c_\epsilon|^2\rangle
\]
\[
=\sum_{\epsilon=\Delta}^E \epsilon\int w_\epsilon d\hbox{Haar}
\]
where the integral is over the complex $M$-dimensional sphere of unit
vectors in the Hilbert space, ${\cal H}_M$, spanned by the vectors of
form (\ref{Psisho}),  coordinatized with $w_\epsilon$ ranging over the
($M$-1)-simplex and $\theta_\epsilon$ ranging over the $M$-torus as
explained at the beginning of Section \ref{Sect:Haar}, where
$w_\epsilon=|c_\epsilon|^2$ etc. Obviously the torus factor of the
integral just gives 1, so the integral has, by (\ref{J1}), the value
$1/M$ for each $\epsilon$.  So we conclude that
$\langle\bar\epsilon^{\mathrm{modern}}_{\mathrm{S}}\rangle$ has the same
value as $\langle\bar\epsilon^{\mathrm{modapprox}}_{\mathrm{S}}\rangle$,
i.e.
\[
\langle\bar\epsilon^{\mathrm{modern}}_{\mathrm{S}}\rangle=\frac{E}{2},
\]
which, of course, has to be the correct value by the symmetry under the
interchange of S and B in this case.

Turning to averages over Haar measure of the trace of the square of the
reduced density operator of S, we have, on the one hand,
\[
\langle{\rm tr} ((\rho^{\mathrm{modapprox}}_{\mathrm{S}})^2)\rangle
=\langle\sum_{\epsilon=\Delta}^E \frac{1}{M^2}\rangle=\sum_{\epsilon=\Delta}^E \frac{1}{M^2}
=\left (\frac{\Delta}{E}\right )^2\left (\frac{E}{\Delta}\right)
\]
\[
= \frac{1}{M} \ \ (= \frac{\Delta}{E}).
\]
Whereas, on the other hand,
\[
\langle{\rm tr} ((\rho^{\mathrm{modern}}_{\mathrm{S}})^2)\rangle
=\langle\sum_{\epsilon=\Delta}^E|c_\epsilon|^4\rangle=\left
(\frac{E}{\Delta}\right)
\int w_1^2 d\hbox{Haar}=M\frac{2}{M(M+1)}\simeq \frac{2}{M} \ \ (\simeq \frac{2\Delta}{E})
\]
(where we have used (\ref{J11}) in calculating the integral)
which differs from the approximate value by a factor of 2!

However, if we turn to calculate the averages over Haar measure of the
von Neumann entropies of the approximate and exact reduced density
operator of S, we find, on the one hand,
\begin{equation}
\label{vonNworstapprox}
\langle S^{\mathrm{modapprox}}_{\mathrm{S}}\rangle=\langle -{\rm tr}
(\rho^{\mathrm{modapprox}}_{\mathrm{S}}\log\rho^{\mathrm{modapprox}}_{\mathrm{S}})\rangle
= \langle \log M \rangle = \log M \quad (= \log(E/\Delta)).
\end{equation}
On the other hand,
\[
\langle S^{\mathrm{modern}}_{\mathrm{S}}\rangle=\langle -{\rm tr}
(\rho^{\mathrm{modern}}_{\mathrm{S}}\log\rho^{\mathrm{modern}}_{\mathrm{S}})\rangle
=-M\int_{\mathrm{Unit \ Sphere \ in \ {\mathbb C}^M}} w_1\log w_1 d\hbox{Haar}
\]
\[
= \ \ \hbox{(by (\ref{Haar}) and (\ref{Simplex}))} \ \
-M(M-1)!\int_{\mathrm{Simplex}}w_1\log w_1 dw_1\dots dw_{M-1}
\]
\[
=-M(M-1)\int_0^1 w\log w(1-w)^{M-2} dw.
\]
One may do this integral by noticing that $w\log
w=(dw^\alpha/d\alpha)|_{\alpha=1}$ -- obtaining for its value,
$d(B(\alpha+1,M-1))/d\alpha|_{\alpha=1}$ where $B$ denotes the beta
function (see e.g.\ \cite{Gradshteyn}).  One finds that
$-M(M-1)$ times this simplifies (using (\ref{betagamma})) to
$\psi(1+M)-\psi(2)$ where $\psi$ denotes the $\psi$ (or `digamma') function defined by
$\psi(x)=d\log\Gamma(x)/dx$, and this \cite{Gradshteyn}, in turn, equals
$\sum_{k=2}^M 1/k$ which, by the standard asymptotic expansion of
Euler's constant, $C$ ($=0.5772\dots$) is equal to $\log M + C -1 + 1/2M + O(1/M^2)$.  So
we conclude that
\begin{equation}
\label{vonNworstexact}
\langle S^{\mathrm{modern}}_{\mathrm{S}}\rangle=\log M + C - 1 + O(1/M)
\quad (=\log(E/\Delta) + C - 1 + O(\Delta/E))
\end{equation}
where $C$ is Euler's constant ($0.5772\dots$)

Comparison of (\ref{vonNworstexact}) and (\ref{vonNworstapprox}) shows
that the use of (\ref{purereduced}) for this `worst case scenario' leads
to a von Neumann entropy which, for large $M$, is very close to the
average over Haar measure of its actual value.  In view of the fact that
both of these values are very close to the maximum possible value,
$\log M$, of the entropy of any density operator on ${\cal H}_{\mathrm
S}$ (which is of course $M$-dimensional in this case) we conclude both
that most totem states, $\Psi$, for this model must have a reduced
density operator on S whose von Neumann entropy is close to $\log M$;
and that the use of (\ref{purereduced}) leads to a good approximation
for this value.  And both of these things hold even though, as we saw
above, our general arguments do not apply to this case and even though,
for this case, as we saw above, (\ref{purereduced}) leads to a  trace of
the square of the reduced density operator of S which is only half of
the average over Haar measure of its actual value.

We will next use the Lubkin-Page asymptotic formula,
(\ref{LubkinPagemeanentropy}), to obtain a result which tends to confirm
the accuracy of our general formula (\ref{pureentropy}), obtained using
(\ref{purereduced}), for the von Neumann entropy for our densities of
states of interest, (\ref{sigmasbpower}), (\ref{sigmasbexponential}),
(\ref{sigmasbexpsquared}).  Our result will show that the value of the
von Neumann entropy obtained with (\ref{purereduced}) well-approximates
a certain restricted average of the exact von Neumann entropy.  Before
we present this result, we shall find it helpful to first explain what
we mean here by a `restricted average' in a different context:

Let us look back at the result
essentially due to Lubkin, (\ref{reLubkin}), which we (re-)obtained above, for
the average over Haar measure on vectors, $\Psi$, belonging to the
tensor product,  ${\cal H}_m\otimes {\cal H}_n$, of two Hilbert spaces,
of the trace of the square of the reduced density operator,
$\rho^{mn}_m$ on ${\cal H}_m$. Averaging over all totem vectors, $\Psi\in
{\cal H}_m\otimes {\cal H}_n$, amounts, as we explained above, to
averaging with the invariant measure on the complex $mn$-sphere over the
coefficients, $c_{ak}$, in the basis-expansion, (\ref{Psimn}), of
$\Psi$, which, in turn, writing $c_{ak}$ as
$w_{ak}^{1/2}e^{i\theta_{ak}}$, we saw, amounts to integrating w.r.t. the
$w_{ak}$ over the ($mn-1$)-simplex and w.r.t. the $\theta_{ak}$ over the
$mn$-torus.  What we now wish to point out is that, if we restrict to $c_{ak}$
which take the form $(1/\sqrt{mn})e^{i\theta_{ak}}$ and just average
over these (i.e.\ by integrating with respect to the $\theta_{ak}$ over
the $mn$-torus) one easily finds -- denoting
our restricted average with the symbol `$[\quad ]$' -- that
\[
[{\rm tr}((\rho^{mn}_m)^2)]=\frac{m+n-1}{mn},
\]
and this is not a bad approximation to the value, $(m+n)/(mn+1)$, of the
full average, $\langle {\rm tr}((\rho^{mn}_m)^2)\rangle$, of ${\rm
tr}((\rho^{mn}_m)^2)$
 -- the two expressions differing, in fact, only by terms of order $1/mn$. In
terms of our geometrical picture, in which averaging w.r.t. Haar measure
amounts to integrating over the ($mn-1$)-simplex times the $mn$-torus,
in passing from the unrestricted average, `$\langle\quad\rangle$', to our
restricted average, `$[\quad ]$', we have replaced the integral over the
($mn-1$)-simplex by the value at its
centroid (where $w_{ak}=1/mn$ for all $a$ and $k$), but continue to integrate
over the $mn$-torus.  Of course
this restricted average, `$[\quad ]$', is a basis-dependent notion, but
what we have learnt is that we can choose any bases ($\lbrace e_1,
\dots, e_m\rbrace$ and $\lbrace f_1, \dots, f_n\rbrace$) we like and we
obtain this reasonably good approximation to the unrestricted average
(at least for the quantity ${\rm tr}((\rho^{mn}_m)^2)$).

We conclude that the corresponding restricted set of totem vectors (cf.\ (\ref{Psimn}),
\[
\Psi=\frac{1}{\sqrt{mn}}e^{i\theta_{ak}}e_a\otimes f_k \quad\hbox{(summed over $a$ and $k$)}
\]
is (for any choice of bases, $\lbrace e_1, \dots, e_m\rbrace$ and
$\lbrace f_1, \dots, f_n\rbrace$ and as the $\theta_{ak}$ range over the $mn$-torus)
a sufficiently representative set of
totem vectors for the restricted average over this set to indicate
sufficiently well the behaviour of a generic totem state, $\Psi$ (at
least as far as ${\rm tr}((\rho^{mn}_m)^2)$ is concerned) .

We shall proceed in a similar spirit, but now for the totem of Section
\ref{Sect:Spec}.  We expect that it won't make too big a
difference if, instead of averaging over the full set of totem vectors,
$\Psi\in {\cal H}_M$, we consider a suitable restricted
average.  To motivate the restriction that we shall make, we notice
first, that, if we expand such vectors, $\Psi$, as in (\ref{Psibigtotem}),
then it follows from (\ref{meanrhoepsilon}) that the (unrestricted) average value
(i.e.\ over Haar measure on the set of all $\Psi\in {\cal H}_M$) of the trace
of each $r^{\mathrm S}_\epsilon$ (defined in the paragraph after
(\ref{reducedrhomodexact})) is given by
\begin{equation}
\label{avtracerhoepsilon}
\langle {\rm tr}(r^{\mathrm S}_\epsilon)\rangle
\ \ \left
(= \left\langle\sum_{i=1}^{n_{\mathrm S}(\epsilon)}\sum_{j=1}^{n_{\mathrm B}(E-\epsilon)}
|c^{ij}_\epsilon|^2\right\rangle\right )
=\frac{n_{\mathrm S}(\epsilon)n_{\mathrm B}(E-\epsilon)}{M} \ \ (=P_{\mathrm S}(\epsilon)\Delta)
\end{equation}
-- the equality in parenthesis following from (\ref{reducedrhomodexact}).

In view of this, we take our restricted average  to be over vectors,
$\Psi\in {\cal H}_M$,  such that, in the expansion, (\ref{Psibigtotem}),
for each $\epsilon$, the coefficients $c^{ij}_\epsilon$ are constrained
to satisfy exactly
\[
\sum_{i=1}^{n_{\mathrm S}(\epsilon)}\sum_{j=1}^{n_{\mathrm B}(E-\epsilon)}
|c^{ij}_\epsilon|^2 \ \ (= {\rm tr}(r^{\mathrm S}_\epsilon))
=\frac{n_{\mathrm S}(\epsilon)n_{\mathrm B}(E-\epsilon)}{M}.
\]
(We remark that, in view of what we explained in the previous two paragraphs,
we could alternatively restrict much further and simply average over
$\Psi$ in (\ref{Psibigtotem}) for which every $c^{ij}_\epsilon$ takes
the form $e^{i\theta^{ij}_\epsilon}/\sqrt M$ and still be able to arrive
at similar conclusions to those below.  However the restriction we adopt
has the advantage of allowing us to directly use the Lubkin-Page approximation
in exactly the form (\ref{LubkinPagemeanentropy}).)
In other words, denoting $n_{\mathrm S}(\epsilon)n_{\mathrm B}(E-\epsilon)/M$
by $\mu_\epsilon$, we average over $\Psi\in {\cal H}_M$ which take the form
$\oplus_{\epsilon=0}^{E} \sqrt\mu_\epsilon \Psi^M_\epsilon$ (each
$\Psi^M_\epsilon$ being normalized) where we
regard ${\cal H}_M$ as the direct sum, $\oplus_{\epsilon=0}^E {\cal H}^M_\epsilon$,
where, for each $\epsilon$ ($=0, \Delta, \dots, E$), ${\cal
H}^M_\epsilon$ denotes the ($n_{\mathrm S}(\epsilon)n_{\mathrm
B}(E-\epsilon)$-dimensional) Hilbert subspace of ${\cal H}_M$ spanned by
the vectors $|\epsilon, i\rangle|E-\epsilon, j\rangle, i=1\dots
n_{\mathrm S}(\epsilon), j=1\dots n_{\mathrm B}(E-\epsilon)$ in ${\cal H}^{\mathrm S}_\epsilon\otimes {\cal H}^{\mathrm B}_{E-\epsilon}$ -- see after equation
(\ref{reducedrhomodexact}).  For such restricted $\Psi$, $\rho^{\mathrm{modern}}_{\mathrm S}$ will take the form
\[
\rho^{\mathrm{modern}}_{\mathrm S}=\sum_{\epsilon=\Delta}^E
\mu_\epsilon{R^{\mathrm S}_\epsilon}
\]
where $R^{\mathrm S}_\epsilon$ is the partial trace of
$|\Psi^M_\epsilon\rangle\langle\Psi^M_\epsilon|$ over ${\cal
H}^B_{E-\epsilon}$ (which will equal $r^{\mathrm S}_\epsilon$ divided by its
trace, which is $\mu_\epsilon$).  Clearly, by the lemma in Section \ref{Sect:Gen}, we therefore have
\begin{equation}
\label{SrhoS}
S(\rho^{\mathrm{modern}}_{\mathrm S})=S \left (\sum_{\epsilon=\Delta}^E
\mu_\epsilon S(R^{\mathrm S}_\epsilon)\right )=\sum_{\epsilon=\Delta}^E \mu_\epsilon S(R^{\mathrm
S}_\epsilon)-\sum_{\epsilon=\Delta}^E \mu_\epsilon\log\mu_\epsilon.
\end{equation}
But now we notice that, if we identify $m$ with
$n_{\mathrm S}(\epsilon)$ and $n$ with $n_{\mathrm B}(E-\epsilon)$, then
we can identify ${\cal H}^M_\epsilon$ with the
Hilbert space, ${\cal H}_m\otimes {\cal H}_n$, of Sections \ref{Sect:Intro} and \ref{Sect:Haar},  and,
under this identification,  $R^{\mathrm S}_\epsilon$ is identified with
$\rho^{mn}_m$, and $S(R^{\mathrm S}_\epsilon)$ with $S(\rho^{mn}_m)$.
Moreover, averaging  $S(R^{\mathrm S}_\epsilon)$ over ${\cal
H}^M_\epsilon$ is, under (the reverse of) this identification, then obviously the same as
taking the unrestricted average of $S(\rho^{mn}_m)$ over Haar measure on
unit vectors in ${\cal H}_m\otimes {\cal H}_n$ and so we may estimate
its value using the Lubkin-Page approximation (\ref{LubkinPagemeanentropy}).
Making these identifications, if we now use `$[\quad ]$' to denote our
restricted average over our restricted totem vectors, $\Psi$, and
`$\langle\quad\rangle$' to denote the unrestricted average over Haar
measure on unit vectors in ${\cal H}^M_\epsilon$ for each $\epsilon$,
we may calculate using the formula (\ref{SrhoS}) in our lemma of
Section \ref{Sect:Gen}:
\[
[S(\rho^{\mathrm{modern}}_{\mathrm S})]=\left[S \left
(\sum_{\epsilon=\Delta}^E \mu_\epsilon S(R^{\mathrm S}_\epsilon)\right
)\right]=\sum_{\epsilon=\Delta}^E \mu_\epsilon \langle S(R^{\mathrm
S}_\epsilon)\rangle -\sum_{\epsilon=\Delta}^E \mu_\epsilon\log\mu_\epsilon
\]
which, recalling that
$\mu_\epsilon=n_{\mathrm S}(\epsilon)n_{\mathrm B}(E-\epsilon)/M$
and using (\ref{LubkinPagemeanentropy}), equals
\begin{eqnarray}\nonumber
&&\sum_{\epsilon=\Delta}^E \frac{n_{\mathrm S}(\epsilon)n_{\mathrm B}(E-\epsilon)}{M}
\left (\log ({\rm min}(n_{\mathrm S}(\epsilon), n_{\mathrm B}(E-\epsilon)))-
\frac{{\rm min}(n_{\mathrm S}(\epsilon), n_{\mathrm B}(E-\epsilon))}
{2{\rm max}(n_{\mathrm S}(\epsilon), n_{\mathrm B}(E-\epsilon))}\right.
\\&&\nonumber
-\left.\log\left(\frac{n_{\mathrm S}(\epsilon)n_{\mathrm B}(E-\epsilon)}{M}\right)
+O\left (\frac{1}{n_{\mathrm S}(\epsilon)n_{\mathrm B}(E-\epsilon)} \right)\right )
\end{eqnarray}
which easily simplifies to $[S(\rho^{\mathrm{modern}}_{\mathrm S})]$
\[
=-M^{-1}\left(\sum_{\epsilon=\Delta}^{E_c} n_{\mathrm{S}}
(\epsilon)n_{\mathrm{B}}(E-\epsilon)\log(M^{-1}n_{\mathrm{B}}(E-\epsilon))
+ \sum_{\epsilon=E_c+\Delta}^E n_{\mathrm{S}}
(\epsilon)n_{\mathrm{B}}(E-\epsilon)\log(M^{-1}n_{\mathrm{S}}(\epsilon))
\right )
\]
\begin{equation}
\label{Switherror}
-M^{-1}\left(\sum_{\epsilon=\Delta}^{E_c} \frac{n_{\mathrm S}(\epsilon)^2}{2} +
\sum_{\epsilon=E_c+\Delta}^E \frac{n_{\mathrm B}(E-\epsilon)^2}{2}\right)
+O(1)
\end{equation}
where $E_c$ is as defined after (\ref{purereduced}).

Comparing (\ref{Switherror}) with (\ref{pureentropy}), we notice that
the first line of (\ref{Switherror}) coincides with the formula,
(\ref{pureentropy}), $S^{\mathrm{modapprox}}_{\mathrm{S}}$ for the von
Neumann entropy  of $\rho_{\mathrm S}^{\mathrm{modapprox}}$ which we
derived from (\ref{purereduced}).  Thus we may conclude that our
restricted average over totem vectors of
$S^{\mathrm{modern}}_{\mathrm{S}}$ will be given by the formula we gave
for  $S_{\mathrm S}^{\mathrm{modapprox}}$ in (\ref{pureentropy}) --
plus an `error term' given by the last line of (\ref{Switherror}).
Moreover, the close agreement found above between  $S_{\mathrm
S}^{\mathrm{modern}}$ and $S_{\mathrm S}^{\mathrm{modapprox}}$ in the
`worst case scenario' discussed above, strongly suggests that the same
statement will be true for the unrestricted average.
In order to conclude that this amounts to an independent check of the
correctness of the approximate formula 
$S^{\mathrm{modern}}_{\mathrm{S}}$ of (\ref{pureentropy}) for our
densities of states of interest, (\ref{sigmasbpower}),
(\ref{sigmasbexponential}), (\ref{sigmasbexpsquared}), it remains to
show that (/investigate when) the `error term' (i.e.\ the second line in
(\ref{Switherror})) is small.  To end this section we turn to this last
question:

It is in fact easy to see (after converting the sum to an integral, using (\ref{continuum})) that: (a) for our power-law densities of states, (\ref{sigmasbpower}), with $A_{\mathrm{S}}=A_{\mathrm{B}}$ and $N_{\mathrm{S}}=N_{\mathrm{B}}=N$ say, the last line of (\ref{Switherror}) (minus the $O(1)$ term) is (using Stirling's approximation -- see Section \ref{Sect:Power}) $1/\sqrt{\pi N}$; (b) for our (equal) exponential densities of states, (\ref{sigmasbexponential}), it is $(1/bE)(1-e^{-bE})$; and (c) for our (equal) quadratic densities of states, (\ref{sigmasbexpsquared}),  it is (approximating the integral with the leading term of the asymptotic formula in Endnote \cite{Ftnt9})  $\exp(-qE^2/2)$.   These terms will all be much smaller than typical values of the first line of (\ref{Switherror}) provided $N$ is large in (a), provided $E\gg 1/b$ (cf.\ Equation (\ref{bEgg1})) in (b), and provided $E\gg 1/\sqrt q$ (cf.\ Equation (\ref{qEsquaredbig})) in (c).

So in all cases of interest here, and, no doubt, in many others too, the last line of (\ref{Switherror}) will be negligibly small.

\begin{acknowledgments}
I thank Michael Kay for useful discussions about traditional statistical mechanics and also for information about string theory and Richard Hall for showing me how to do integrals involving logarithms.
\end{acknowledgments}


\begin{thebibliography}{99}
\bibitem{fictTemp} In the case the energy probability density is sharply peaked, $T(\epsilon)$ would then be interpretable as the temperature of a small subsystem or of a small system placed in thermal contact with the given system. See Section \ref{Sect:peak}.
\bibitem{Ftnt1} Actually Lubkin was by no means the first to adopt the
natural invariant measure (`Haar measure') in
a related context.  See the discussion and references in \cite{GoldsteinLebowitzetal}.
\bibitem{modern} Throughout the paper, we will refer to explanations of the origin of thermality, based on the assumption of a total pure state, as being `modern' (in contradistinction to the `traditional' assumption of a microcanonical ensemble).
\bibitem{spurious} Throughout this paper, we shall often consider, side-by-side with our finite-sum formulae, continuum approximations where we make the replacement (\ref{continuum}).
Usually, the formulae are essentially interchangeable and one can (and it can be helpful to) check they give the same answers.   However, we caution that, for example, the continuum versions of the entropy formulae which we obtain, using the prescription (\ref{continuum}), in Section \ref{Sect:Entropy} can end up having energy-intervals on which the `number of energy levels' is greater than zero but less than 1 and this can result in spurious negative terms which go like the logarithm of (a constant with the dimensions of inverse energy times the) total energy $E$.   An example of such a term occurs in Section \ref{Sect:ExpEsq} as is mentioned there.
\bibitem{Ftnt4} $\beta$ arises, in general as a suitable `large-size' limit of
$d\log\sigma_{\mathrm B}(E)/dE$.  For details on   how the passage from
microcanonical to canonical, which we have illustrated in detail in the
text only for power-law energy-bath densities of states, may be
generalized to a wider class of energy-bath models, we refer again to
\cite{GoldsteinLebowitzetal} and to \cite{Thirring}.
\bibitem{rigour} Our proposition in Section \ref{Sect:Ans} is not quite a proposition in the sense of pure mathematics and, indeed, it is, in some respects (deliberately) vague.  In particular, the phrase ``as far as physical quantities of interest are concerned'' needs clarifying.  What we do expect (and give evidence for in Part~2) is that we may (with high probability and to a very good approximation) at least replace the actual entropy
$-{\rm tr}
(\rho^{\mathrm{modern}}_{\mathrm{S}}\log\rho^{\mathrm{modern}}_{\mathrm{S}}))$ by $-{\rm tr}
(\rho^{\mathrm{modapprox}}_{\mathrm{S}}\log\rho^{\mathrm{modapprox}}_{\mathrm{S}}))$ and the moments of energy -- i.e.\  $\overline{\epsilon^n}={\rm tr}
(\rho^{\mathrm{modern}}_{\mathrm{S}}H_{\mathrm{S}}^n)$ by ${\rm tr}
(\rho^{\mathrm{modapprox}}_{\mathrm{S}}H_{\mathrm{S}}^n)$ (=$\int_0^E \epsilon^n P_{\mathrm S}(\epsilon)d\epsilon$ $={\rm tr}
(\rho^{\mathrm{microc}}_{\mathrm{S}}H_{\mathrm{S}}^n)$ -- cf.\ Equation \ref{moments} in Section \ref{Sect:Power}).  It remains a significant challenge to remove the vagueness and to provide a rigorous proof.  However, we expect that such a rigorous proof could be based on the arguments we give in Part~2.
\bibitem{comparable} Once our `system', `S', and `energy-bath', `B',  are of comparable size, these notions of course become interchangeable and we might alternatively, e.g., call them `subsystem A' and `subsystem B'.  However we retain the system and energy-bath terminology and notation to maintain contact with the literature in which S is assumed smaller than B.  One drawback is that we use the same letter for both system and entropy but we do make them distinguishable by reserving a roman font (S) for system and italic ($S$) for entropy.
\bibitem{Ftnt7} Our Gaussian approximation (\ref{binomapprox}) to
$b(k;n,p)$ (\ref{binomial}), is not the same as the more usual approximation
\cite{Feller} which relates the Binomial and Gaussian distributions but
it has an amusing relationship to it.  The latter, more usual,
approximation is, when (for simplicity) we again assume that $pn$ is an
integer, the statement
\begin{equation}
\label{Gaussapprox}
b(pn+\kappa;n,p)\simeq \frac{1}{\sqrt{2\pi
np(1-p)}}\exp\left(-\frac{\kappa^2}{2np(1-p)}\right ).
\end{equation}
To understand the relationship between (\ref{binomapprox}) and
(\ref{Gaussapprox}) consider, for example the case where $n=100$ and
$p=\frac{1}{2}$.  Then (\ref{Gaussapprox}) with, say, $\kappa=2$ tells
us that if one tosses a fair coin a hundred times, then the probability
of getting 52 heads and 48 tails (or vice versa) is approximately
$\exp(-2/25)$ times the probability of getting 50 heads and 50 tails.
On the other hand, (\ref{binomapprox}) with $x=\pm 1/50$ tells us that
the same quantity  (i.e.\ $\exp(-2/25)$ times the probability of getting
50 heads and 50 tails when tossing a fair coin a hundred times) is also
an approximation to the probability of getting exactly 50 heads and 50
tails when one tosses a hundred times a biased coin which has a
probability $52/100$ of landing heads on each throw (or of landing
tails on each throw). The fact that these two distinct probabilities are
approximately equal is, arguably, intuitively obvious.
\bibitem{Ftnt8} In fact, allowing different values of $c$
($c_{\mathrm{S}}$ and $c_{\mathrm{B}}$) in the obvious places in
(\ref{sigmasbexponential}) will, provided $c_{\mathrm{S}}$ and
$c_{\mathrm{B}}$ remain of comparable size, only introduce small
corrections to the quantities we calculate.  In particular, as may be
readily checked.  (Hint: write $c_{\mathrm{S}}=ce^\mu$ and
$c_{\mathrm{B}}=ce^{-\mu}$.  Then $E_c$ will be $E/2 - \mu$) the
expected energies of S and B will remain exactly unchanged as $E/2$,
while there will be only small corrections to the entropies.  In
particular, $S_{\mathrm{S}}^{\mathrm{modern}}$ generalizes from
$bE/4+\log(cE)$ to $bE/4+\log(c_{\mathrm{S}}c_{\mathrm{B}}E^2)/2
-(\log(c_{\mathrm{S}}/c_{\mathrm{B}}))^2/4E$. Different values of $b$
will, as might be expected, have a more drastic effect. See also
\cite{Kayprefactor} where we will consider densities of states of the
form $\sigma(\epsilon) = c \epsilon^{-p} e^{b\epsilon}$ -- i.e.\ with inverse-power
prefactors such as occur in string theory.
\bibitem{Ftnt9} In fact we have $M=EK^2\Delta e^{qE^2}(1/qE^2 +
O(1/q^2E^4))$ whereupon
$P_{\mathrm{S}}(\epsilon)=E^{-1}(qE^2+O(1))\exp(-2q\epsilon(E-\epsilon))$.
 This may easily be seen by a direct calcution of $M$ as given by
(\ref{intnorm}) with (\ref{Mexpenergysquared}), on using the well-known
asymptotic formula $\int_{-a}^a\exp(x^2) dx = \exp(a^2)(1/a +
O(1/a^3))$.
\bibitem{guess} Before we can take the logarithm of the long-string density of states, we would need to multiply it by a constant with the dimensions of energy so that the argument of the logarithm is dimensionless.  Our criticism is that the Horowitz Polchinski theory does not seem to naturally supply any such constant.  It might be countered that our own work does not single out a preferred value of $\Delta$.  But this would not be a comparably serious criticism: The role of $\Delta$ in our work is only to help us to make an `educated guess' as to the totem state $\Psi$ by choosing a $\Psi$ at random from our energy band $[E, E+\Delta]$.  That the choice of $\Delta$ is `harmless' is confirmed by the fact (see Section \ref{Sect:Gen}) that our formulae for mean energy and entropy are independent of $\Delta$.
\bibitem{Ftnt3}  The equality of $S(\rho^{mn}_m)$ and $S(\rho^{mn}_{\ \ n})$ follows by the well-known general result that, for a pure total
state on any bipartite system, the von Neumann entropies of the two
reduced density operators are necessarily always equal.  As is
well-known, this is easily proven from the Schmidt decomposition which,
by the way, we shall recall, in passing, in another context, in Section \ref{Sect:Haar}.
In fact, by (\ref{Schmidt}), given an arbitrary state vector, 
$\Psi\in {\cal H}_m\otimes {\cal H}_n$ (${\cal H}_m$ and ${\cal H}_n$ any $m$- and $n$-dimensional Hilbert spaces) we clearly have that the reduced density operator, $\rho^{mn}_m$,
of $|\Psi\rangle\langle\Psi|$ on ${\cal H}_m$ is
\[
\sum_{i=1}^\nu\lambda_i|\tilde e_i\rangle\langle \tilde e_i|,
\]
while the reduced density operator, $\rho^{mn}_{\ \ n}$, of $|\Psi\rangle\langle\Psi|$ on ${\cal H}_n$
is
\[
\sum_{i=1}^\nu\lambda_i|\tilde f_i\rangle\langle \tilde f_i|,
\]
where $\nu={\mathrm{min}}(m,n)$.  Clearly, the von Neumann entropy of each of these reduced density operators is $-\sum_{i=1}^\nu \lambda_i\log\lambda_i$.
\end{thebibliography}
\end{document}